\documentclass{ucetd}
\usepackage[utf8]{inputenc}
\usepackage[T1]{fontenc}
\usepackage{subcaption,graphicx}
\usepackage{natbib}
\usepackage{mathtools}  
\usepackage{amssymb}    
\usepackage{amsthm}

\usepackage[braket, qm]{qcircuit}
\usepackage{amsmath}
\usepackage{tikz}

\newcommand*\qcontrolcolor[2]{\push{\footnotesize \tikz[baseline=(char.base)]{
                              \node[shape=circle,draw,inner sep=0.6pt,color=#1] (char) {#2};}}
                              \qw}  
\newcommand*\qcontrol[1]{\qcontrolcolor{black}{#1}}
\newcommand*\onecontrol{\qcontrolcolor{red!80!black}{1}}
\newcommand*\twocontrol{\qcontrolcolor{blue!100!white}{2}}
\newcommand*\threecontrol{\qcontrolcolor{green!50!black}{3}}

\usepackage{pgfplots}
\pgfplotsset{compat=1.14}
\usetikzlibrary{patterns}
\usetikzlibrary{arrows}
\usetikzlibrary{external}
\usetikzlibrary{shapes}
\usetikzlibrary{fadings,shapes.arrows,shadows}
\usetikzlibrary{shadows.blur}
\usetikzlibrary{shapes.symbols}
\usetikzlibrary{shapes.multipart}
\usetikzlibrary{positioning}
\usepackage{xstring}

\usepackage[braket]{qcircuit}
\usepackage{physics}
\usepackage[boxed]{algorithm2e}
\usepackage{braket}
\usepackage{xcolor}
\usepackage{tabularx}
\usepackage{commath}
\usepackage{textcomp}
\usepackage{booktabs}
\usepackage{makecell}

\usepackage{placeins} 

\usepackage{rotating}
\usepackage{afterpage}
\usepackage{atbegshi}
\usepackage{floatpag}
\usepackage{zref-abspage,zref-user}
\makeatletter
\newcommand{\rotateFigurePageForLabel}[1]{%
  \zlabel{#1}%
  \AtBeginShipout{%
  \ifnum\c@page=\zref@extractdefault{#1}{abspage}{0}
    \pdfpageattr{/Rotate 90}
  \fi}
}
\makeatother

\usepackage{tikz}
\usetikzlibrary{automata, positioning, arrows, decorations.markings}
\tikzset{myptr/.style={decoration={markings,mark=at position 1 with %
    {\arrow[scale=2,>=stealth]{>}}},postaction={decorate}}}

\newcommand{\colorQubit}{blue}
\newcommand{\colorQubitBB}{gray}
\newcommand{\colorQutrit}{orange!70!red}
\newcommand{\colorLightness}{60}
\newcommand{\colorQubitLight}{\colorQubit!\colorLightness}

\newcommand{\colorQutritLight}{\colorQutrit!\colorLightness}
\definecolor{brown}{RGB}{77, 51, 0}

\newcommand{\hide}[1]{}
\tikzfading[name=arrowfading, top color=transparent!0, bottom color=transparent!95]
\tikzset{arrowfill/.style={#1,
    }}
\tikzset{arrowstyle1/.style n args={0}{%
    arrowfill={top color=plotBlue!20!white,bottom color=plotBlue!70!white, shape border rotate=270},
    single arrow,
    minimum height=1.4cm,
    minimum width=1.6cm,
    single arrow head extend=0.2cm,}}
\tikzset{arrowstyle2/.style n args={0}{%
    arrowfill={top color=plotGreen!20!white,bottom color=plotGreen!70!white, shape border rotate=270},
    single arrow,
    minimum height=1.4cm,
    minimum width=1.6cm,
    single arrow head extend=0.2cm,}}
\tikzset{arrowstyle3/.style n args={0}{%
    arrowfill={top color=plotRed!20!white,bottom color=plotRed!70!white, shape border rotate=270},
    single arrow,
    minimum height=1.4cm,
    minimum width=1.6cm,
    single arrow head extend=0.2cm,}}
\definecolor{plotBlue}{rgb}{0.37894736842105264, 0.6947368421052632, 0.9947368421052631}
\definecolor{plotGreen}{rgb}{0.5894736842105263, 0.8894736842105263, 0.23684210526315788}
\definecolor{plotRed}{rgb}{0.9473684210526315, 0.29473684210526313, 0.531578947368421}

\newcommand\figscale{1}
\newcommand\figfirstextrascale{1}
\newcommand\figcompactextrascale{1}
\newcommand\figscaleplot{0.95}
\newcommand\sensitivitytwowidth{0.322}

\usepackage[final]{changes}  
\newcommand{\addcolor}{blue!70!green}
\setaddedmarkup{\textcolor{\addcolor}{\uwave{#1}}}
\setdeletedmarkup{\textcolor{red!70!black}{\sout{#1}}}
\setdeletedmarkup{}

\newcommand{\thesistitle}{New Abstractions for Quantum Computing}
\newcommand{\thesisauthor}{Casey Duckering}
\department{Computer Science}
\division{Physical Sciences}
\degree{Doctor of Philosophy}
\date{December 2022}

\title{\thesistitle}
\author{\thesisauthor}

\dedication{Dedication Text}
\epigraph{Epigraph Text}

\usepackage{doi}
\usepackage{xurl}
\hypersetup{bookmarksnumbered,
            colorlinks,
            linkcolor=blue,
            citecolor=blue,
            urlcolor=blue,
            linktoc=page,
            pdftitle={\thesistitle},
            pdfauthor={\thesisauthor},
            pdfsubject={Casey Duckering's PhD dissertation at the University of Chicago, Physical Science Division, Computer Science Department, December 2022},
            pdfkeywords=quantum quantum-computing abstraction qubit qutrit qudit ququart quantum-memory compiler nisq fault-tolerant,
            pdfborder={0 0 0}}
\makeatletter
\let\ORG@hyper@linkstart\hyper@linkstart
\protected\def\hyper@linkstart#1#2{%
  \lowercase{\ORG@hyper@linkstart{#1}{#2}}}
\makeatother

\floatpagestyle{plain}

\begin{document}
\maketitle

\makecopyright{}

\tableofcontents
\listoffigures
\listoftables

\acknowledgments{}
I would like to thank my advisor, Fred Chong, for his constant mentorship and support throughout my PhD.
Thanks also to my dissertation committee members Hank Hoffman and Ken Brown for their time and valuable feedback on my systems and error correction ideas.
I am grateful to Craig Gidney at my summer internships for teaching me all his quantum tricks that have served me well during my PhD.

None of this research would exist without the friendship, collaboration, and casual conversations with my research group and other co-authors: Adam, Adrian, Alex, Andrew, Ben, Claire, Dan, David, Gokul, Hele, Jonathan, Josh, Kartik, Kate, Kunal, Max, Natalie, Pranav, Reza, Rohan, Ryan, Siddharth, Sophia, Soumik, Yongshan, and Yunong.
All the friends I've made during my time in Chicago, especially everyone at Tea Time, the Ministry, and Team Beer, have made my time in Chicago worthwhile.
Finally, thanks to my family for their love and support.

~

This research is funded in part by EPiQC, an NSF Expedition in Computing, under grants CCF-1730449/1832377; in part by STAQ, under grant NSF Phy-1818914; in part by DOE grants DE-SC0020289 and DE-SC0020331; and in part by NSF OMA-2016136 and the Q-NEXT DOE NQI Center.
This research used resources of the Oak Ridge Leadership Computing Facility, which is a DOE Office of Science User Facility supported under Contract DE-AC05-00OR22725.
Disclosure: F. Chong is also Chief Scientist for Quantum Software at ColdQuanta and an advisor to Quantum Circuits, Inc.

\abstract{}
The field of quantum computing is at an exciting time where we are constructing novel hardware, evaluating algorithms, and finding out what works best.
As qubit technology grows and matures, we need to be ready to design and program larger quantum computer systems.
An important aspect of systems design is layered abstractions to reduce complexity and guide intuition.
Classical computer systems have built up many abstractions over their history including the layers of the hardware stack and programming abstractions like loops.
Researchers initially ported these abstractions with little modification when designing quantum computer systems and only in recent years have some of those abstractions been broken in the name of optimization and efficiency.

We argue that new or quantum-tailored abstractions are needed to get the most benefit out of quantum computer systems.
We keep the benefits gained through breaking old abstraction by finding abstractions aligned with quantum physics and the technology.
This dissertation is supported by three examples of abstractions that could become a core part of how we design and program quantum computers: third-level logical state as scratch space, memory as a third spacial dimension for quantum data, and hierarchical program structure.

\mainmatter{}

\chapter{Introduction}

Moore's Law and the expectation that computers double in speed every 18 months is at an end, so hard problems in chemistry, physics simulation, and combinatorial optimization cannot be solved by waiting for a faster computer.
Since the end of Moore's Law, researchers have been developing special-purpose accelerators to squeeze better performance out of each transistor.
However once fully realized, quantum computers can solve specific classes of problems in simulation and cryptography exponentially faster.

Quantum computers work by harnessing quantum physics instead of classical Newtonian physics.
Because quantum physics is a superset of classical physics, we often treat quantum computers as classical computers with the additional features of \emph{superposition}, \emph{entanglement}, and \emph{interference}.
This view is apparent in Shor's algorithm (\cite{shor}) which creates a quantum superposition, followed by classical arithmetic, and finishes with quantum phase estimation.

Seeing quantum programming through a classical lens can be limiting and sometimes harmful.
It is common for programmers who are new to quantum to invent a ``quantum algorithm'' that is simply a randomized classical algorithm run on a quantum computer, using quantum measurements as random number generators.
More subtly, concepts such as binary representation of data, random access memory, and hierarchical modularity of programs when used in the design of quantum computers limit the performance due to mismatches with the underlying technology.
Even classical concepts of causality and movement of data can be limiting; quantum teleportation, a quantum protocol described in~\cite{mikeike}, moves quantum data long distances by pre-transferring another resource \emph{before} the data exists.

When we design quantum architectures and compilers, the abstractions we use are key to a good design.
The abstraction of two-level \emph{bits} is very beneficial for classical computer reliability but is yet to be decided for quantum.
Early classical computers used base-10 addresses and arithmetic until early computer architects settled on binary as the most efficient and reliable design.
This history informs the general assumption that binary (base-2) is best for quantum computers, but that is not necessarily the case.
We discuss this further in Chapter~\ref{chapter:qudits}.

Because quantum computing is a rapidly developing field with many competing technologies there is no clear ``best'' for any use case.
Each quantum technology has capabilities and constraints that inform a variety of hardware designs and architectures that show how to turn a qubit technology into a practical quantum computer.
The principles of abstraction and modularity we use to build any complex system still apply when we design a quantum computer hardware layout, instruction set, compiler, and programming language, but we must tailor the abstractions to best fit the physics and the technology or we will limit future efficiency.

This dissertation presents three cases of new or old abstractions that we have tailored for quantum computing.
We discuss the methodologies to select these abstractions and how we use them with a particular class of quantum architectures.
We show that good abstractions can allow more space efficient algorithms and more effective compilers.

This dissertation is comprised of three core papers introducing three abstractions covered in the following chapters.
Additional content from other work is included that shows further benefits and refinement to the abstractions.
We start in Chapter~\ref{chapter:qudits} by introducing three-level quantum \emph{trits} and other d-level quantum \emph{dits}: \emph{Asymptotic Improvements to Quantum Circuits via Qutrits},~\cite{gokhale2019asymptotic} and \emph{Efficient Quantum Circuit Decompositions via Intermediate Qudits},~\cite{baker2020compress}.
These abstractions replace and augment the use of binary qubits with three-level \emph{qutrits} or d-level \emph{qudits}, but require us to completely rethink how algorithms and compilers allocate and use scratch space.
Most quantum technologies can reliably support three or more quantum states with minor changes to the control signal design and no change to the hardware design.
Supported technologies include superconducting transmon, ion trap, and neutral atom, but notably not some types of photonic qubits.

Chapter~\ref{chapter:memory} considers abstractions that spatially separate quantum data storage or memory from computation on that data: \emph{Virtualized Logical Qubits: A 2.5D Architecture for Error-Corrected Quantum Computing},~\cite{vlq}.
Classical computers contain high speed buses that can transfer data between memory (RAM) and computation (CPU), but this extreme separation of memory from compute does not make sense either for current small (NISQ) or for future (fault-tolerant) quantum computers.
The typical abstraction for both kinds is a monolithic 2D array of qubits because NISQ computers cannot sacrifice the data-parallelism and fault-tolerance requires constant error correction to prevent errors.
Compiler design is simple in this monolithic model because there is no heterogeneity; compilers can place related data nearby in the plane.
But we compare an alternative to the monolithic model.
We redesign the surface code to use small amounts of distributed memory and find that it improves the space efficiency of fault-tolerant algorithms.

Classical programmers have used a hierarchy of function calls and modules in the design of a program to great effect.
Hierarchy gives structure to what would otherwise be a very long list of primitive instructions.
Compilers use this structure to guide optimizations and to avoid duplicate work of repeated components.
However, quantum programmers currently trend toward highly hand-optimized programs with no hierarchy; they use optimization passes that erase any hierarchy and perform flat, program-wide optimizations.
Chapter~\ref{chapter:modular} introduces \emph{Orchestrated Trios: Compiling for Efficient Communication in Quantum Programs with 3-Qubit Gates},~\cite{trios},
to show that hierarchy can guide quantum compiler heuristics even for small- to mid-size programs.
Program hierarchy enables sequences of compiler passes to repeat for each level, improving heuristic performance and allowing new kinds of passes like our connectivity-aware split pass.
This is key for quantum where data locality constraints restrict data movement and can inform program structure.

Picking the right abstractions are crucial for quantum programming, compiling, and execution.
Chapter~\ref{chapter:conclusion} concludes with a discussion and other places where we still need better abstractions.

\chapter{Beyond Binary}%
\label{chapter:qudits}

\section{Introduction}

Recent advances in both hardware and software for quantum computation have demonstrated significant progress towards practical outcomes.
While early research efforts focused on longer-term systems employing full error correction to execute large programs for algorithms like~\cite{shor} and~\cite{grover}, recent work has focused on NISQ (Noisy Intermediate Scale Quantum,~\cite{nisq}) computation.
The NISQ regime considers near-term machines with just tens to hundreds of quantum bits (qubits) and moderate errors.

In the NISQ regime, quantum programs rely directly on the individual qubits in the quantum device and severe resource constraints prohibit the use of error correction.
Given the severe constraints on quantum resources, it is critical to fully optimize the compilation of a quantum program in order to have successful computation.
Prior architectural research on techniques such as mapping, scheduling, and parallelism (\cite{adam-magic-estimates, Parallelism, scheduling1}) have helped to extend the amount of useful computation possible, but without error correction, programs are exposed to noise and errors in their qubits.
On the flip side, programs in the NISQ regime can directly take advantage of typically unused technology capabilities.

This chapter shows how to greatly reduce resource requirements by replacing the binary abstraction required by two-level qubits with a new abstraction enabled by three-level \emph{qutrits} or multi-level \emph{qudits}.
Qutrits and qudits are natural features of technologies in the NISQ regime, which we evaluate, but the takeaways from this chapter may require further research to apply in an error-corrected setting.

While quantum computation is typically expressed as a two-level binary abstraction of qubits, the underlying physics of quantum systems are not intrinsically binary.
Whereas classical computers operate in binary states at the physical level (e.g.\ clipping above and below a threshold voltage), quantum computers have natural access to an infinite spectrum of discrete energy levels. In fact, hardware must actively suppress higher level states in order to achieve the two-level qubit approximation.
Hence, using three-level qutrits is simply a choice of including an additional discrete energy level, albeit at the cost of more opportunities for error.

Prior work on qutrits (or more generally, d-level \textit{qudits}) identified only constant factor gains from extending beyond qubits.
In general, the prior work~\cite{Pavlidis} has emphasized the information compression advantages of qutrits.
For example, $N$ qubits can be expressed in base-3 ternary as $\frac{N}{\log_2(3)}$ qutrits, which leads to $\log_2(3)\approx 1.6$-constant factor improvements in space and runtime.

This chapter evaluates the benefits of a novel abstraction that uses qutrits in a novel fashion.
We use the first two states as usual to represent computed values in binary but use the third state as temporary storage when needed.
The per-operation error rate of qutrit operations is higher but the runtime (i.e.\ circuit depth or critical path) is \textit{asymptotically} faster, and the overall reliability of computations is improved due to the novel temporary storage.
Moreover, this abstraction only applies qutrit operations in an intermediary stage: the input and output are still qubits, which is important for initialization and measurement on real devices (\cite{HesingA, HesingB}) and reduces the burden to transition to the new abstraction.

We consider the benefits of different applications of this temporary qutrit abstraction.
The first application we consider is a novel implementation of the generalized Toffoli circuit by~\cite{gokhale2019asymptotic}, a subroutine used in many quantum algorithms.
By cleverly storing intermediate computations in the unused third state of input qubits, our implementation avoids the use of costly additional temporary qubits (called ancilla), but it achieves the speed of the fastest implementations that require many ancilla qubits.

In contrast, we also consider potential automated uses of temporary qutrits.
The hand-designed generalized Toffoli implementation makes excellent use of one additional logical state and, while hand-optimization can be a good way to squeeze performance out of resource-constrained devices, codifying manual strategies into our compilers can have wider performance benefit and free most programmers to think at a higher level.
By intelligently ``compressing'' the data in groups of idle qubits into smaller groups of qutrits as in~\cite{baker2020compress}\footnotemark{} (using the $\log_2(3)$ compression ratio) or qudits ($\log_2(d)$ ratio), similar benefits for resource-constrained quantum computers can be achieved for a wider range of quantum programs.
\footnotetext{CD's contributions to the works that comprise this chapter, \cite{gokhale2019asymptotic} and \cite{baker2020compress}, include the novel circuit designs (in addition to contributions from PG and JMB for the ancilla-free Generalized Toffoli and with equal contributions from JB for all others), the qudit circuit implementations, numerical simulations, validation, and simulation results.}

The main benefit of compression is to produce ancilla, specifically clean ancilla, \textit{generated} locally during the compilation of an algorithm into a quantum circuit.
That is, we propose a new circuit which performs qubit-qudit compression storing the information of many qubits as a small number of qudits at the cost of some gate overhead.
These compression circuits produce clean ancilla in the $\ket{0}$ state. The stored data can be retrieved later when needed since all quantum operations are reversible (this is technically a re-encoding, not compression).
Essentially, when certain groups of qubits will be unused for a long period of time, we can repurpose them by compressing them and using the produced ancilla.
This ``compression'' is a rearrangement of the stored binary values into higher states, letting us store more information into the same number of physical quantum devices and free up qubits for computation.
We evaluate this compression strategy in the design of an improved quantum adder circuit.

The key result of this chapter is that use of this abstraction by quantum subroutines or compilers extends the frontier of what limited-size quantum computers can compute.
In particular, the frontier is defined by the zone in which every machine qubit is a data qubit, for example a 100-qubit program running on a 100-qubit machine.
In this frontier zone, we do not have room for non-data workspace qubits known as ancilla.
The lack of ancilla in the frontier zone is a costly constraint that generally leads to inefficient circuits.
For this reason, typical circuits instead operate well below the frontier zone, with many machine qubits used as ancilla.
This chapter demonstrates that ancilla can be substituted with qutrits, enabling us to operate efficiently within the ancilla-free frontier zone.

We highlight the primary contributions of this chapter:
\begin{enumerate}
    \item A circuit construction for the generalized Toffoli subroutine that uses temporary qutrits and no ancilla qubits.  This is an asymptotically faster circuit ($633N \rightarrow 38 \log_2 N$) than equivalent qubit-only ancilla-free constructions.
    \item Qutrit and qudit ``compression'' circuit designs.
    \item A circuit construction for arithmetic addition in binary using qudit compression and no ancilla qubits.
    \item An open-source qudit circuit library and simulator, now a core feature of Google's Cirq (\cite{Cirq}).
\end{enumerate}

This chapter is organized as follows: Section~\ref{sec:qudits-background} presents relevant background about quantum computation and Section~\ref{sec:qudits-related_work} outlines related prior work that we benchmark our work against.
Section~\ref{sec:qudits-circuit-constructions} demonstrates our key circuit construction, and Section~\ref{sec:qudits-application-to-algorithms} surveys applications of this construction toward important quantum algorithms.
Section~\ref{sec:qudits-simulator} introduces our open-source qudit circuit simulator. Section~\ref{sec:qudits-error-modeling} explains our noise modeling methodology, and Section~\ref{sec:qudits-results} presents simulation results for the generalized Toffoli circuits under these noise models.

In the remainder of the chapter, we present an application of this technique to give logarithmic depth decompositions of quantum arithmetic circuits---a carry lookahead adder and, by extension, addition by a constant.
In Section~\ref{sec:qudits-compression} we present two compression circuits for qubit-qutrit and qubit-ququart ($d=4$) compression and evaluate advantages of various compression schemes.
In Section~\ref{sec:qudits-decompositions} we present our decomposition of the zero-ancilla, in-place $A+B$ adder which takes as input two registers $A$ and $B$ of qubits and possibly carry-in and carry-out; any fresh $\ket{0}$ states used are generated locally.
We then evaluate the costs of this decomposition.
We end with extensions to our arithmetic decomposition in Sections~\ref{sec:qudits-carryin_carryout} and~\ref{sec:qudits-plus_k} and finish with a discussion and summary in Section~\ref{sec:qudits-discussion}.

\FloatBarrier{}



\section{Background}%
\label{sec:qudits-background}

A qubit is the fundamental unit of quantum computation. Compared to their classical counterparts which take values of either 0 and 1, qubits may exist in a superposition of the two states. We designate these two basis states as $\ket{0}$ and $\ket{1}$ and can represent any qubit as $\ket{\psi} = \alpha\ket{0} + \beta\ket{1}$ with $\|\alpha\|^2 + \|\beta\|^2 = 1$. $\|\alpha\|^2$ and $\|\beta\|^2$ correspond to the probabilities of measuring $\ket{0}$ and $\ket{1}$ respectively.

Quantum states can be acted on by quantum gates which (a) preserve valid probability distributions that sum to 1 and (b) guarantee reversibility. For example, the X gate transforms a state $\ket{\psi} = \alpha\ket{0} + \beta\ket{1}$ to $X\ket{\psi} = \beta\ket{0} + \alpha\ket{1}$. The X gate is also an example of a classical reversible operation, equivalent to the NOT operation. In quantum computation, we have a single irreversible operation called measurement that transforms a quantum state into one of the two basis states with a given probability based on $\alpha$ and $\beta$.

In order to interact different qubits, two-qubit operations are used. The CNOT gate appears both in classical reversible computation and in quantum computation. It has a control qubit and a target qubit. When the control qubit is in the $\ket{1}$ state, the CNOT performs a NOT operation on the target. The CNOT gate serves a special role in quantum computation, allowing quantum states to become entangled so that a pair of qubits cannot be described as two individual qubit states. Any operation may be conditioned on one or more controls that act like the conditions of an if-statement, only performing the operation on the states where all controls are $\ket{1}$.

\begin{figure}[h]
    \centering
    \input{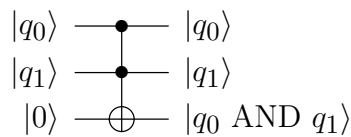}
    \caption{Reversible AND circuit using a single ancilla bit. The inputs are on the left, and time flows rightward to the outputs. This AND gate is implemented using a Toffoli (CCNOT) gate with inputs $q_0$, $q_1$ and a single ancilla initialized to 0. At the end of the circuit, $q_0$ and $q_1$ are preserved, and the ancilla bit is set to 1 if and only if both other inputs are 1.}%
    \label{fig:reversible_AND}
\end{figure}

Many classical operations, such as AND and OR gates, are irreversible and therefore cannot directly be executed as quantum gates. For example, consider the output of 1 from an OR gate with two inputs. With only this information about the output, the value of the inputs cannot be uniquely determined. These operations can be made reversible by the addition of extra, temporary workspace bits initialized to 0. Using a single additional ancilla, the AND operation can be computed reversibly as in Figure~\ref{fig:reversible_AND}.

Classical operations are fed an input state and produce an output state but quantum operations do more.
Quantum operations take a \emph{superposition} state, a complex linear combination of some or all $2^n$ classical (binary) input states and produce an output \emph{superposition} state.
For example, the quantum CNOT gate applied to a pair of control and target qubits $\ket{ct}$ transforms an input superposition $\alpha_1\ket{00}+\alpha_2\ket{01}+\alpha_3\ket{10}+\alpha_4\ket{11}$ to the output superposition $\alpha_1\ket{00}+\alpha_2\ket{01}+\alpha_3\ket{11}+\alpha_4\ket{10}=\alpha_1\ket{00}+\alpha_2\ket{01}+\alpha_4\ket{10}+\alpha_3\ket{11}$.

Physical systems in classical hardware are typically binary. However, in common quantum hardware, such as in superconducting and trapped ion computers, there is an infinite spectrum of discrete energy levels. The qubit abstraction is an artificial approximation achieved by suppressing all but the lowest two energy levels. Instead, the hardware may be configured to manipulate the lowest three energy levels by operating on qutrits. In general, such a computer could be configured to operate on any number of $d$ levels, except as $d$ increases the number of opportunities for error, termed error channels, increases. Here, we focus on $d = 3$ and later $d = 4$ with which we achieve the desired improvements to the Generalized Toffoli gate and qudit ``compression''.
For a complete guide to superconducting qubits we refer to~\cite{sc_errors}.

In a three level system, we consider the computational basis states $\ket{0}$, $\ket{1}$, and $\ket{2}$ for qutrits. A qutrit state $\ket{\psi}$ may be represented analogously to a qubit as $\ket{\psi} = \alpha\ket{0} + \beta\ket{1} + \gamma\ket{2}$, where $\norm{\alpha}^2 + \norm{\beta}^2 + \norm{\gamma}^2 = 1$. Qutrits are manipulated in a similar manner to qubits; however, there are additional gates which may be performed on qutrits.

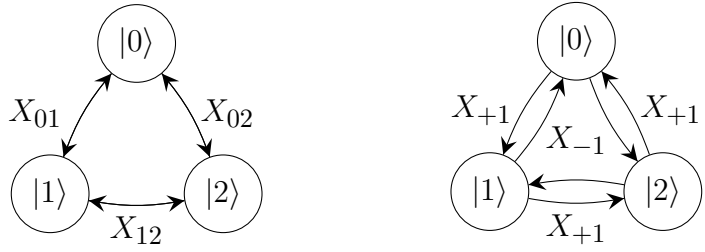
\begin{figure}[h]
    \centering
        \begin{tikzpicture}[scale=1, every node/.style={scale=1}]
            \node[state] at (0,0) (0) {$\ket{0}$};
            \node[state] at (-1.15, -2) (1) {$\ket{1}$};
            \node[state] at (1.15, -2) (2) {$\ket{2}$};
            \draw   (0) edge[myptr, bend right=10, above, left=1] node{$X_{01}~$} (1)
                    (1) edge[myptr, bend left=10, above, right=1] (0)
                    (1) edge[myptr, bend right=10, below] node{$X_{12}$} (2)
                    (2) edge[myptr, bend left=10, above]  (1)
                    (0) edge[myptr, bend left=10, below, right=1] node{$X_{02}$} (2)
                    (2) edge[myptr, bend right=10, below, right=1] (0);
        \end{tikzpicture}
        \hspace{5em}
        \begin{tikzpicture}[scale=1, every node/.style={scale=1}]
            \node[state] at (0,0) (0) {$\ket{0}$};
            \node[state] at (-1.15, -2) (1) {$\ket{1}$};
            \node[state] at (1.15, -2) (2) {$\ket{2}$};
            \node at (0,-1.333) {$X_{-1}$};
            \draw   (0) edge[myptr, above, bend right=10, left=1] node{$X_{+1}$} (1)
                    (1) edge[myptr, above, bend right=10, right=1] (0)
                    (1) edge[myptr, below, bend right=10] node{$X_{+1}$} (2)
                    (2) edge[myptr, above, bend right=10] (1)
                    (0) edge[myptr, below, bend right=10, left=1] (2)
                    (2) edge[myptr, above, bend right=10, right=1] node{$X_{+1}$} (0);
        \end{tikzpicture}
    \caption{The five nontrivial permutations on the basis elements for a qutrit. (Left) Each operation here switches two basis elements while leaving the third unchanged. These operations are self-inverses. (Right) These two operations permute the three basis elements by performing a $+1\mod{3}$ and $-1\mod{3}$ operation. They are each other's inverses.}%
    \label{fig:qutrit_xops}
\end{figure}

For instance, in quantum binary logic, there is only a single X gate. In ternary, there are three X gates denoted $X_{01}$, $X_{02}$, and $X_{12}$. Each of these $X_{ij}$ for $i \neq j$ can be viewed as swapping the amplitudes of $\ket{i}$ and $\ket{j}$ and leaving the third basis element unchanged. For example, for a qutrit $\ket{\psi} = \alpha\ket{0} + \beta\ket{1} + \gamma\ket{2}$, applying $X_{02}$ produces $X_{02}\ket{\psi} = \gamma\ket{0} + \beta\ket{1} + \alpha\ket{2}$. Each of these operations' actions can be found in the left state diagram in Figure~\ref{fig:qutrit_xops}.

There are two additional non-trivial operations on a single trit. They are the $+1$ and $-1$ (sometimes referred to as a $+2$) operations (with $+$ meaning addition modulo 3). These operations can be written as $X_{01}X_{12}$ and $X_{12}X_{01}$, respectively; however, for simplicity, we will refer to them as $X_{+1}$ and $X_{-1}$ operations. A summary of these gates' actions can be found in the right state diagram in Figure~\ref{fig:qutrit_xops}.

When we use qudits with more than three levels, there are many mores gates which can be used depending on $d$.
For a single qudit we have access to every permutation of the $d$ basis states, or $d! - 1$ nontrivial operations, but in practice, many of these operations are unnecessary and only a small number are needed for universal computation.
We make use of the increment permutations, denoted $X_{+k}$ where $+$ is addition modulo $d$, which rotates a state $\ket{i}$ to $\ket{i + k \mod d}$ and the flip permutations denoted $X_{ij}$ which flip or switch the states $\ket{i}$ and $\ket{j}$, leaving all others unchanged. $X_{01}$ is equivalent to the qubit $X$ gate.

Other, non-classical, operations may be performed on a single qudit. For example, the Hadamard gate (\cite{mikeike}) can be extended to work on qudits in a similar fashion as the X gate was extended. In fact, all single qubit gates, like rotations, may be extended to operate on qudits. In order to distinguish qubit, qutrit, and qudit gates, all non-qubit gates will appear with an appropriate subscript.

Each of these operations can be extended to two qudits as a controlled operation that applies the single-qudit operation conditioned on the control qudit being in a certain state.
For example, consider applying an $X_{+2}$ operation on a $d=4$ level system conditioned on a control qudit being in the $\ket{3}$ state.
These controlled qudits have been physically realized and they are universal for qudit computation, as shown by~\cite{MS2000}.
This can be extended to any number of controls but only two-qudit gates can be directly executed on typical quantum hardware; any use of a multi-controlled gate has a decomposition into one and two qudit gates since these gates are universal.
We only require a single 3-qubit, 2-controlled gate (Toffoli-like) whose decomposition is given by~\cite{Di} into basic one- and two-qubit gates.
We represent these gates in circuit diagrams with the control types indicated by circles with the control values inside.
The applied gates, specifically the increment ($X_{+i}$) and flip gates ($X_{ij}$) will be given as a square labeled with the name of the gate.

One question concerning the feasibility of using higher states beyond the standard two is whether these gates can be implemented and perform the desired manipulations.
Qudit gates have been successfully implemented by~\cite{Di, MS2000, Klimov2003, Chi2022}, indicating that it is possible to consider higher level systems apart from qubit only systems.

In order to evaluate an implementation of a quantum circuit, we consider quantum circuit costs.
Quantum circuits consist of a sequence of operations, also called gates, applied to a set of input qubits.
These circuits do not have fan-in or fan-out and so when represented each horizontal line in the circuit diagram corresponds to a single qubit and time flows from left to right from inputs to outputs.
The space cost of a circuit is therefore the number of qubits (or qudits) and this cost is referred to as circuit \textit{width}.
Requiring ancilla increases the circuit width and therefore the space cost of a circuit.
The time cost for a circuit is the \textit{depth} of a circuit.
The depth is the length of the critical path (in number of gates) from input to output.

\FloatBarrier{}

\section{Prior Work}%
\label{sec:qudits-related_work}

\subsection{Qudits}
Qutrits, and more generally qudits, have been been studied in past work both experimentally and theoretically. Experimentally, $d$ as large as 10 has been achieved (including with two-qudit operations) by~\cite{Kues}, and $d=3$ qutrits are commonly used internally in many quantum systems, including~\cite{Zinner2018, Wallraff2018}.

However, in past work, qudits have conferred only an information compression advantage. For example, $n$ qubits can be compressed to $\frac{n}{\log_2(d)}$ qudits, giving only a constant-factor advantage in~\cite{Pavlidis} at the cost of greater errors from operating qudits instead of qubits. Under the assumption of linear cost scaling with respect to $d$, \cite{Greentree, Khan} demonstrated that $d=3$ is optimal, although as we show in Section~\ref{sec:qudits-error-modeling} the cost is generally superlinear in $d$.

The information compression advantage of qudits has been applied specifically to Grover's search algorithm by~\cite{YaleFan, Li, Wang, Ivanov} and to Shor's factoring algorithm by~\cite{Bocharov}. Ultimately, the trade-off between information compression and higher per-qudit errors has not been favorable in past work. As such, the past research towards building practical quantum computers has focused on qubits.

We introduce qutrit-based, ancilla-free circuits which are \textit{asymptotically} better than equivalent qubit-only, ancilla-free circuits. Unlike prior work, we demonstrate a compelling advantage in both runtime and reliability, thus justifying the use of qutrits.

\subsection{Generalized Toffoli Gate Circuits}
We start by focusing on the Generalized Toffoli gate, which simply adds more controls to the Toffoli circuit in Figure~\ref{fig:reversible_AND}. The Generalized Toffoli gate is an important primitive used across a wide range of quantum algorithms, and it has been the focus of extensive past optimization work. Table~\ref{tab:n_controlled} compares past circuit constructions for the Generalized Toffoli gate to our construction, which is presented in full in Section~\ref{subsec:generalized_toffoli_construction}.

\begin{table*}[h]
\begin{tabular}{lcccc}
    & Depth & Ancilla & Qudit Types & Constants \\ \toprule
    \textbf{This Work} & $\log{n}$ & 0 & Controls are qutrits & Small \\ \midrule
    \cite{GidneyBlogPost} & $n$ & 0 & Qubits & Large \\ \midrule
    \cite{He} & $\log{n}$ & $n$ & Qubits & Small \\ \midrule
    \cite{barenco} & $n^2$ & 0 & Qubits & Small \\ \midrule
    \cite{Wang} & $n$ & 0 & Controls are qutrits & Small \\ \midrule
    \makecell[l]{\cite{Lanyon},\\\cite{Ralph}} & $n$ & 0 & Target is $d=n$-level qudit & Small \\
\end{tabular}
\caption{Asymptotic comparison of $n$-controlled gate decompositions. The total gate count for all circuits scales linearly (except for \cite{barenco}, which scales quadratically). Our construction uses qutrits to achieve logarithmic depth without ancilla. We benchmark our circuit construction against \cite{GidneyBlogPost}, which is the asymptotically best ancilla-free qubit circuit.}%
\label{tab:n_controlled}
\end{table*}

Among prior work, the \cite{GidneyBlogPost}, \cite{He}, and \cite{barenco} designs are all qubit-only. The three circuits have varying trade-offs. While Gidney and Barenco operate at the ancilla-free frontier, they have large circuit depths: linear with a large constant for Gidney and quadratic for Barenco. The Gidney design also requires rotation gates for very small angles, which can pose an experimental challenge. While the He circuit achieves logarithmic depth, it requires an ancilla for each data qubit, effectively halving the effective potential of any given quantum hardware. Nonetheless, in practice, most circuit implementations use these linear-ancilla constructions due to their small depths and gate counts.

As in our approach, circuit constructions from \cite{Lanyon}, \cite{Ralph}, and \cite{Wang} have attempted to improve the ancilla-free Generalized Toffoli gate by using qudits. Both the \cite{Lanyon} and \cite{Ralph} constructions, which have been demonstrated experimentally, achieve linear circuit depths by operating the target as a $d=n$-level qudit. \cite{Wang} also achieves a linear circuit depth but by operating each control as a qutrit.

Our circuit construction, presented in Section~\ref{subsec:generalized_toffoli_construction}, has similar structure to the He design, which can be represented as a binary tree of gates. However, instead of storing temporary results with a linear number of ancilla qubits, our circuit temporarily stores information directly in the qutrit $\ket{2}$ state of the controls. Thus, no ancilla are needed.

In our simulations, we benchmark our circuit construction against the \cite{GidneyBlogPost} construction because it is the asymptotically best qubit circuit in the ancilla-free frontier zone. We label these two benchmarks as \textcolor{black}{QUTRIT} and \textcolor{black}{QUBIT}. The \textcolor{black}{QUBIT} circuit handles the lack of ancilla by using \textit{dirty} ancilla, which unlike \textit{clean} (initialized to $\ket{0}$) ancilla, can have an unknown initial state. Dirty ancilla can therefore be bootstrapped internally from a quantum circuit. However, this technique requires a large number of Toffoli gates which makes the decomposition particularly expensive in gate count.

Augmenting the base Gidney construction with a single ancilla or dirty ancilla does reduce the constants for the decomposition significantly, although the asymptotic depth and gate counts are maintained. For completeness, we also benchmark our circuit against this augmented construction,  \textcolor{black}{QUBIT+ANCILLA}. However, the augmented circuit does not operate at the ancilla-free frontier, and it can conflict with parallelism.

\FloatBarrier{}

\section{Circuit Construction}%
\label{sec:qudits-circuit-constructions}

In order for quantum circuits to be executable on hardware, they are typically decomposed into single- and two-qudit gates. Performing efficient, low depth, and low gate count decompositions is important in both the NISQ regime and beyond. Our circuits assume all-to-all connectivity, same as prior work.
If the quantum device does not support all-to-all connectivity, additional gates would be inserted by the compiler as needed, but this should not change our results.

\subsection{Key Intuition}

To develop intuition for our technique, we first present a Toffoli gate decomposition (the first step of implementation) which lays the foundation for our generalization to multiple controls. In each of the following constructions, all inputs and outputs are qubits, but we may occupy the $\ket{2}$ state temporarily during computation. Maintaining binary input and output allows these circuit constructions to be inserted into any preexisting qubit-only circuits.

\begin{figure}[h]
\[
    \Qcircuit @R=0.5em @C=0.25em {
    \lstick{\ket{q_{0}}} & \onecontrol & \qw  & \onecontrol & \qw & \\
    \lstick{\ket{q_{1}}} & \gate{X_{+1}}\qwx  & \twocontrol & \gate{X_{-1}}\qwx  &  \qw & \\
    \lstick{\ket{q_{2}}} & \qw  & \gate{X} \qwx  & \qw  &  \qw & \\
    }
\]
\caption{A Toffoli decomposition via qutrits. Each input and output is a qubit. The red controls activate on $\ket{1}$ and the blue controls activate on $\ket{2}$. The first gate temporarily elevates $q_1$ to $\ket{2}$ if both $q_0$ and $q_1$ were $\ket{1}$. We then perform the qubit-X operation only if $q_1$ is $\ket{2}$. The final gate restores $q_1$ to its original state.}%
\label{fig:toffoli_decomposition}
\end{figure}
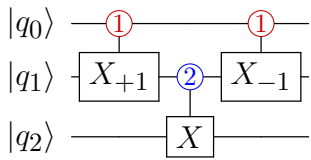

In Figure~\ref{fig:toffoli_decomposition}, a Toffoli decomposition using qutrits is given. A similar construction for the Toffoli gate is known from past work~\cite{Lanyon, Ralph}. The goal is to perform an X operation on the last (target) input qubit $q_2$ if and only if the two control qubits, $q_0$ and $q_1$, are both $\ket{1}$. First a $\ket{1}$-controlled $X_{+1}$ is performed on $q_0$ and $q_1$. This elevates $q_1$ to $\ket{2}$ iff $q_0$ and $q_1$ were both $\ket{1}$. Then a $\ket{2}$-controlled qubit-$X$ gate is applied to $q_2$. Therefore, $X$ is performed only when both $q_0$ and $q_1$ were $\ket{1}$, as desired. The controls are restored to their original states by a $\ket{1}$-controlled $X_{-1}$ gate, which undoes the effect of the first gate. The key intuition in this decomposition is that the qutrit $\ket{2}$ state can be used to store temporary information.

\subsection{Generalized Toffoli Gate Using Temporary Qutrits}%
\label{subsec:generalized_toffoli_construction}

\begin{figure}[h]
  \input{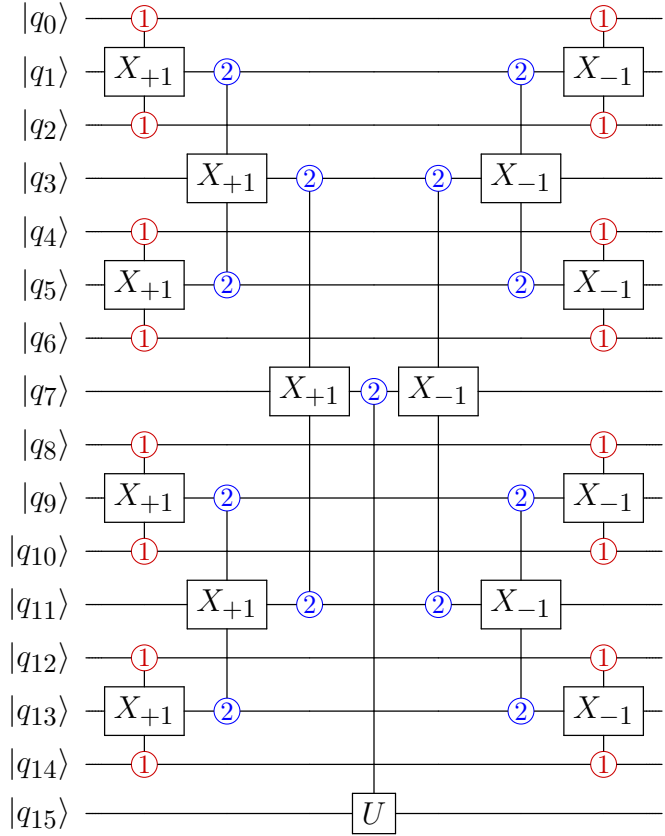}
  \caption{Our circuit decomposition for the Generalized Toffoli gate is shown for 15 controls and 1 target. The inputs and outputs are both qubits, but we allow occupation of the $\ket{2}$ qutrit state in between. The circuit has a tree structure and maintains the property that the root of each subtree can only be elevated to $\ket{2}$ if all of its control leaves were $\ket{1}$. Thus, the $U$ gate is only executed if all controls are $\ket{1}$. The right half of the circuit performs uncomputation to restore the controls to their original state. This construction applies more generally to any multiply-controlled $U$ gate. Note that the three-input gates are decomposed into 6 two-input and 7 single-input gates in our actual simulation, as based on the decomposition in~\cite{Di}.}%
  \label{fig:btb_cnu}
\end{figure}

We now present our circuit decomposition for the Generalized Toffoli in Figure~\ref{fig:btb_cnu}. The decomposition is expressed in terms of three-qutrit gates (two controls, one target) instead of single- and two-qutrit gates, because the circuit can be understood as purely classical reversible operations at this granularity. For implementation and in our simulation, we use a decomposition from~\cite{Di} that requires 6 two-qutrit and 7 single-qutrit physically implementable quantum gates.

\FloatBarrier{}

Our circuit decomposition is most intuitively understood by treating the left half of the circuit as a tree structure. The desired property is that the root of the tree, $q_7$, is $\ket{2}$ if and only if each of the 15 controls was originally in the $\ket{1}$ state. To verify this property, we observe the root $q_7$ can only become $\ket{2}$ iff $q_7$ was originally $\ket{1}$ and $q_3$ and $q_{11}$ were both previously $\ket{2}$. At the next level of the tree, we see $q_3$ could have only been $\ket{2}$ if $q_3$ was originally $\ket{1}$ and both $q_1$ and $q_5$ were previously $\ket{2}$, and similarly for the other triplets. At the bottom level of the tree, the triplets are controlled on the $\ket{1}$ state, which are only activated when the even-index controls are all $\ket{1}$. Thus, if any of the controls were not $\ket{1}$, the $\ket{2}$ states would fail to propagate to the root of the tree. The right half of the circuit performs \textit{uncomputation} to restore the controls to their original state.

After each subsequent level of the tree structure, the number of qubits under consideration is reduced by a factor of $\sim2$. Thus, the circuit depth is logarithmic in $n$. Moreover, each qutrit is operated on by a constant number of gates, so the total number of gates is linear in $n$.

Our circuit decomposition still works in a straightforward fashion when the control type of the top qubit, $q_0$, activates on $\ket{2}$ or $\ket{0}$ instead of activating on $\ket{1}$. These two constructions are necessary for the Incrementer circuit in Section~\ref{sec:qudits-incrementer}.

We verified our circuits, both formally and via simulation. Our verification scripts are available on our GitHub (\cite{GithubQutrits}).

\FloatBarrier{}

\section{Application to Algorithms}%
\label{sec:qudits-application-to-algorithms}

The Generalized Toffoli gate is an important primitive in a broad range of quantum algorithms. In this section, we survey some of the applications of our circuit decomposition.

\subsection{Grover's Algorithm}

\begin{figure}[ht]
  \input{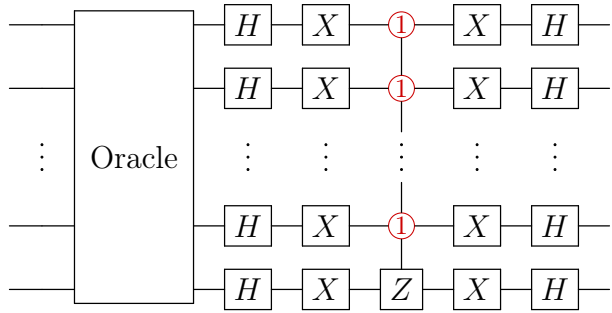}
  \caption{Each iteration of Grover Search has a multiply-controlled $Z$ gate. Our logarithmic depth decomposition, reduces a $\log M$ factor in Grover's algorithm to $\log\log M$.}%
  \label{fig:grover_search}
\end{figure}

Grover's Algorithm for search over $M$ unordered items requires just $O(\sqrt{M})$ oracle queries. However, each oracle query is followed by a post-processing step which requires a multiply-controlled gate with $N = \lceil \log_2 M \rceil$ controls (\cite{mikeike}). The explicit circuit diagram is shown in Figure~\ref{fig:grover_search}.

Our log-depth circuit construction directly applies to the multiply-controlled gate. Thus, we reduce a $\log M$ factor in Grover search time complexity to $\log \log M$ via our ancilla-free qutrit decomposition.

\subsection{Incrementer}%
\label{sec:qudits-incrementer}

The Incrementer circuit performs the $+ 1 \mod 2^N$ operation to a register of $N$ qubits. While logarithmic circuit depth is achieved with linear ancilla qubits by~\cite{Draper}, the best ancilla-free incrementers require either linear depth with large linearity constants as in ~\cite{Gidney} or quadratic depth in~\cite{barenco}. Using alternate control activations for our Generalized Toffoli gate decomposition, the incrementer circuit is reduced to $O(\log^2 N)$ depth with no ancilla, a significant improvement over past work.

\begin{figure}[ht]
  \input{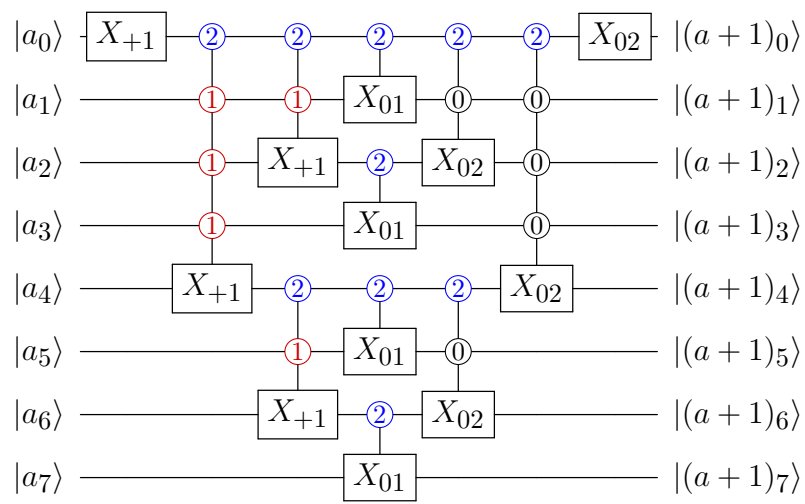}
  \caption{Our circuit decomposition for the Incrementer. At each subcircuit in the recursive design, multiply-controlled gates are used to efficiently propagate carries over half of the subcircuit. The $\ket{2}$ control checks for carry generation and the chain of $\ket{1}$ controls checks for carry propagation. The circuit depth is $\log^2 N$, which is only possible because of our log depth multiply-controlled gate primitive.}%
  \label{fig:incrementer}
\end{figure}

Our incrementer circuit construction is shown in Figure~\ref{fig:incrementer} for an $N=8$ wide register. The multiple-controlled $X_{+1}$ gates perform the job of computing carries: a carry is performed iff the least significant bit generates (represented by the $\ket{2}$ control) and all subsequent bits propagate (represented by the consecutive $\ket{1}$ controls). We present an $N=8$ incrementer here and have verified the general construction, both by formal proof and by explicit circuit simulation for larger $N$.

The critical path of this circuit is the chain of $\log{N}$ multiply-controlled gates (of width $\frac{N}{2}$, $\frac{N}{4}$, $\frac{N}{8}$, \ldots) which act on $\ket{a_0}$. Since our  multiply-controlled gate decomposition has log-depth, we arrive at a total circuit depth circuit scaling of $\log^2 N$.

\subsection{Arithmetic Circuits and Shor's Algorithm}

The Incrementer circuit is a key subcircuit in many other arithmetic circuits such as constant addition.
By adding a control to the first and last X gates, this circuit can be used for addition, modular multiplication, and modular exponentiation.
Modular exponentiation was shown to be a bottleneck in the runtime for executing Shor's algorithm for factorization by~\cite{Gidney, Microsoft}.
While a shallower Incrementer circuit alone is not sufficient to reduce the asymptotic cost of modular exponentiation (and therefore Shor's algorithm), it does reduce constants relative to qubit-only circuits.
Qudit arithmetic circuits using qudit ``compression'' are discussed in Section~\ref{sec:qudits-compression}.

\subsection{Error Correction and Fault Tolerance}

One possible benefit for qutrits in error correction and error mitigation is as an error \emph{flag} (\cite{flag-qubits}).
Flag qubits are extra qubits that are toggled by carefully placed extra gates in a way that does not change program outcome.
If no errors occur during execution, these qubits are always returned to the $\ket{0}$ state but if an error occurs, this often leads to a flag qubit measured in the $\ket{1}$ state, indicating an error occurred.
Qutrits can be used in the same way but on resource constrained-devices with no qubits to spare.

The Generalized Toffoli gate has applications to circuits for both error correction (\cite{Cory}) and fault tolerance (\cite{Dennis}). We foresee two paths of applying these circuits. First, our circuit construction can be used to construct error-resilient \textit{logical qubits} more efficiently. This is critical for quantum algorithms like Grover's and Shor's which are expected to require such logical qubits. In the nearer-term, NISQ algorithms are likely to make use of limited error mitigation. For instance, recent results have demonstrated that error correcting a single qubit at a time for the Variational Quantum Eigensolver algorithm can significantly reduce total error (\cite{Otten}). Thus, our circuit construction is also relevant for NISQ-era error correction.

\FloatBarrier{}

\section{Simulator}%
\label{sec:qudits-simulator}

To simulate our circuit constructions, we developed a qudit simulation library, built on Google's Cirq Python library (\cite{Cirq}). Cirq is a qubit-based quantum circuit library and includes a number of useful abstractions for quantum states, gates, circuits, and scheduling.

Our work extends Cirq by discarding the assumption of two-level qubit states.
Instead, all state vectors and gate matrices are expanded to apply to $d$-level qudits, where $d$ is a circuit parameter.
We include a library of common gates for $d=3$ qutrits.
Our software adds a comprehensive noise simulator, detailed below in Section~\ref{subsec:noise-simulation}.

In order to verify our circuits are logically correct, we first simulated them with noise disabled.
We wrote a Cirq simulator for classical subcircuits that allows gates to specify their action on classical non-superposition input states without considering full state vectors.
Therefore, each classical input state can be verified in space and time proportional to the circuit width.
By contrast, Cirq's default simulator relies on a dense state vector representation requiring space and time exponential in the circuit width.
Reducing this scaling from exponential to linear dramatically improved our verification procedure, allowing us to verify circuit constructions for all possible classical inputs across circuit sizes up to widths of 14.

Our software is fully open source (\cite{GithubQutrits}) and the core qudit support has also been added to Cirq.

\subsection{Noise Simulation}%
\label{subsec:noise-simulation}

\begin{figure}[ht]
  \input{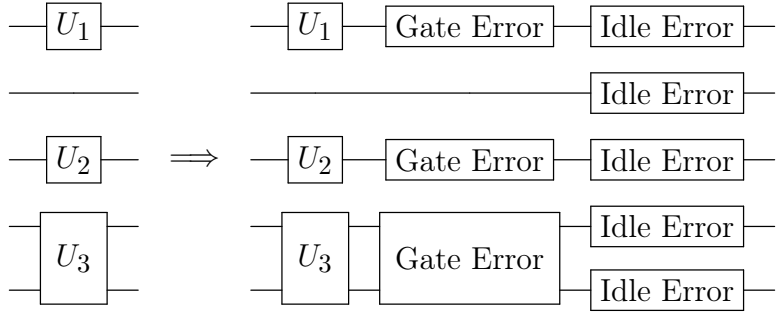}
  \caption{This \texttt{Moment} comprises three gates executed in parallel. To simulate with noise, we first apply the ideal gates, followed by a gate error noise channel on each affected qudit. This gate error noise channel depends on whether the corresponding gate was single- or two-qudit. Finally, we apply an idle error to every qudit. The idle error noise channel depends on the duration of the \texttt{Moment}.}%
  \label{fig:noise_simulation_methodology}
\end{figure}

Figure~\ref{fig:noise_simulation_methodology} depicts a schematic view of our noise simulation procedure which accounts for both gate errors and idle errors, described below. To determine when to apply each gate and idle error, we use Cirq's scheduler which schedules each gate as early as possible, creating a sequence of \texttt{Moment}'s of simultaneous gates. During each \texttt{Moment}, our noise simulator applies a gate error to every qudit acted on. Finally, the simulator applies an idle error to every qudit. This noise simulation methodology is consistent with previous simulation techniques such as~\cite{Miller} which accounted for gate errors and~\cite{QX_Simulator} for idle errors.

Gate errors arise from the imperfect application of quantum gates. Two-qudit gates are noisier than single-qudit gates (\cite{ibmq}), so we apply different noise channels for the two. Our specific gate error probabilities are given in Section~\ref{sec:qudits-error-modeling}.

Idle errors arise from the continuous decoherence of a quantum system due to energy relaxation and interaction with the environment. The idle errors differ from gate errors in two ways which require special treatment:
\begin{enumerate}
    \item Idle errors depend on duration, which in turn depend on the schedule of simultaneous gates (\texttt{Moment}s). In particular, two-qudit gates take longer to apply than single-qudit gates. Thus, if a \texttt{Moment} contains a two-qudit gate, the idling errors must be scaled appropriately. Our specific scaling factors are given in Section~\ref{sec:qudits-error-modeling}.
    \item For the generic model of gate errors, the error channel is applied with probability independent of the quantum state. This is not true for idle errors such as $T_1$ amplitude damping, which only applies when the qudit is in an excited state. This is treated in the simulator by computing idle error probabilities during each \texttt{Moment}, for each qutrit.
\end{enumerate}

Gate errors are reduced by performing fewer \textit{total gates}, and idle errors are reduced by decreasing the circuit \textit{depth}. Since our circuit constructions asymptotically decrease the depth, this means our circuit constructions scale favorably in terms of asymptotically fewer idle errors.

Our full noise simulation procedure is summarized in Algorithm~\ref{alg:pseudocode}. The ultimate metric of interest is the mean \textit{fidelity}, which is defined as the squared overlap between the ideal (noise-free) and actual output state vectors. Fidelity expresses the probability of overall successful execution. We do not consider initialization errors and readout errors, because our circuit constructions maintain binary input and output, only occupying the qutrit $\ket{2}$ states during intermediate computation. Therefore, the initialization and readout errors for our circuits are identical to those for conventional qubit circuits.

\SetEndCharOfAlgoLine{}
\begin{algorithm}
\SetAlgoLined%
$\ket{\Psi} \leftarrow$ random initial state vector\;
$\ket{\Psi}_{\text{ideal}} =$ circuit applied to $\ket{\Psi}$ without noise\;
\BlankLine%
\ForEach{\textup{\texttt{Moment}}}{%
 \ForEach{\textup{\texttt{Gate} $\in$ \texttt{Moment}}} {%
   $\ket{\psi} \leftarrow$ \texttt{Gate} applied to $\ket{\psi}$\;
   \texttt{GateError} $\leftarrow$ DrawRand(\texttt{GateError Prob.})\;
   $\ket{\psi} \leftarrow$ \texttt{GateError} applied to $\ket{\psi}$\;
 }
 \BlankLine%
 \ForEach{\textup{\texttt{Qutrit}}}{%
   \eIf{\textup{\texttt{Moment}} has 2-qudit gate}{%
     \texttt{IdleErrors} $\leftarrow$ long-duration idle errors\;
   }{%
     \texttt{IdleErrors} $\leftarrow$ short-duration idle errors\;
  }
  \texttt{Prob.} $\leftarrow [ \|\texttt{M} \ket{\Psi}\|^2 \text{ for } \texttt{M} \in \texttt{IdleErrors}]$ \;
  \texttt{IdleError} $\leftarrow$ DrawRand(\texttt{Prob.})\;
  $\ket{\psi} \leftarrow$ \texttt{IdleError} applied to $\ket{\psi}$\;
  Renormalize($\ket{\psi}$)\;
 }
}
 \Return{} $\braket{\Psi_{\text{ideal}} | \Psi}^2$, fidelity between ideal \& actual output;
 \caption{Pseudocode for each simulation trial, given a particular circuit and noise model.}%
 \label{alg:pseudocode}
\end{algorithm}

We also do not consider crosstalk errors, which occur when gates are executed in parallel. The effect of crosstalk is very device-dependent and difficult to generalize. Moreover, crosstalk can be mitigated by breaking each \texttt{Moment} into a small number of sub-moments and then scheduling two-qutrit operations to reduce crosstalk, as demonstrated in~\cite{Crosstalk1, Crosstalk2}.

\subsection{Simulator Efficiency}

Simulating a quantum circuit with a classical computer is, in general, exponentially difficult in the size of the input because the state of $N$ qudits is represented by a state vector of $d^N$ complex numbers. For 14 qutrits, with complex numbers stored as two 8-byte floats (\texttt{complex128} in NumPy), a state vector occupies 77 megabytes.

A naive circuit simulation implementation would treat every quantum gate or \texttt{Moment} as a $d^N \times d^N$ matrix. For 14 qutrits, a single such matrix would occupy 366 terabytes---out of range of simulability. While the exponential nature of simulating our circuits is unavoidable, we mitigate the cost by using a variety of techniques which rely only on state vectors, rather than full square matrices. For example, we maintain Cirq's approach of applying gates by Einstein Summation (\cite{TensorNetworks}), which obviates computation of the $d^N \times d^N$ matrix corresponding to every gate or \texttt{Moment}.

Our noise simulator only relies on state vectors, by adopting the quantum trajectory methodology of~\cite{QuantumTrajectories, CppQuantumTrajectories}, which is also used by the Rigetti PyQuil noise simulator (\cite{rigetti}). At a high level, the effect of noise channels like gate and idle errors is to turn a coherent quantum state into an incoherent mix of classical probability-weighted quantum states (for example, $\ket{0}$ and $\ket{1}$ with 50\% probability each). The most complete description of such an incoherent quantum state is called the density matrix and has dimension $d^N \times d^N$. The quantum trajectory methodology is a stochastic approach---instead of maintaining a density matrix, only a single state is propagated and the error term is drawn randomly at each timestep. Over repeated trials, the quantum trajectory methodology converges to the same results as from full density matrix simulation (\cite{rigetti}). Our simulator employs this technique---each simulation in Algorithm~\ref{alg:pseudocode} constitutes a single quantum trajectory trial. At every step, a specific \texttt{GateError} or \texttt{IdleError} term is picked, based on a weighted random draw.

Finally, our random state vector generation function was also implemented in $O(d^N)$ space and time. This is an improvement over other open source libraries, \cite{QuTiP, QuTiP2}, which perform random state vector generation by generating full $d^N \times d^N$ unitary matrices from a Haar-random distribution and then truncating to a single column. Our simulator directly computes the first column and circumvents the full matrix computation.

With optimizations, our simulator is able to simulate circuits up to 14 qutrits in width. This is in the range as other state-of-the-art noisy quantum circuit simulations (since 14 qutrits $\approx$ 22 qubits, \cite{28QubitNoisySimulation}). While each simulation trial took several minutes (depending on the particular circuit and noise model), we were able to run trials in parallel over multiple processes and multiple machines, as described in Section~\ref{sec:qudits-results}.

\FloatBarrier{}

\section{Noise Models}%
\label{sec:qudits-error-modeling}

In this section, we describe our noise models at a high level. We chose noise models which represent realistic near-term machines. We first present a generic, parametrized noise model roughly applicable to all quantum systems. We then present specific parameters, under the generic noise model, which apply to near-term superconducting quantum computers. Finally, we present a specific noise model for trapped ion quantum computers.

\subsection{Generic Noise Model}%
\label{subsec:generic}

\subsubsection{Gate Errors}
The scaling of gate errors for a $d$-level qudit can be roughly summarized as increasing as $d^4$ for two-qudit gates and $d^2$ for single-qudit gates. For $d=2$, there are 4 single-qubit gate error channels and 16 two-qubit gate error channels. For $d=3$ there are 9 and 81 single- and two-qutrit gate error channels respectively. Consistent with other simulators, \cite{rigetti, QX_Simulator}, we use the symmetric depolarizing gate error model, which assumes equal probabilities between each error channel. Under these noise models, two-qutrit gates are $(1-80p_2) / (1-15p_2)$ times less reliable than two-qubit gates, where $p_2$ is the probability of each two-qubit gate error channel. Similarly, single-qutrit gates are $(1-8p_1) / (1 - 3p_1)$ times less reliable than single-qubit gates, where $p_1$ is the probability of each single-qubit gate error channel.

\subsubsection{Idle Errors}%
\label{subsubsec:idle-errors}

Our treatment of idle errors focuses on the relaxation from higher to lower energy states in quantum devices. This is called amplitude damping or $T_1$ relaxation. This noise channel irreversibly takes qudits to lower states. For qubits, the only amplitude damping channel is from $\ket{1}$ to $\ket{0}$, and we denote this damping probability as $\lambda_1$. For qutrits, we also model damping from $\ket{2}$ to $\ket{0}$, which occurs with probability $\lambda_2$.

\subsection{Superconducting QC}%
\label{subsec:superconducting}

We chose four noise models based on superconducting quantum computers expected in the next few years. These noise models comply with the generic noise model above and are thus parametrized by $p_1$, $p_2$, $\lambda_1$ and $\lambda_2$. The $\lambda_i$ probabilities are derived from two other experimental parameters: the gate time $\Delta t$ and $T_1$, a timescale that captures how long a qudit persists coherently.

As a starting point for representative near-term noise models, we consider parameters for \textit{current} superconducting quantum computers. For IBM's public cloud-accessible superconducting quantum computers, we have $3p_1 \approx 10^{-3}$ and $15p_2 \approx 10^{-2}$. The duration of single- and two-qubit gates is $\Delta t \approx 100ns$ and $\Delta t \approx 300ns$ respectively, and the IBM devices have $T_1 \approx 100 \mu s$ (\cite{ibmq, Linke}).

However, simulation for these current parameters indicates an error is almost certain to occur during execution of a modest size 14-input Generalized Toffoli circuit. This motivates us to instead consider noise models for better devices which are a few years away. Accordingly, we adopt a baseline superconducting noise model, labeled as \textcolor{black}{SC}, corresponding to a superconducting device which has 10x lower gate errors and 10x longer $T_1$ duration than the current IBM hardware. This range of parameters has already been achieved experimentally in superconducting devices for gate errors by~\cite{LowGateError1, LowGateError2} and for $T_1$ duration by~\cite{LongT1_1, LongT1_2} independently. Faster gates (shorter $\Delta t$) are yet another path towards greater noise resilience. We do not vary gate speeds, because errors only depend on the $\Delta t / T_1$ ratio, and we already vary $T_1$. In practice however, faster gates could also improve noise-resilience.

We also consider three additional near-term device noise models, indexed to the \textcolor{black}{SC} noise model. These three models further improve gate errors, $T_1$, or both, by a 10x factor. The specific parameters are given in Table~\ref{tab:sc_noise_models}. Our 10x improvement projections are realistic extrapolations of progress in hardware. In particular, Schoelkopf's Law---the quantum analogue of Moore's Law---has observed that $T_1$ durations have increased by 10x every 3 years for the past 20 years~\cite{LesHouchesQEDNotes}. Hence, 100x longer $T_1$ is a reasonable projection for devices that are $\sim 6$ years away.

\begin{table}[h]
\centering
\begin{tabular}{l|ll|l}
  Noise Model & $3 p_1$ & $15 p_2$ & $T_1$ \\ \hline
 \textcolor{black}{SC} & $10^{-4}$ & $10^{-3}$ & 1 ms \rule{0pt}{2.6ex}
 \\ \textcolor{black}{SC+T1} & $10^{-4}$ & $10^{-3}$ &  10 ms
 \\ \textcolor{black}{SC+GATES} & $10^{-5}$ & $10^{-4}$ & 1 ms
 \\ \textcolor{black}{SC+T1+GATES} & $10^{-5}$ & $10^{-4}$ & 10 ms
\end{tabular}
\caption{Noise models simulated for superconducting devices. Current publicly accessible IBM superconducting quantum computers have single- and two-qubit gate errors of $3p_1 \approx 10^{-3}$ and $15p_2 \approx 10^{-2}$, as well as $T_1$ lifetimes of 0.1 ms (\cite{ibmq, Linke}). Our baseline benchmark, \textcolor{black}{SC}, assumes 10x better gate errors and $T_1$. The other three benchmarks add a further 10x improvement to $T_1$, gate errors, or both.}%
\label{tab:sc_noise_models}
\end{table}

\subsection{Trapped Ion $^{171}$Yb$^+$ QC}%
\label{subsec:trapped-ion}
We also simulated noise models for trapped ion quantum computing devices.  Trapped ion devices are well matched to our qutrit-based circuit constructions because they feature all-to-all connectivity (\cite{brown2016co}), and many ions that are ideal candidates for QC devices are naturally multi-level systems.

We focus on the $^{171}$Yb${^+}$ ion, which has been experimentally demonstrated as both a qubit and qutrit by \cite{HesingA, HesingB}. Trapped ions are often favored in QC schemes due to their long $T_1$ times. One of the main advantages of using a trapped ion is the ability to take advantage of magnetically insensitive states known as ``clock states''. By defining the computational subspace on these clock states, idle errors caused from fluctuations in the magnetic field are minimized---this is termed a \textcolor{black}{DRESSED\_QUTRIT}, in contrast with a \textcolor{black}{BARE\_QUTRIT}. However, compared to superconducting devices, gates are much slower. Thus, gate errors are the dominant error source for ion trap devices. We modelled a fundamental source of these errors: the spontaneous scattering of photons originating from the lasers used to drive the gates. The duration of single- and two-qubit gates used in this calculation was $\Delta t \approx 1$ $\mu$s and $\Delta t \approx 200$ $\mu$s respectively (\cite{PhysRevA.97.052301}). The single- and two-qudit gate error probabilities are given in Table~\ref{tab:ti_noise_models}.

\begin{table}[h]
\centering
\begin{tabular}{l|ll}
  Noise Model & $p_1$ & $p_2$ \\ \hline
 \textcolor{black}{TI\_QUBIT} & $6.4 \times 10^{-4}$ & $1.3 \times 10^{-4}$  \rule{0pt}{2.6ex}
 \\ \textcolor{black}{BARE\_QUTRIT} & $2.2 \times 10^{-4}$ & $4.3 \times 10^{-4}$
 \\ \textcolor{black}{DRESSED\_QUTRIT} & $1.5 \times 10^{-4}$ & $3.1 \times 10^{-4}$
\end{tabular}
\caption{Noise models simulated for trapped ion devices. The single- and two-qutrit gate error channel probabilities are based on calculations from experimental parameters. For all three models, we use single- and two-qudit gate times of $\Delta t \approx 1$ $\mu s$ and $\Delta t \approx 200$ $\mu s$ respectively.}%
\label{tab:ti_noise_models}
\end{table}

\FloatBarrier{}

\section{Simulation Results}%
\label{sec:qudits-results}

Figure~\ref{fig:circuit_depths} plots the exact circuit depths for all three benchmarked circuits. The qubit-based circuit constructions from past work are linear in depth and have a high linearity constant. Augmenting with a single borrowed ancilla reduces the circuit depth by a factor of 8. However, both circuit constructions are surpassed significantly by our qutrit construction, which scales logarithmically in $N$ and has a relatively small leading coefficient.

Figure~\ref{fig:two_qudit_gate_counts} plots the total number of two-qudit gates for all three circuit constructions. As noted in Section~\ref{sec:qudits-circuit-constructions}, our circuit construction is not asymptotically better in total gate count---all three plots have linear scaling. However, as emphasized by the logarithmic vertical axis, the linearity constant for our qutrit circuit is 70x smaller than for the equivalent ancilla-free qubit circuit and 8x smaller than for the borrowed-ancilla qubit circuit.

Our simulations under realistic noise models were run in parallel on over 100 n1-standard-4 Google Cloud instances. These simulations represent over 20,000 CPU hours, which was sufficient to estimate mean fidelity to an error of $2 \sigma < 0.1\%$ for each circuit-noise model pair.

\begin{figure*}[ht]
    \centering%
        \scalebox{0.91}{
        \begin{tikzpicture}[baseline]
\pgfplotsset{every axis/.append style={thick}}
\pgfplotsset{every tick label/.append style={font=\small}}
\pgfplotsset{every axis label/.append style={font=\small}}

\newcommand{\extraBarHeight}{1}

\begin{axis}[
    title=Fidelity for Superconducting Models,
    symbolic x coords={left,SC,a,SC+T1,b,SC+GATES,c,SC+T1+GATES,right},
    width=.75\textwidth,
    height=220pt,
    ybar=5pt,
    bar width=15pt,
    xmin=left, xmax=right,
    ymin=0, ymax=120,
    ytick={0, 25, 50, 75, 100},
    ytick distance=25,
    xtick=data,
    ,
    legend style={draw=none, fill=none, at={(0.8,1.03)},anchor=north,font=\small},
    legend columns=-1,
    legend image code/.code={\draw[#1, draw=none] (0em,-0.2em) rectangle (0.6em,0.4em);},
    axis line style={draw=black!20!white},
    axis on top,
    y axis line style={draw=none},
    axis x line*=bottom,
    tick style={draw=none},
    yticklabel={\pgfmathparse{\tick*1}\pgfmathprintnumber{\pgfmathresult}\%},
    clip=false,
    enlarge y limits=0,
    ,
    nodes near coords always on top/.style={
        every node near coord/.append style={
            anchor=south,
            rotate=0,
            font=\small,
            inner sep=0.2em,
        },
    },
    nodes near coords={
        \StrPosition{\pgfplotspointmeta}{Y}[\Result]%
        \StrGobbleLeft{\pgfplotspointmeta}{\Result}[\Result]%
        \StrGobbleRight{\Result}{1}[\Result]%
        \pgfmathparse{\Result<10}%
        \ifnum\pgfmathresult=1
            \pgfmathparse{\Result-\extraBarHeight}%
        \else
            \pgfmathparse{\Result}%
        \fi%
        \pgfmathprintnumber[fixed,precision=2]{\pgfmathresult}\%
    },
    nodes near coords always on top,
]

\addplot[style={color=transparent, draw=none, fill=\colorQubitLight, mark=none}] coordinates {
    (SC, 1.01)  
    (SC+T1, 1.56)  
    (SC+GATES, 1.01)  
    (SC+T1+GATES, 26.1)  
};
\addlegendentry{QUBIT\:\:\:\:};

\addplot[style={draw=none, pattern color=\colorQubitBB, pattern=north east lines, mark=none}] coordinates {
    (SC, 18.5)  
    (SC+T1, 52.3)  
    (SC+GATES, 30.2)  
    (SC+T1+GATES, 84.1)  
};
\addlegendentry{QUBIT+ANCILLA\:\:\:\:};

\addplot[style={draw=none, fill=\colorQutritLight, mark=none}] coordinates {
    (SC, 56.8)  
    (SC+T1, 65.9)  
    (SC+GATES, 83.1)  
    (SC+T1+GATES, 94.7)  
};
\addlegendentry{QUTRIT};

\end{axis}

%
%
%
%
\end{tikzpicture}%
%
        \kern-5em%
        \begin{tikzpicture}[baseline]
\pgfplotsset{every axis/.append style={thick}}
\pgfplotsset{every y tick label/.append style={font=\small}}
\pgfplotsset{every x tick label/.append style={font=\small}}
\pgfplotsset{every axis label/.append style={font=\small}}

\newcommand{\beginCustomBarAxisFirst}[1]{
\begin{axis}[
    title=Fidelity for Trapped Ion Models,
    symbolic x coords={#1},
    width=0.33\textwidth,
    height=220pt,
    ybar=5pt,
    bar width=15pt,
    xmin=left, xmax=right,
    ymin=0, ymax=120,
    ytick={0, 25, 50, 75, 100},
    ytick distance=25,
    xtick=data,
    enlarge x limits=0,
    ,
    legend style={draw=none, fill=none, at={(0.5,1.03)},anchor=north,font=\small},
    legend columns=2,
    legend image code/.code={\draw[##1, draw=none] (0em,-0.2em) rectangle (0.6em,0.4em);},
    axis line style={draw=black!20!white},
    axis on top,
    y axis line style={draw=none},
    axis x line*=bottom,
    tick style={draw=none},
    yticklabel={~},
    clip=false,
    enlarge y limits=0,
    ,
    nodes near coords always on top/.style={
        every node near coord/.append style={
            anchor=south,
            rotate=0,
            font=\small,
            inner sep=0.2em,
        },
    },
    nodes near coords={
        \pgfmathprintnumber[fixed,fixed zerofill,precision=1]{\pgfplotspointmeta}\%
    },
    nodes near coords always on top,
]
}

\newcommand{\beginCustomBarAxis}[1]{
\begin{axis}[
    title=,
    symbolic x coords={#1},
    width=0.33\textwidth,
    height=220pt,
    ybar=5pt,
    bar width=15pt,
    xmin=left, xmax=right,
    ymin=0, ymax=120,
    ytick={0, 25, 50, 75, 100},
    ytick distance=25,
    xtick=data,
    enlarge x limits=0,
    x tick label style={rotate=90, anchor=west, yshift=-0.1em, xshift=0.4em, color=black},
    ,
    legend style={draw=none, fill=none, at={(0.5,1.03)},anchor=north,font=\small},
    legend columns=2,
    legend image code/.code={\draw[##1, draw=none] (0em,-0.2em) rectangle (0.6em,0.4em);},
    axis line style={draw=none},
    axis on top,
    tick style={draw=none},
    yticklabel={~},
    clip=false,
    enlarge y limits=0,
    ,
    nodes near coords always on top/.style={
        every node near coord/.append style={
            anchor=south,
            rotate=0,
            font=\small,
            inner sep=0.2em,
        },
    },
    nodes near coords={
        \pgfmathprintnumber[fixed,fixed zerofill,precision=1]{\pgfplotspointmeta}\%
    },
    nodes near coords always on top,
]
}

\beginCustomBarAxisFirst{left,q,TI\textunderscore QUBIT,a,b,x,c,d,y,right}
\addplot[style={draw=none, fill=\colorQubitLight, mark=none}] coordinates {
    (TI\textunderscore QUBIT, 44.66)
};

\addplot[style={draw=none, pattern color=\colorQubitBB, pattern=north east lines, mark=none}] coordinates {
    (TI\textunderscore QUBIT, 89.85)
};
\end{axis}

\beginCustomBarAxis{left,q,v,u,c,d,BARE\textunderscore QUTRIT,e,f,y,right}
\addplot[style={draw=none, fill=\colorQutritLight, mark=none}] coordinates {
    (BARE\textunderscore QUTRIT, 94.92)
};
\end{axis}

\beginCustomBarAxis{left,q,u,c,d,x,e,f,DRESSED\textunderscore QUTRIT,right}
\addplot[style={draw=none, fill=\colorQutritLight, mark=none}] coordinates {
    (DRESSED\textunderscore QUTRIT, 96.08)
};
\end{axis}

\end{tikzpicture}%
        }%
    \caption{Circuit simulation results for all possible pairs of circuit constructions and noise models. Each bar represents 1000+ trials, so the error bars are all $2\sigma < 0.1\%$. Our QUTRIT construction significantly outperforms the QUBIT construction. The QUBIT+ANCILLA bars are drawn with dashed lines to emphasize that it has access to an extra ancilla bit, unlike our construction.}%
    \label{fig:simulation_results}
\end{figure*}
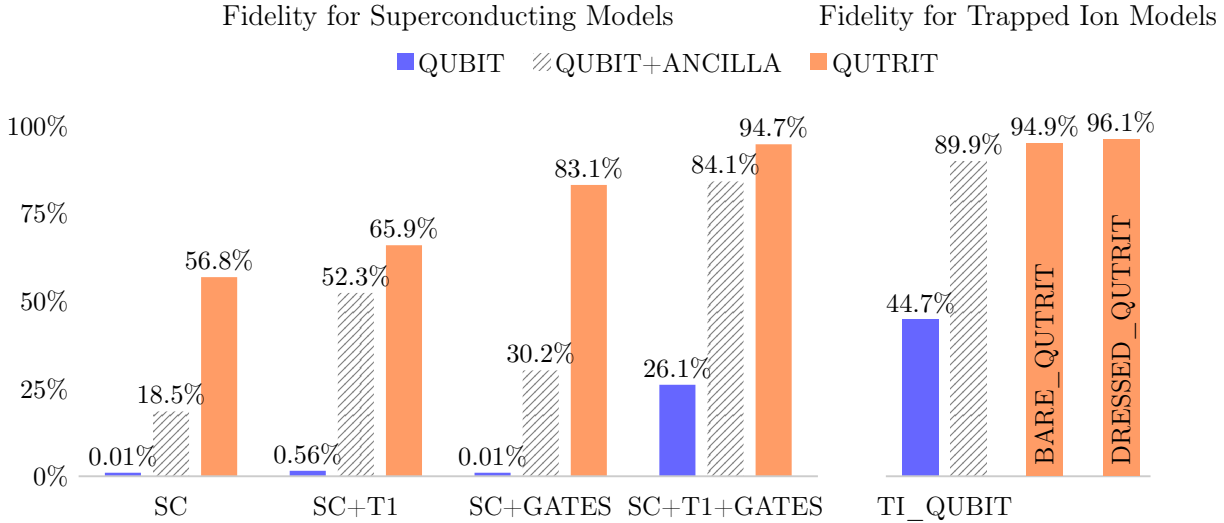

The full results of our circuit simulations are shown in Figure~\ref{fig:simulation_results}. All simulations are for the 14-input (13 controls, 1 target) Generalized Toffoli gate. We simulated each of the three circuit benchmarks against each of our noise models (when applicable), yielding the 16 bars in the figure.

\begin{figure}[]
    \begin{tikzpicture}
\pgfplotsset{every axis/.append style={thick}}
\pgfplotsset{every tick label/.append style={font=\small}}
\pgfplotsset{every axis label/.append style={font=\small}}

\begin{semilogyaxis}[
    title=Circuit Depth,
    xlabel=Number of Qudits,
    width=\columnwidth + 20pt,
    height=175pt,
    xmin=4, xmax=200,
    ymin=10,
    restrict x to domain=4:200,
    xtick distance=25,
    ,
    legend style={draw=none, fill=none, at={(0.5,1.03)},anchor=north,font=\small},
    legend columns=-1,
    axis line style={draw=none},
    tick style={draw=none},
    clip=false,
]

\addplot[color=\colorQubit] table[x=Qubits, y=Depth, col sep=comma]
    {chapters/qudits/data/gateCounts-CnXLinear.csv}
node [anchor=north east] {$\sim 633N$};
\addlegendentry{QUBIT\:\:\:\:};

\addplot[color=\colorQubitBB] table[x=Qubits, y=Depth, col sep=comma]
    {chapters/qudits/data/gateCounts-CnXLinearBB.csv}
node [anchor=north east] {$\sim 76N$};
\addlegendentry{QUBIT+ANCILLA\:\:\:\:};

\addplot[color=\colorQutrit] table[x=Qubits, y=Depth, col sep=comma]
    {chapters/qudits/data/gateCounts-CnXQutrit.csv}
node [anchor=north east] {$\sim 38\log_2(N)$};
\addlegendentry{QUTRIT};

\end{semilogyaxis}
\end{tikzpicture}
    \caption{Exact circuit depths for all three benchmarked circuit constructions for the N-controlled Generalized Toffoli up to $N=200$. Both QUBIT and QUBIT+ANCILLA scale linearly in depth and both are bested by QUTRIT's logarithmic depth.}%
    \label{fig:circuit_depths}
\end{figure}
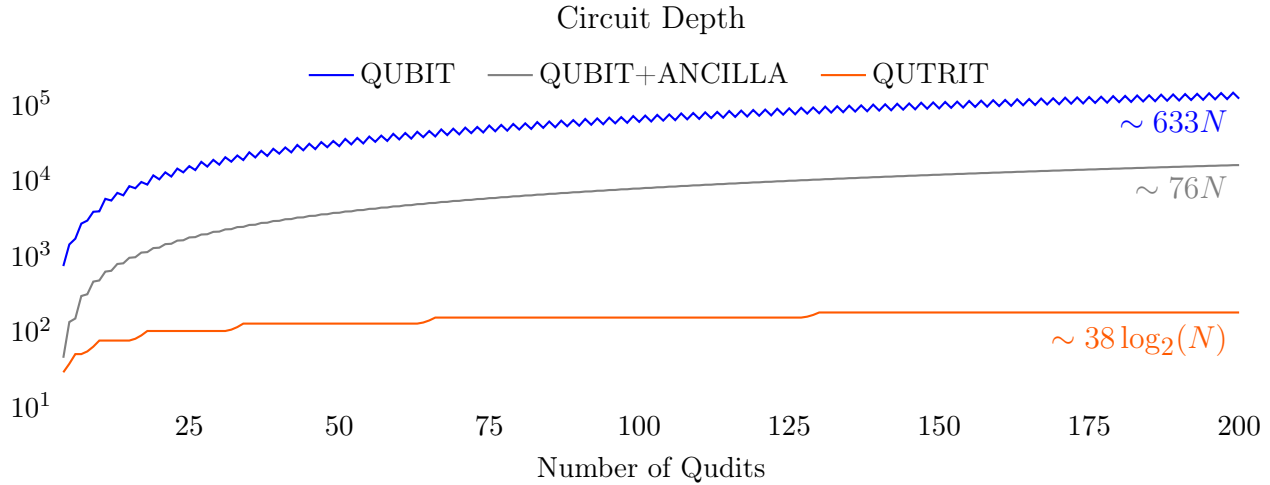
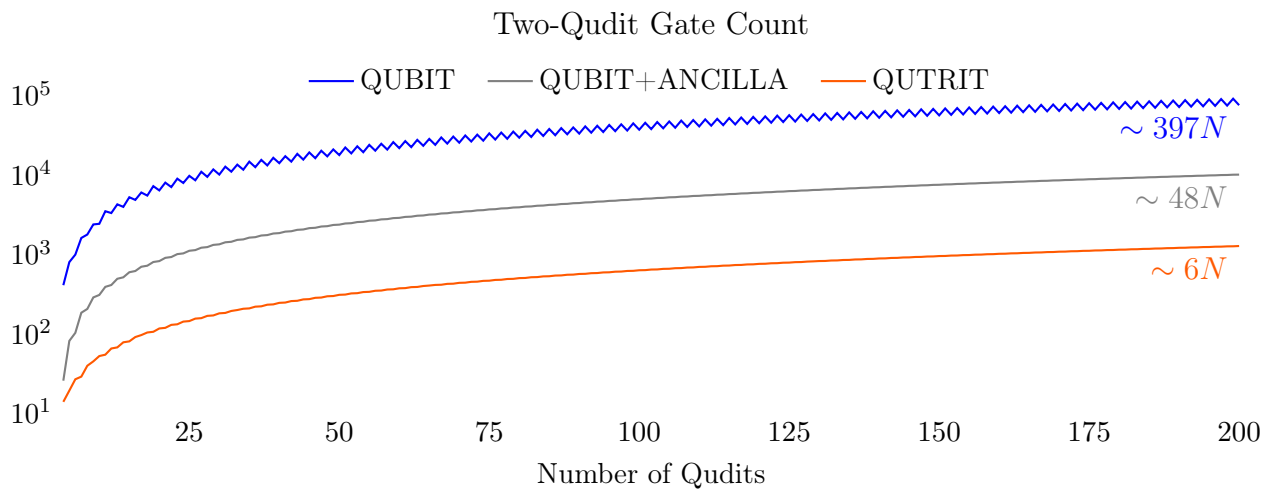
\begin{figure}[]
    \begin{tikzpicture}
\pgfplotsset{every axis/.append style={thick}}
\pgfplotsset{every tick label/.append style={font=\small}}
\pgfplotsset{every axis label/.append style={font=\small}}

\begin{semilogyaxis}[
    title=Two-Qudit Gate Count,
    xlabel=Number of Qudits,
    width=\columnwidth + 20pt,
    height=175pt,
    xmin=4, xmax=200,
    ymin=10,
    restrict x to domain=4:200,
    xtick distance=25,
    ,
    legend style={draw=none, fill=none, at={(0.5,1.03)},anchor=north,font=\small},
    legend columns=-1,
    axis line style={draw=none},
    tick style={draw=none},
    clip=false,
]

\addplot[color=\colorQubit] table[x=Qubits, y=Two Qubit Gate Count, col sep=comma]
    {chapters/qudits/data/gateCounts-CnXLinear.csv}
node [anchor=north east] {$\sim 397N$};
\addlegendentry{QUBIT\:\:\:\:};

\addplot[color=\colorQubitBB] table[x=Qubits, y=Two Qubit Gate Count, col sep=comma]
    {chapters/qudits/data/gateCounts-CnXLinearBB.csv}
node [anchor=north east] {$\sim 48N$};
\addlegendentry{QUBIT+ANCILLA\:\:\:\:};

\addplot[color=\colorQutrit] table[x=Qubits, y=Two Qubit Gate Count, col sep=comma]
    {chapters/qudits/data/gateCounts-CnXQutrit.csv}
node [anchor=north east] {$\sim 6N$};
\addlegendentry{QUTRIT};

\end{semilogyaxis}
\end{tikzpicture}
    \caption{Exact two-qudit gate counts for the three benchmarked circuit constructions for the N-controlled Generalized Toffoli. All three plots scale linearly; however the QUTRIT construction has a substantially lower linearity constant.}%
    \label{fig:two_qudit_gate_counts}
\end{figure}

Figure~\ref{fig:simulation_results} demonstrates that our \textcolor{black}{QUTRIT} construction (orange bars) significantly outperforms the ancilla-free \textcolor{black}{QUBIT} benchmark (blue bars) in fidelity (success probability) by more than 10,000x.

For the \textcolor{black}{SC}, \textcolor{black}{SC+T1}, and \textcolor{black}{SC+GATES} noise models, our qutrit constructions achieve between 57--83\% mean fidelity, whereas the ancilla-free qubit constructions all have almost 0\% fidelity. Only the lowest-error model, \textcolor{black}{SC+T1+GATES} achieves modest fidelity of 26\% for the \textcolor{black}{QUBIT} circuit, but in this regime, the qutrit circuit is close to 100\% fidelity.

The trapped ion noise models achieve similar results---the \linebreak \textcolor{black}{DRESSED\_QUTRIT} and the \textcolor{black}{BARE\_QUTRIT} achieve approximately 95\% fidelity via the \textcolor{black}{QUTRIT} circuit, whereas the \textcolor{black}{TI\_QUBIT} noise model has only 45\% fidelity. Between the dressed and bare qutrits, the dressed qutrit exhibits higher fidelity than the bare qutrit, as expected. Moreover, the dressed qutrit is resilient to leakage errors, so the simulation results should be viewed as a lower bound on its advantage over the qubit and bare qutrit.

Based on these results, trapped ion qutrits are a particularly strong match to our qutrit circuits. In addition to attaining the highest fidelities, trapped ions generally have all-to-all connectivity (\cite{brown2016co}) within each ion chain, which is critical as our circuit construction requires operations between distant qutrits.

The superconducting noise models also achieve good fidelities. They exhibit a particularly large advantage over ancilla-free qubit constructions because idle errors are significant for superconducting systems, and our qutrit construction significantly reduces idling (circuit depth). However, most superconducting quantum systems only feature nearest-neighbor or short-range connectivity. Accounting for data movement on a nearest-neighbor-connectivity 2D architecture would expand the qutrit circuit depth from $\log{N}$ to $\sqrt{N}$ (since the distance between any two qutrits would scale as $\sqrt{N}$). However, \cite{Quantum_RAM} has experimentally demonstrated fully-connected superconducting quantum systems via random access memory. Such systems would also be well matched to our circuit construction.

For completeness, Figure~\ref{fig:simulation_results} also shows fidelities for the \linebreak \textcolor{black}{QUBIT+ANCILLA} circuit benchmark, which augments the ancilla-free \textcolor{black}{QUBIT} circuit with a single dirty ancilla. Since \textcolor{black}{QUBIT+ANCILLA} has linearity constants $\sim 10$x better than the ancilla-free qubit circuit, it exhibits significantly better fidelities. While our \textcolor{black}{QUTRIT} circuit still outperforms the \textcolor{black}{QUBIT+ANCILLA} circuit, we expect a crossing point where augmenting a qubit-only Generalized Toffoli with enough ancilla would eventually outperform \textcolor{black}{QUTRIT}. However, we emphasize that the gap between an ancilla-free and constant-ancilla construction for the Generalized Toffoli is actually a fundamental rather than an incremental gap, because:
\begin{itemize}
    \item Constant-ancilla constructions prevent circuit parallelization. For example, consider the parallel execution of $N/k$ disjoint Generalized Toffoli gates, each of width $k$ for some constant $k$. An ancilla-free Generalized Toffoli would pose no issues, but an ancilla-augmented Generalized Toffoli would require $\Theta(N/k)$ ancilla. Thus, constant-ancilla constructions can impose a choice between serializing to linear depth or regressing to linear ancilla count. The Incrementer circuit in Figure~\ref{fig:incrementer} is a concrete example of this scenario---any multiply-controlled gate decomposition requiring a single clean ancilla or more than 1 dirty ancilla would contradict the parallelism and reduce runtime.
    \item Even if we only consider serial circuits, given the exponential advantage of certain quantum algorithms, there is a significant practical difference between operating at the ancilla-free frontier and operating just a few data qubits below the frontier.
\end{itemize}

While we only performed simulations up to 14 inputs in width, we would see an even bigger advantage in larger circuits because our construction has asymptotically lower depth and therefore asymptotically lower idle errors. We also expect to see an advantage for the circuits in Section~\ref{sec:qudits-application-to-algorithms} that rely on the Generalized Toffoli, although we did not explicitly simulate these circuits.

\FloatBarrier{}

\section{Qubit-Qudit Compression}%
\label{sec:qudits-compression}

We see, using qudits as temporary storage can benefit our circuit for the Generalized Toffoli but can the abstraction of using extra levels as temporary storage be useful in other contexts?
In this section, we show how to re-encode, or ``compress'' idle data to free-up extra workspace ancilla, without requiring a larger quantum computer.

Typically, when using a higher radix computing paradigm, we express a circuit entirely in the specified base, that is all inputs and outputs are in the designated radix. An alternative approach is to fix the input and output radix but allow the use of higher level states temporarily during the computation, i.e.\ we are permitted to occupy any level up to a specified $d$ during a computation with the guarantee that we return to the specified radix.

What does this gain for us? It is known that by simply fully encoding a computation into a higher radix we obtain a constant space and time advantage over binary-only circuits.
However, as we showed earlier, the use of these higher states can act as temporary storage, similar to the use of an ancilla, and can convey an asymptotic reduction in circuit depth.
This circuit construction suggests we can obtain better circuits while using fewer qubits by accessing higher states temporarily.

We take this a step further and \textit{generate} ancilla temporarily out of input qubits in order to take advantage of previously known efficient binary circuit decompositions like that of~\cite{draper2006logarithmic}. Using this method, we can reduce the number of external ancilla needed from $O(n)$ to $0$ while keeping the same asymptotic circuit depth. To do this, we allow subsets of qubits to temporarily store higher values, becoming qudits, to store the information of many qubits within a few qudits. As a concrete example, consider three qubits. There are $2^3 = 8$ total basis states while for two qutrits there are $3^2 = 9$ basis states. Therefore all the information of 3 qubits can be stored in two qutrits and the third qubit can be left in a chosen state, $\ket{0}$, a clean ancilla. We refer to this process as \textit{compression}, that is storing the information of many qu\textit{bits} in a smaller number of qu\textit{dits}.
While a better term for this process might be \textit{re-encoding} with a different radix, its behavior from a systems perspective is similar to lossless compression of data to save memory.

We consider various reversible compression schemes labeled x-y-z compression, where $x$ is the radix of the input qudits, $y$ is the radix of the output qudits, and $z$ is the number of ancilla generated. Such operations exist if $x^m \le y^n$ with $0 < n < m$ and $m - n = z$ for some integers $m, n$, the number of input qudits and the number of non-ancilla outputs, respectively. Put more simply, these proposed compression circuits exist if the number of basis states of the inputs is fewer than the number of basis states of the non-ancilla outputs and the number of non-ancilla outputs is strictly smaller than the number of inputs. Ideally, we choose compression schemes with a good compression ratio, i.e.\ those for which $x^m / y^n \approx 1$.

In this section, we consider 2-3-1 and 2-4-1 compression as methods of generating ancilla for simplicity.  Many other schemes such as 2-8-2 and 3-9-1 are possible but require increasingly complex compression circuits.

\subsection{Qubit to Qutrit Compression}%
\label{sec:qudits-231compression}

\begin{table}[]
    \centering
    \bgroup{}
    \setlength\tabcolsep{1.5em}
    \begin{tabular}{ccc|ccc}
        A & B & C & A' & B' & C' \\\hline
        0 & 0 & 0 & 0  & 0  & 0 \\
        0 & 0 & 1 & 2  & 2  & 0 \\
        0 & 1 & 0 & 0  & 1  & 0 \\
        0 & 1 & 1 & 0  & 2  & 0 \\
        1 & 0 & 0 & 1  & 0  & 0 \\
        1 & 0 & 1 & 2  & 1  & 0 \\
        1 & 1 & 0 & 1  & 1  & 0 \\
        1 & 1 & 1 & 1  & 2  & 0 \\
    \end{tabular}
    \egroup{}
    \vspace{1.5em}
    \caption{Truth table for 2-3-1 Compression}%
    \label{tab:231_compression_table}
\end{table}
\begin{figure}[]
    \centering
    \[\input{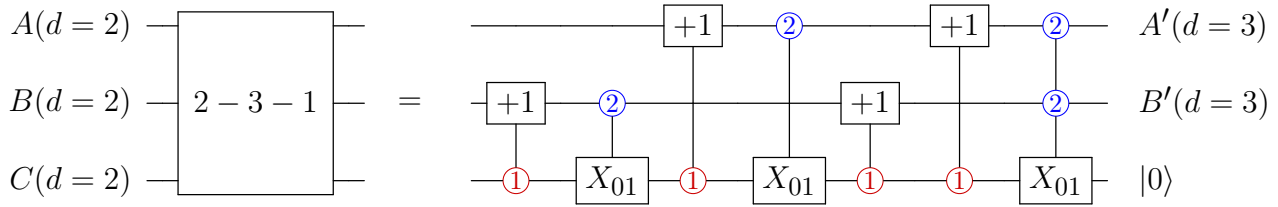}\]
    \caption{The compression of 3 qubits into 2 qutrits and an ancilla, $\ket{0}$. All $+1$ gates are done modulo 3. Using a sequence of qutrit gates, we can transform three input qubits into the desired ancilla. When A, B and C are not going to be used for a long time in the circuit, they can be temporarily repurposed as an ancilla bit elsewhere in the circuit. When we want to operate on these stored bits, we run the inverse of this circuit using \textit{any} ancilla for the third qubit.}%
    \label{fig:23compression}
\end{figure}

In 2-3-1 compression we take as input three qubits and output 2 qutrits and a single ancilla, a qubit guaranteed to be in the $\ket{0}$ state. First, consider the truth table of Table~\ref{tab:231_compression_table}. We note the partial function represented by this truth table is invertible, implying there exists a reversible circuit that realizes it. The third output, C', is guaranteed to be in the $\ket{0}$ state, an ancilla. By storing qubit information used infrequently we can generate an extremely useful ancilla to be used elsewhere in the circuit.
Because we ensure all inputs are binary, we do not need to consider the inputs with value $2$ to the ternary circuit. An example circuit realizing this truth table is given in Figure~\ref{fig:23compression}. When a compression circuit of this type is applied, we need to keep track of which pair of qutrits encodes the three qubits, in order.
When the compressed data is needed, we can decompress by applying the inverse of this function. The inverse circuit is simply the gates in reverse order with $+1$ replaced with $-1$. Notably, this inversion requires an ancilla as input.
To retrieve the information, the inverse should be applied taking in any free ancilla and then the stored bits can be computed on as normal.

This circuit, while accomplishing what is desired, can be rather inefficient. For example, in architectures with limited connectivity this circuit requires some number of expensive communication operations since every input qubit must be adjacent at some point. Furthermore, this circuit requires the use of a two-controlled qutrit gate which is typically decomposed into a sequence of 6 two-qutrit gates and 10 single-qutrit gates as in~\cite{Di}. In total this compression requires 22 gates, 12 two-qutrit and 10 single-qutrit gates.

\FloatBarrier{}

\subsection{Qubit to Ququart Compression}%
\label{sec:qudits-241compression}

While 2-3-1 compression required a fairly substantial number of gates, the 2-4-1 compression circuit can convert qubit inputs into ancilla more simply and with few gates. This does not come for free. In quantum computing,  we subject our computation to a greater probability of error by using higher radix gates and by persisting for longer durations in higher energy states. In Table~\ref{tab:241_compression_table}, we show that two qubits can be compressed into a single ququart and one ancilla.  2-4-1 compression is simpler than 2-3-1 compression because $2^2$ states fit evenly in a single ququart with $4$ states.  In Figure~\ref{fig:241_compression}, we show a compression circuit using only 3 two-ququart gates in total, a substantial reduction over the 2-3-1 counterpart.  In the next section, we show how compression and decompression can be used to design efficient circuits requiring no ancilla.

\begin{table}[h]
    \vspace{1.5em}  
    \centering
    \bgroup{}
    \setlength\tabcolsep{1.5em}
    \begin{tabular}{cc|cc}
        A & B & A' & B' \\\hline
        0 & 0 & 0  & 0  \\
        0 & 1 & 2  & 0  \\
        1 & 0 & 1  & 0  \\
        1 & 1 & 3  & 0  \\
    \end{tabular}
    \egroup{}
    \vspace{1.5em}
    \caption{Truth table for 2-4-1 Compression}%
    \label{tab:241_compression_table}
    \vspace{-1.5em}
\end{table}
\begin{figure}[h]
    \centering
    \[\input{chapters/qudits/figs/24compression.qcircuit}\]
    \caption{The compression of 2 qubits into a single ququart and generating an ancilla, $\ket{0}$. The $+2$ gate here is done modulo 4. This operation takes as input two qubits, A and B, and produces a single ququart and an ancilla $\ket{0}$. To do this, we need only 3 two-ququart gates. Similarly, to retrieve the stored information, we can do the inverse of this operation using \textit{any} ancilla for the second qubit.}%
    \label{fig:241_compression}
\end{figure}

\FloatBarrier{}

\section{A+B Adder}%
\label{sec:qudits-decompositions}

We now present our $A+B$ adder. This circuit takes as input two equal-sized registers of qubits, $A$ and $B$, and optionally carry-in or carry-out bits. This decomposition uses no ancilla and instead generates ancilla locally when needed by sub-components. In prior work, to achieve a logarithmic depth decomposition, $O(n)$ many ancilla were required where $n$ is the size of the input register. We will demonstrate how this efficient decomposition can be used along with our new compression technique to obtain an $O(\log n)$ depth decomposition of the same adder in-place without the extra use of ancilla.

\begin{figure}
    \centering
    \[
    \input{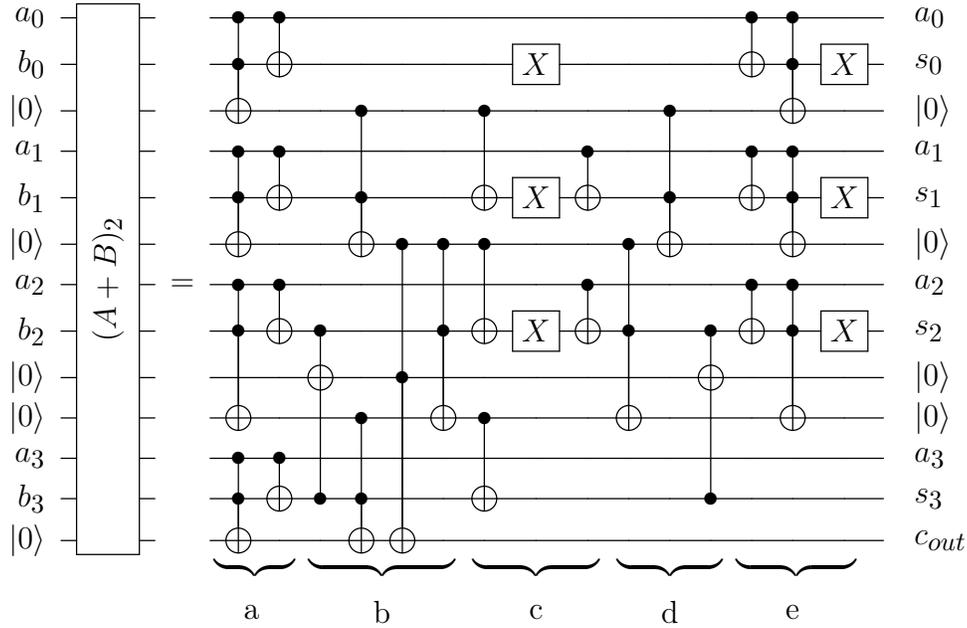}
    \]
    \caption{An adder circuit from~\cite{draper2006logarithmic} on two four-bit registers $A$ and $B$ with a carry-out bit using ancilla. The sum $S$ is computed in-place on register $B$ while $A$ is untouched and the ancilla are restored to $\ket{0}$. We use this as a sub-component of our general decomposition. Each of the ancilla in this circuit can be generated from other input qubits not shown here via our compression circuits. Part a of the circuit computes carry, generate, and propagate for each bit position. Part b computes the carry-in for every bit position. Part c does the addition, storing the output in register $B$. Parts d and e uncompute b and a respectively, restoring the ancilla back to $\ket{0}$.}%
    \label{fig:a_plus_b_prior}
\end{figure}

We first briefly review the work of~\cite{draper2006logarithmic} which gives a qubit-only in-place adder with ancilla which we will refer to as ${(A + B)}_2$. We give the decomposition for registers of size $4$ in Figure~\ref{fig:a_plus_b_prior}. One of the key contributions of this prior work is to demonstrate how, in logarithmic depth, the carry bits could be computed and used (and subsequently uncomputed to restore input ancilla back to the $\ket{0}$ state). This decomposition requires $2m - w(m) - \lfloor\log{m}\rfloor$ ancilla, where $w(m)$ is the number of $1$'s appearing in the binary expansion of the number of inputs, $m$. We will use this number later to determine how many ancilla to generate via compression. This same prior work demonstrates several variants of this circuit.  We require those with either a carry-in bit, a carry-out bit, or both.

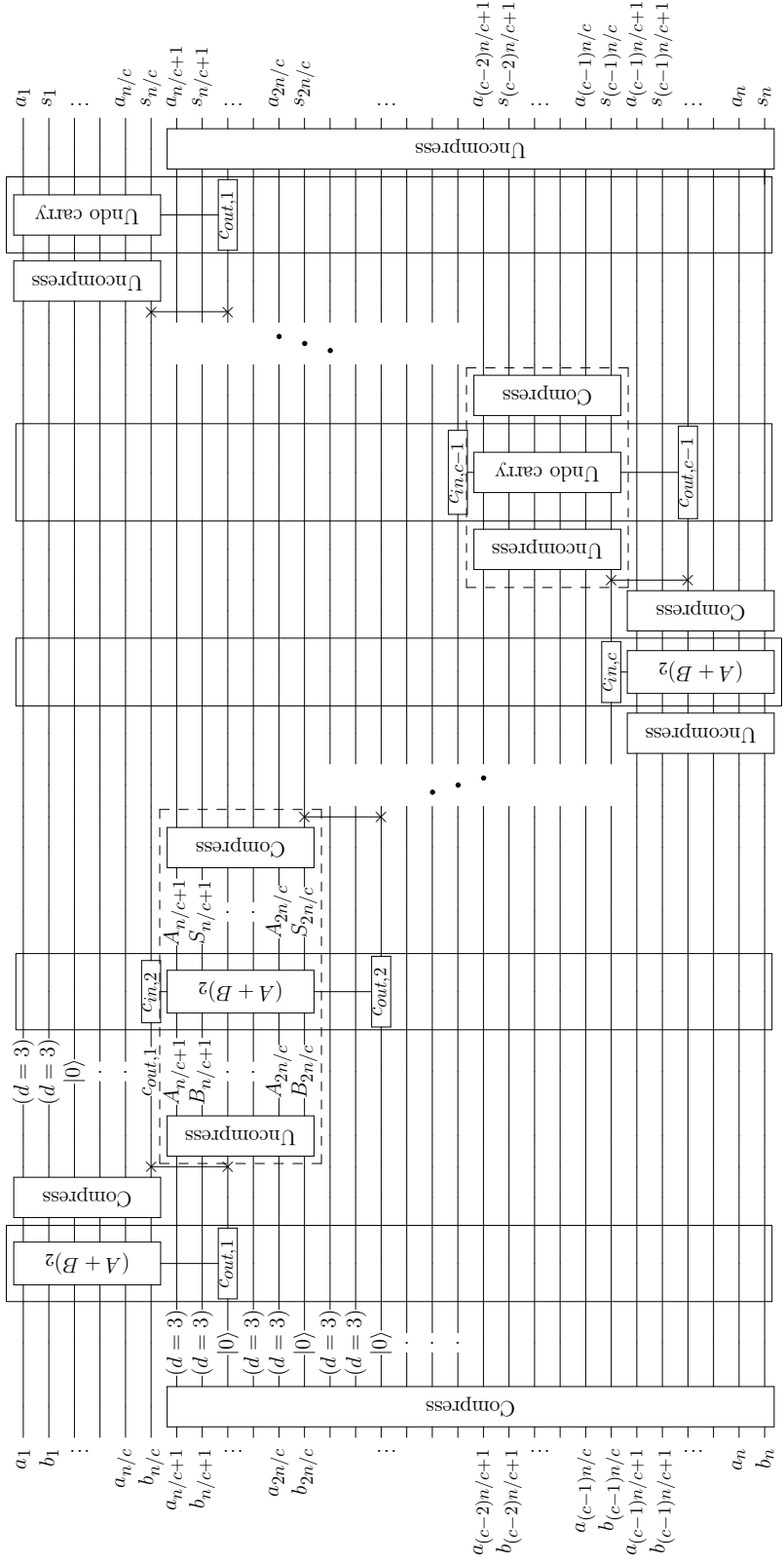
\begin{sidewaysfigure*}[p!]
    \centering
    \scalebox{0.7}{%
    \newcommand*\comp{{\mbox{\rotatebox[origin=c]{90}{Compress}}}}
\newcommand*\unc{{\mbox{\rotatebox[origin=c]{90}{Uncompress}}}}
\newcommand*\abgate{{\mbox{\rotatebox[origin=c]{90}{$(A+B)_2$}}}}
\newcommand*\uac{{\mbox{\rotatebox[origin=c]{90}{Undo carry}}}}
\newcommand*\bigdot{{\tikz{\filldraw (0,0) circle (0.1em);}}}
\Qcircuit @R=0.01em @C=0.5em @!R {
\lstick{a_1}	&	\qw	&	\qw	&	\multigate{5}{\abgate}	&	\multigate{5}{\comp}	&	\qw	&	\qw	&	\push{(d=3)}\qw	&	\push{{\rule{2.9em}{0em}}}	&	\qw[-2]	&	\qw	&	\qw	&	\qw	&	\push{{\rule{1em}{0em}}}	&	\qw[-2]	&	\qw			&	\push{{\rule{2.5em}{0em}}}	&	\qw[-2]	&	\qw	&	\qw	&	\push{{\rule{3.9em}{0em}}}	&	\qw[-2]	&	\qw	&	\push{{\rule{1em}{0em}}}	&	\qw[-2]	&	\qw	&	\multigate{5}{\unc}	&	\multigate{5}{\uac}	&	\qw	&	\rstick{a_1}	\qw	&		\\
\lstick{b_1}	&	\qw	&	\qw	&	\ghost{\abgate}	&	\ghost{\comp}	&	\qw	&	\qw	&	\push{(d=3)}\qw	&	\qw	&	\qw	&	\qw	&	\qw	&	\qw	&	\qw	&	\qw	&	\qw			&	\qw	&	\qw	&	\qw	&	\qw	&	\qw	&	\qw	&	\qw	&	\qw	&	\qw	&	\qw	&	\ghost{\unc}	&	\ghost{\uac}	&	\qw	&	\rstick{s_1}	\qw	&		\\
\lstick{\vdots}	&	\qw	&	\qw	&	\ghost{\abgate}	&	\ghost{\comp}	&	\qw	&	\qw	&	\push{\ket{0}}\qw	&	\qw	&	\qw	&	\qw	&	\qw	&	\qw	&	\qw	&	\qw	&	\qw			&	\qw	&	\qw	&	\qw	&	\qw	&	\qw	&	\qw	&	\qw	&	\qw	&	\qw	&	\qw	&	\ghost{\unc}	&	\ghost{\uac}	&	\qw	&	\rstick{\vdots}	\qw	&		\\
	&	\qw	&	\qw	&	\ghost{\abgate}	&	\ghost{\comp}	&	\qw	&	\qw	&	\push{~.~}\qw	&	\qw	&	\qw	&	\qw	&	\qw	&	\qw	&	\qw	&	\qw	&	\qw			&	\qw	&	\qw	&	\qw	&	\qw	&	\qw	&	\qw	&	\qw	&	\qw	&	\qw	&	\qw	&	\ghost{\unc}	&	\ghost{\uac}	&	\qw	&		\qw	&		\\
\lstick{a_{n/c}}	&	\qw	&	\qw	&	\ghost{\abgate}	&	\ghost{\comp}	&	\qw	&	\qw	&	\push{~.~}\qw	&	\qw	&	\qw	&	\qw	&	\qw	&	\qw	&	\qw	&	\qw	&	\qw			&	\qw	&	\qw	&	\qw	&	\qw	&	\qw	&	\qw	&	\qw	&	\qw	&	\qw	&	\qw	&	\ghost{\unc}	&	\ghost{\uac}	&	\qw	&	\rstick{a_{n/c}}	\qw	&		\\
\lstick{b_{n/c}}	&	\qw	&	\qw	&	\ghost{\abgate}	&	\ghost{\comp}	&	\qswap	&	\qw	&	\push{c_{out,1}}\qw	&	\multigate{0}{c_{in,2}}\qwx[1]	&	\qw	&	\qw	&	\qw	&	\qw	&	\qw	&	\qw	&	\qw			&	\qw	&	\qw	&	\qw	&	\qw	&	\qw	&	\qw	&	\qw	&	\qw	&	\qw	&	\qswap	&	\ghost{\unc}	&	\ghost{\uac}	&	\qw	&	\rstick{s_{n/c}}	\qw	&		\\
\lstick{a_{n/c+1}}	&	\multigate{23}{\comp}	&	\push{(d=3)}\qw	&	\qw	&	\qw	&	\qw	&	\multigate{5}{\unc}	&	\push{A_{n/c+1}}\qw	&	\multigate{5}{\abgate}	&	\push{A_{n/c+1}}\qw	&	\multigate{5}{\comp}	&	\qw	&	\qw	&	\qw	&	\qw	&	\qw			&	\qw	&	\qw	&	\qw	&	\qw	&	\qw	&	\qw	&	\qw	&		&		&	\qw	&	\qw	&	\qw	&	\multigate{23}{\unc}	&	\rstick{a_{n/c+1}}	\qw	&		\\
\lstick{b_{n/c+1}}	&	\ghost{\comp}	&	\push{(d=3)}\qw	&	\qw	&	\qw	&	\qw	&	\ghost{\unc}	&	\push{B_{n/c+1}}\qw	&	\ghost{\abgate}	&	\push{S_{n/c+1}}\qw	&	\ghost{\comp}	&	\qw	&	\qw	&	\qw	&	\qw	&	\qw			&	\qw	&	\qw	&	\qw	&	\qw	&	\qw	&	\qw	&	\qw	&		&		&	\qw	&	\qw	&	\qw	&	\ghost{\unc}	&	\rstick{s_{n/c+1}}	\qw	&		\\
\lstick{\vdots}	&	\ghost{\comp}	&	\push{\ket{0}}\qw	&	\multigate{0}{c_{out,1}}\qwx[-3]	&	\qw	&	\qswap\qwx[-3]	&	\ghost{\unc}	&	\push{~.~}\qw	&	\ghost{\abgate}	&	\push{~.~}\qw	&	\ghost{\comp}	&	\qw	&	\qw	&	\qw	&	\qw	&	\qw			&	\qw	&	\qw	&	\qw	&	\qw	&	\qw	&	\qw	&	\qw	&		&		&	\qswap\qwx[-3]	&	\qw	&	\multigate{0}{c_{out,1}}\qwx[-3]	&	\ghost{\unc}	&	\rstick{\vdots}	\qw	&		\\
	&	\ghost{\comp}	&	\push{(d=3)}\qw	&	\qw	&	\qw	&	\qw	&	\ghost{\unc}	&	\push{~.~}\qw	&	\ghost{\abgate}	&	\push{~.~}\qw	&	\ghost{\comp}	&	\qw	&	\qw	&	\qw	&	\qw	&	\qw			&	\qw	&	\qw	&	\qw	&	\qw	&	\qw	&	\qw	&	\qw	&		&		&	\qw	&	\qw	&	\qw	&	\ghost{\unc}	&		\qw	&		\\
\lstick{a_{2n/c}}	&	\ghost{\comp}	&	\push{(d=3)}\qw	&	\qw	&	\qw	&	\qw	&	\ghost{\unc}	&	\push{A_{2n/c}}\qw	&	\ghost{\abgate}	&	\push{A_{2n/c}}\qw	&	\ghost{\comp}	&	\qw	&	\qw	&	\qw	&	\qw	&	\qw			&	\qw	&	\qw	&	\qw	&	\qw	&	\qw	&	\qw	&	\qw	&	~~\bigdot	&		&	\qw	&	\qw	&	\qw	&	\ghost{\unc}	&	\rstick{a_{2n/c}}	\qw	&		\\
\lstick{b_{2n/c}}	&	\ghost{\comp}	&	\push{\ket{0}}\qw	&	\qw	&	\qw	&	\qw	&	\ghost{\unc}	&	\push{B_{2n/c}}\qw	&	\ghost{\abgate}	&	\push{S_{2n/c}}\qw	&	\ghost{\comp}	&	\qswap	&	\qw	&	\qw	&	\qw	&	\qw			&	\qw	&	\qw	&	\qw	&	\qw	&	\qw	&	\qw	&	\qw	&	~\bigdot~	&		&	\qw	&	\qw	&	\qw	&	\ghost{\unc}	&	\rstick{s_{2n/c}}	\qw	&		\\
	&	\ghost{\comp}	&	\push{(d=3)}\qw	&	\qw	&	\qw	&	\qw	&	\qw	&	\qw	&	\qw	&	\qw	&	\qw	&	\qw	&	\qw	&		&		&	\qw			&	\qw	&	\qw	&	\qw	&	\qw	&	\qw	&	\qw	&	\qw	&	\bigdot~~	&		&	\qw	&	\qw	&	\qw	&	\ghost{\unc}	&		\qw	&		\\
	&	\ghost{\comp}	&	\push{(d=3)}\qw	&	\qw	&	\qw	&	\qw	&	\qw	&	\qw	&	\qw	&	\qw	&	\qw	&	\qw	&	\qw	&		&		&	\qw			&	\qw	&	\qw	&	\qw	&	\qw	&	\qw	&	\qw	&	\qw	&		&		&	\qw	&	\qw	&	\qw	&	\ghost{\unc}	&		\qw	&		\\
\lstick{\vdots}	&	\ghost{\comp}	&	\push{\ket{0}}\qw	&	\qw	&	\qw	&	\qw	&	\qw	&	\qw	&	\multigate{0}{c_{out,2}}\qwx[-3]	&	\qw	&	\qw	&	\qswap\qwx[-3]	&	\qw	&		&		&	\qw			&	\qw	&	\qw	&	\qw	&	\qw	&	\qw	&	\qw	&	\qw	&		&		&	\qw	&	\qw	&	\qw	&	\ghost{\unc}	&	\rstick{\vdots}	\qw	&		\\
	&	\ghost{\comp}	&	\push{~.~}\qw	&	\qw	&	\qw	&	\qw	&	\qw	&	\qw	&	\qw	&	\qw	&	\qw	&	\qw	&	\qw	&		&		&	\qw			&	\qw	&	\qw	&	\qw	&	\qw	&	\qw	&	\qw	&	\qw	&		&		&	\qw	&	\qw	&	\qw	&	\ghost{\unc}	&		\qw	&		\\
	&	\ghost{\comp}	&	\push{~.~}\qw	&	\qw	&	\qw	&	\qw	&	\qw	&	\qw	&	\qw	&	\qw	&	\qw	&	\qw	&	\qw	&	\bigdot~~	&		&	\qw			&	\qw	&	\qw	&	\qw	&	\qw	&	\qw	&	\qw	&	\qw	&		&		&	\qw	&	\qw	&	\qw	&	\ghost{\unc}	&		\qw	&		\\
	&	\ghost{\comp}	&	\push{~.~}\qw	&	\qw	&	\qw	&	\qw	&	\qw	&	\qw	&	\qw	&	\qw	&	\qw	&	\qw	&	\qw	&	~\bigdot~	&		&	\qw			&	\qw	&	\qw	&	\qw	&	\qw	&	\multigate{0}{c_{in,c-1}}\qwx[1]	&	\qw	&	\qw	&		&		&	\qw	&	\qw	&	\qw	&	\ghost{\unc}	&		\qw	&		\\
\lstick{a_{(c-2)n/c+1}}	&	\ghost{\comp}	&	\qw	&	\qw	&	\qw	&	\qw	&	\qw	&	\qw	&	\qw	&	\qw	&	\qw	&	\qw	&	\qw	&	~~\bigdot	&		&	\qw			&	\qw	&	\qw	&	\qw	&	\multigate{5}{\unc}	&	\multigate{5}{\uac}	&	\multigate{5}{\comp}	&	\qw	&	\qw	&	\qw	&	\qw	&	\qw	&	\qw	&	\ghost{\unc}	&	\rstick{a_{(c-2)n/c+1}}	\qw	&		\\
\lstick{b_{(c-2)n/c+1}}	&	\ghost{\comp}	&	\qw	&	\qw	&	\qw	&	\qw	&	\qw	&	\qw	&	\qw	&	\qw	&	\qw	&	\qw	&	\qw	&		&		&	\qw			&	\qw	&	\qw	&	\qw	&	\ghost{\unc}	&	\ghost{\uac}	&	\ghost{\comp}	&	\qw	&	\qw	&	\qw	&	\qw	&	\qw	&	\qw	&	\ghost{\unc}	&	\rstick{s_{(c-2)n/c+1}}	\qw	&		\\
\lstick{\vdots}	&	\ghost{\comp}	&	\qw	&	\qw	&	\qw	&	\qw	&	\qw	&	\qw	&	\qw	&	\qw	&	\qw	&	\qw	&	\qw	&		&		&	\qw			&	\qw	&	\qw	&	\qw	&	\ghost{\unc}	&	\ghost{\uac}	&	\ghost{\comp}	&	\qw	&	\qw	&	\qw	&	\qw	&	\qw	&	\qw	&	\ghost{\unc}	&	\rstick{\vdots}	\qw	&		\\
	&	\ghost{\comp}	&	\qw	&	\qw	&	\qw	&	\qw	&	\qw	&	\qw	&	\qw	&	\qw	&	\qw	&	\qw	&	\qw	&		&		&	\qw			&	\qw	&	\qw	&	\qw	&	\ghost{\unc}	&	\ghost{\uac}	&	\ghost{\comp}	&	\qw	&	\qw	&	\qw	&	\qw	&	\qw	&	\qw	&	\ghost{\unc}	&		\qw	&		\\
\lstick{a_{(c-1)n/c}}	&	\ghost{\comp}	&	\qw	&	\qw	&	\qw	&	\qw	&	\qw	&	\qw	&	\qw	&	\qw	&	\qw	&	\qw	&	\qw	&		&		&	\qw			&	\qw	&	\qw	&	\qw	&	\ghost{\unc}	&	\ghost{\uac}	&	\ghost{\comp}	&	\qw	&	\qw	&	\qw	&	\qw	&	\qw	&	\qw	&	\ghost{\unc}	&	\rstick{a_{(c-1)n/c}}	\qw	&		\\
\lstick{b_{(c-1)n/c}}	&	\ghost{\comp}	&	\qw	&	\qw	&	\qw	&	\qw	&	\qw	&	\qw	&	\qw	&	\qw	&	\qw	&	\qw	&	\qw	&		&		&	\qw			&	\multigate{0}{c_{in,c}}\qwx[1]	&	\qw	&	\qswap	&	\ghost{\unc}	&	\ghost{\uac}	&	\ghost{\comp}	&	\qw	&	\qw	&	\qw	&	\qw	&	\qw	&	\qw	&	\ghost{\unc}	&	\rstick{s_{(c-1)n/c}}	\qw	&		\\
\lstick{a_{(c-1)n/c+1}}	&	\ghost{\comp}	&	\qw	&	\qw	&	\qw	&	\qw	&	\qw	&	\qw	&	\qw	&	\qw	&	\qw	&	\qw	&	\qw	&	\qw	&	\qw	&	\multigate{5}{\unc}			&	\multigate{5}{\abgate}	&	\multigate{5}{\comp}	&	\qw	&	\qw	&	\qw	&	\qw	&	\qw	&	\qw	&	\qw	&	\qw	&	\qw	&	\qw	&	\ghost{\unc}	&	\rstick{a_{(c-1)n/c+1}}	\qw	&		\\
\lstick{b_{(c-1)n/c+1}}	&	\ghost{\comp}	&	\qw	&	\qw	&	\qw	&	\qw	&	\qw	&	\qw	&	\qw	&	\qw	&	\qw	&	\qw	&	\qw	&	\qw	&	\qw	&	\ghost{\unc}			&	\ghost{\abgate}	&	\ghost{\comp}	&	\qw	&	\qw	&	\qw	&	\qw	&	\qw	&	\qw	&	\qw	&	\qw	&	\qw	&	\qw	&	\ghost{\unc}	&	\rstick{s_{(c-1)n/c+1}}	\qw	&		\\
\lstick{\vdots}	&	\ghost{\comp}	&	\qw	&	\qw	&	\qw	&	\qw	&	\qw	&	\qw	&	\qw	&	\qw	&	\qw	&	\qw	&	\qw	&	\qw	&	\qw	&	\ghost{\unc}			&	\ghost{\abgate}	&	\ghost{\comp}	&	\qswap\qwx[-3]	&	\qw	&	\multigate{0}{c_{out,c-1}}\qwx[-3]	&	\qw	&	\qw	&	\qw	&	\qw	&	\qw	&	\qw	&	\qw	&	\ghost{\unc}	&	\rstick{\vdots}	\qw	&		\\
	&	\ghost{\comp}	&	\qw	&	\qw	&	\qw	&	\qw	&	\qw	&	\qw	&	\qw	&	\qw	&	\qw	&	\qw	&	\qw	&	\qw	&	\qw	&	\ghost{\unc}			&	\ghost{\abgate}	&	\ghost{\comp}	&	\qw	&	\qw	&	\qw	&	\qw	&	\qw	&	\qw	&	\qw	&	\qw	&	\qw	&	\qw	&	\ghost{\unc}	&		\qw	&		\\
\lstick{a_{n}}	&	\ghost{\comp}	&	\qw	&	\qw	&	\qw	&	\qw	&	\qw	&	\qw	&	\qw	&	\qw	&	\qw	&	\qw	&	\qw	&	\qw	&	\qw	&	\ghost{\unc}			&	\ghost{\abgate}	&	\ghost{\comp}	&	\qw	&	\qw	&	\qw	&	\qw	&	\qw	&	\qw	&	\qw	&	\qw	&	\qw	&	\qw	&	\ghost{\unc}	&	\rstick{a_{n}}	\qw	&		\\
\lstick{b_{n}}	&	\ghost{\comp}	&	\qw	&	\push{{\rule{2.9em}{0em}}}	&	\qw[-2]	&	\qw	&	\qw	&	\qw	&	\qw	&	\qw	&	\qw	&	\qw	&	\qw	&	\qw	&	\qw	&	\ghost{\unc}			&	\ghost{\abgate}	&	\ghost{\comp}	&	\qw	&	\qw	&	\qw	&	\qw	&	\qw	&	\qw	&	\qw	&	\qw	&	\qw	&	\push{{\rule{2.9em}{0em}}}	&	\ghost{\unc}\qw[-2]	&	\rstick{s_{n}}	\qw	&	\push{{\rule{2em}{0em}}}	%
						{\gategroup{1}{4}{30}{4}{.7em}{-}}						{\gategroup{7}{7}{12}{12}{.7em}{--}}				{\gategroup{1}{9}{30}{9}{.7em}{-}}																		{\gategroup{1}{17}{30}{17}{.7em}{-}}				{\gategroup{19}{19}{24}{22}{.7em}{--}}				{\gategroup{1}{21}{30}{21}{.7em}{-}}														{\gategroup{1}{28}{30}{28}{.7em}{-}}								%
}%
    }
    \caption{Our $A + B$ adder that uses no external ancilla.  The variant shown here for $c=5$ uses 2-3-1 compression to generate one ancilla (marked as $\ket{0}$) for every three unused qubits, storing their values in two qutrits (marked as $d=3$).  A box is drawn around every ${(A+B)}_2$ and Undo carry gate to indicate that they use all the generated ancilla across the circuit.  $c_{out,i}$ or $c_{in,i}$ is included on some of the gates to indicate when the carry-in and carry-out versions are used and on which ancilla the carry-out is stored.  The SWAP gates (pairs of $\times$ in the diagram) simply move a carry-out bit to another ancilla where it is used as the next carry-in. The two blocks of gates shown with dashed lines are repeated $c-2 = 3$ times along the diagonal indicated.  If 2-4-1 compression is used, an ancilla is generated for every two unused qubits so only $c=4$ blocks are needed.  The depth of this circuit is $O(\log{n})$.}%
    \label{aplusbblocks}%
\end{sidewaysfigure*}

We will now present our decomposition shown in Figure~\ref{aplusbblocks}. Let $A = (a_1a_2\dots a_{n})$ and $B = (b_1b_2\dots b_n)$ be the input registers with $a_1, b_1$ the least significant bits of each register. We divide these registers into $c$ blocks $R_1, \dots, R_c$ each of size $2n / c$.  We assume for clarity that $n$ is a multiple of $c$ but our constructions will work for any $n$, with one additional block containing the remaining $2(n \mod c)$ qubits. Take $$R_i = (a_{(i-1)(c/n)+1}b_{(i-1)(c/n)+1}\dots a_{i(c/n)}b_{i(c/n)})$$ then notice for $i > 1$ we can perform an addition circuit ${(A+B)}_2$ with carry-in and carry-out on block $R_i$ in $O(\log(n/c)) = O(\log n)$ depth by generating the proper number of ancilla out of the other input qubits, specifically $2(n/c) - w(n/c) - \lfloor\log{n/c}\rfloor$ ancilla. We will assume a worst case scenario of $2n/c$ ancilla to simplify the analysis. Suppose we are performing ${(A + B)}_2$ on block $R_i$ while every other block is unused. We can perform compression on the currently unused qubits in all other blocks $\{R_j | j \neq i\}$ to obtain generated ancilla which can then be used by the current adder subcircuit.

Recall 2-3-1 compression takes 3 qubits and outputs a single ancilla. Let a 2-3-1 \textit{Compress circuit} be a circuit which takes any number of qubits $m$ as input and applies 2-3-1 compression to triplets resulting in $\lfloor m / 3 \rfloor$ ancilla. Then applying 2-3-1 compression to all qubits in $\{R_j | j \neq i\}$ we obtain $\lfloor (c - 1)2n/3c \rfloor$ ancilla. We now have constraints on what the constant $c$ should be for our decomposition to be feasible. That is we must have $\lfloor (c-1)2n/3c\rfloor \ge 2n/c$. Because we must store intermediate carry values between each ${(A+B)}_2$, we will actually require an additional $c-1$ ancilla, giving us $\lfloor (c - 1)2n/3c \rfloor \ge 2n/c + c - 1$.  By solving the inequality, this implies our construction is feasible for $c = 5$ and $n \ge 30$.  An alternative adder that is ancilla-free but does not scale well asymptotically, like the $O(n)$-depth adder by \cite{cuccaro}, may be used where our construction is infeasible on small problem sizes with $n < 30$.

The circuit construction now goes as follows, first considering the case when we have no carry-in and no carry-out. To add in these additional features requires only minor adjustments, discussed later. First, we compress the qubits in blocks $\{R_j | j \neq 1\}$. Then we apply ${(A + B)}_2$ with carry-out to the block $R_1$ using the newly generated ancilla. The compression block is constant depth ($O(1)$) and the adder is logarithmic depth ($O(\log(n/c)) = O(\log n)$). The qubits $b_1, \dots, b_{n/c}$ now store the first $n/c$ bits of the addition, $s_1, \dots, s_{n/c}$.  Also note the adder circuit restores all ancilla (except the carry-out) to $\ket{0}$. Then, apply a compression block to $R_1$. Swap the carry-out, $c_{out,1}$, to any of the ancilla generated to hold on to whether a carry should be applied to the next block (these carries are where the additional $c-1$ term come from above). Next, we uncompress all of the bits in $R_2$
so we can apply ${(A+B)}_2$ with carry-out \textit{and} carry-in ($c_{in} = c_{out,1}$) to block $R_2$ using the other generated ancilla. We repeat this process until the last block, $R_c$. In this case, since we do not have a carry-out bit we apply ${(A+B)}_2$ with only carry-in ($c_{in} = c_{out, c-1}$).

We have now computed the sum $A+B$ and now must cleanup the intermediate carry bits.  This can be done by working in reverse to uncompute each carry-out without undoing the addition.
One intuitive way would be to simply apply the inverse of the ${(A+B)}_2$ circuit we applied to block $R_{c-1}$ which will uncompute the addition and $c_{out, c-1}$ and then re-apply it \textit{without} carry-out. Now the ancilla storing $c_{out, c-1}$ is restored to $\ket{0}$. We repeat this process on each of the blocks in reverse order. Finally, after $c_{out,1}$ has been uncomputed and the ancilla restored to $\ket{0}$, we uncompress all of the qubits. The resulting output will be the sum $S$ in register $B$ with register $A$ left unchanged from the input.

Uncomputing the intermediate carry-out bits can be improved dramatically by noticing that by applying the inverse of ${(A+B)}_2$ with carry-in and carry-out and the subsequently applying ${(A+B)}_2$ with only carry-in is unnecessary. Instead we can uncompute the carry-out by only applying the inverse of the second half of ${(A+B)}_2$ with carry-out and then executing the second half of ${(A+B)}_2$ with a few extra gates in Figure~\ref{fig:a_plus_b_prior}d to cancel the carry-out.

Earlier, we show our decomposition only works when $c = 5$ using 2-3-1 compression. However, due to page size constraints, we do not show some of the repeated blocks in Figure \ref{aplusbblocks}. The block of gates surrounded by a dashed line is simply repeated in a block diagonal pattern indicated by the ellipsis. If we instead used 2-4-1 compression, the factor of $3$ in the earlier inequality would be replaced with $2$ making $c=4$ feasible with a constraint of $n \geq 12$.

Our decomposition performs addition in-place with zero ancilla, taking advantage of qutrits (qudits in general) to obtain ancilla instead of extra qubits for ancilla. Each of the ${(A+B)}_2$ blocks has depth $O(\log n)$ for input register size $n$ and we perform only a constant $2c-1$ of them so our decomposition also has $O(\log n)$ depth.

\subsection{Carry-in and Carry-out}%
\label{sec:qudits-carryin_carryout}

We can extend the above decomposition to allow for carry-in quite simply. When computing the ${(A+B)}_2$ and Undo carry on $R_1$ we simply use the ${(A+B)}_2$ circuit with carry-in. Similarly, we can allow for carry-out by simply substituting an ${(A+B)}_2$ with carry-in \textit{and} carry-out on block $R_c$.

\subsection{+K Adder}%
\label{sec:qudits-plus_k}

The method used to construct the $A+B$ adder shown above can be applied to any circuit that can be divided into blocks while only needing to pass a constant number of bits to the input of the following block.  One example that follows from $A+B$ is the $+K$ adder.  The $+K$ adder acts on a single register of qubits $B$ and computes the sum $B+K$ in-place where $K$ is a classical constant known when creating the circuit.

The design of our $+K$ adder will use as subcircuits the ${(+K)}_2$ circuit derived from ${(A+B)}_2$ from~\cite{draper2006logarithmic} and described earlier.  The design of ${(+K)}_2$ is the same as ${(A+B)}_2$ except the qubits of register $A$ are removed and all CNOT gates with a control on $a_i$ are removed and only replaced with $X$ gates if $k_i = 1$.  Similarly, the Toffoli gates (controlled-controlled-not gates) are removed and replaced with CNOT gates in the same way.  Depending on the value of $K$, some of the ancilla may also be removed but in the worst case, ${(+K)}_2$ may still require $2n/c - w(n/c) - \lfloor\log{n/c}\rfloor - 1$ ancilla for input size $n/c$ which we upper bound by $2n/c$.  The circuit still has $O(\log n)$ depth.

We use the same diagonal block structure as $A+B$ but now we define $$R_i = (b_{(i-1)(c/n)+1}\dots b_{i(c/n)})$$
At step $i$, the number of ancilla generated by applying 2-3-1 compression to all qubits in $\{R_j | j \neq i\}$ is $\lfloor (c - 1)n/3c \rfloor$.  From this, we obtain the inequality $\lfloor (c - 1)n/3c \rfloor \geq 2n/c + c - 1$ which determines when there are enough unused qubits to generate the required ancilla.  The extra $c - 1$ ancilla are needed to store intermediate carry values.  When we solve this inequality, we find that $c=8$ blocks are required and the circuit will only have enough ancilla when $n \geq 168$.  Both the number of blocks and the minimum $n$ are larger than for $A+B$ because the input to $+K$ is only a single register so the ancilla required per input qubit is doubled, resulting in a higher minimum $n$.

2-3-1 compression is not the only option.  If we use 2-4-1 compression instead, more ancilla can be generated per input qubit and we obtain the inequality $\lfloor (c - 1)n/2c \rfloor \geq 2n/c + c - 1$.  The solution to this tells us that the minimum $c=6$ and we can use the circuit for $n \geq 60$.

\FloatBarrier{}

\section{Discussion and Summary}%
\label{sec:qudits-discussion}

We have shown a new use of qudits in circuit designs, to generate ancilla in-place, and its application to the class of quantum circuits that can be split into blocks. We give a new construction for an in-place addition circuit that uses no ancilla but still obtains the same $O(\log n)$ asymptotic depth as the qubit circuit it was based on that needed $O(n)$ ancilla.  The new circuit can be used as a drop-in replacement in algorithms to use significantly fewer total qubits.  These results should encourage further use of the temporary-qudit abstraction in qudit-assisted quantum computing.

A number of useful quantum circuits, especially arithmetic circuits, make extensive use of multiply-controlled gates. However, these circuits are typically pre-compiled into single- and two-qubit gates using one of the decompositions from prior work, usually one that involves ancilla qubits. Revisiting these arithmetic circuits from first principles, with our qutrit circuit as a new tool, could yield novel and improved circuits like our Incrementer circuit in Section~\ref{sec:qudits-incrementer} and Adder circuit in Section~\ref{sec:qudits-decompositions}.

It still remains to be seen what the most intuitive way for quantum programmers to use qudits.
We have only shown hand-designed subroutines and compression strategies to use temporary qutrits.
The hand-designed generalized Toffoli implementation makes excellent use of one additional logical state and, while hand-optimization can be a good way to squeeze performance out of resource-constrained devices, codifying manual strategies into our compilers can have wider performance benefit and free most programmers to think at a higher level.
The qubit ``compression'' strategy shows benefit in the design of the quantum adder arithmetic circuit, indicating that this strategy could have wider uses.
For example, a compiler could intelligently ``compress'' binary quantum data (stored in groups of idle qubits) into smaller groups of qutrits (using the $\log_2(3)$ compression ratio) or qudits ($\log_2(d)$ ration).
This has the potential to automate the benefits of temporary qutrits to all quantum programs.

Our circuit constructions and compression strategies point towards the benefit of temporary qudit abstraction.
Researchers will undoubtedly find more uses for qudits as a form of temporary data storage, enabled by this new way of thinking about quantum data beyond the binary representation.

\FloatBarrier{}



\chapter{Spatially Local Memory}%
\label{chapter:memory}

\graphicspath{{chapters/memory/svg-inkscape/}}

\section{Introduction}%
\label{sec:memory-introduction}

From the previous chapter, when we abstract multi-level qudits as a two-level qubit with additional storage we enable improvements in new circuit designs and design strategies.
In this chapter, we look at a different form of storage, viewed through the lens of spacial constrains in 2D and 3D.
These constraints come from 2D chip limitations and the 3D world we live in due to the fragility of qubit-to-qubit communications.

Here we focus on error correction, the underlying architecture needed for future fault-tolerant quantum computing.
We are currently in the NISQ (Noisy Intermediate-Scale Quantum, \cite{nisq}) era where great progress has been made at the software level such as improved compilation procedures reducing required overhead for program execution.
However, these machines will be too small for error correction and unable to run large-scale programs due to unreliable qubits.
The ultimate goal is to construct fault-tolerant machines capable of executing thousands of gates and, in the long-term, to execute large-scale algorithms such as \cite{shor} and \cite{grover} with speedups over classical algorithms.
There are a number of promising error correction schemes which have been proposed such as the color code from~\cite{color-code} or the surface code from~\cite{fowler-braid,fowler-lattice,gidney2019factor}.
The surface code is a particularly appealing candidate because of its low overhead, high error threshold, and its reliance on few nearest-neighbor interactions in a 2D array of qubits, a common feature of superconducting transmon qubit hardware.
In fact, Google's next milestone is to demonstrate error corrected qubits (\cite{quantum-supremacy,martinis-caltech}).

Current architectures for both NISQ and fault-tolerant quantum computers make no distinction between the memory and processing of quantum information (typically represented in qubits).
While monolithic designs are currently viable, the engineering challenges of scaling up to hundreds of qubits become readily apparent as larger devices are considered.
For transmon technology used by Google, IBM, and Rigetti, some of these issues include fabrication consistency and crosstalk during parallel operations.
Every qubit needs dedicated control wires and signal generators which fill the refrigerator the device runs in.
To scale to the millions of qubits needed for useful fault-tolerant machines like in \cite{gidney2019factor}, we need some kind of memory-like abstraction to manage the massive scale these devices will need to achieve.
Memory, in essence, decouples the amount of data storage (qubit-count) from the scale of the instruction processor (transmon-count or laser waveguide size, this is classically the CPU).
Finding the most appropriate memory-like abstraction will enable quantum computers to scale more effectively.

In this chapter, we evaluate a recently realized qubit memory technology which stores quantum data in a superconducting cavity local to each qubit (\cite{Quantum_RAM}).
This technology, while new, is expected to become competitive with existing transmon devices.
Stored in cavity, qubits have a significantly longer lifetime (coherence time) but must be loaded into a transmon for computation.
This longer lifetime can increase the total amount of computation before data is lost to errors.
Although the basic concept of a compute qubit and associated memory has been demonstrated experimentally, the contribution of the chapter is to design and evaluate a system-level organization of these components within the context of a novel surface code embedding and fault-tolerant quantum operations.
We provide a proof of concept in the form of a practical use case motivating more complex experimental demonstrations of larger systems using this technology.

\begin{figure}[t]
    \centering
    \scalebox{\figfirstextrascale}{%
    \scalebox{\figscale}{%
    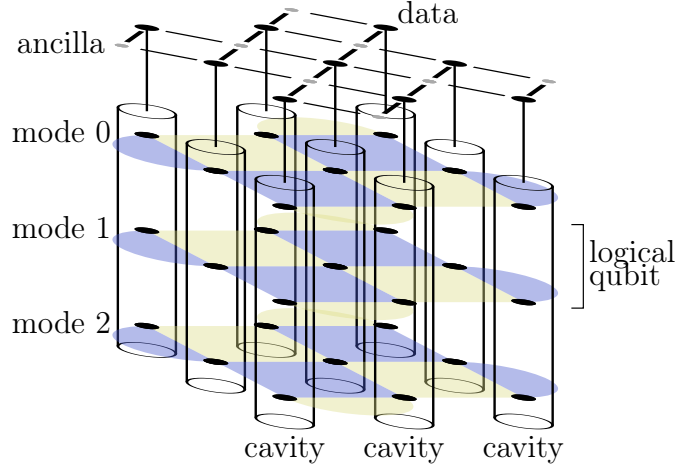%
    }}
    \caption{Our fault-tolerant architecture with random-access memory local to each transmon.
    On top is the typical 2D grid of transmon qubits.
    Attached below each data transmon is a resonant cavity storing error-prone data qubits (shown as black circles).
    This pattern is tiled in 2D to obtain a 2.5D array of logical qubits.
    Our key innovation here is storing the qubits that make up each logical qubit (shown as checkerboards) across many cavities to enable efficient computation.
    }%
    \label{fig:stacked-surfaces}
\end{figure}

Our proposed 2.5D memory-based design (originally presented in \cite{vlq}\footnotemark{}) is a typical 2D grid of transmons with memory added as shown in Figure~\ref{fig:stacked-surfaces}.
This can be compared with the traditional 2D error correction implementation in Figure~\ref{fig:regular-surface-code}, where the checkerboards represent error-corrected logical qubits.
The logical qubits in this system are stored at unique virtual addresses in memory cavities when not in use.
They are loaded to a physical address in the transmons and made accessible for computation on request and are periodically loaded to correct errors, similar to DRAM refresh.
This design allows for more efficient operations such as the transversal CNOT between logical qubits sharing the same physical address, i.e.\ co-located in the same cavities.
This is not possible on the surface code in 2D which requires methods such as braiding or lattice surgery for a CNOT operation.
\footnotetext{CD and JMB contributed equally to the work that comprises this chapter.  CD's contributions include refinements to the surface code mapping, compact embedding and CNOT sequence, numerical results, and magic state analysis.}

We introduce two embeddings of the 2D surface code to this new architecture that spread logical qubits across many cavities.
Despite serialization due to memory access, we are able to store and error-correct stacks of these logical qubits.
Furthermore, we show surface code operations via lattice surgery can be used unchanged in this new architecture while also enabling a more efficient CNOT operation.
Similarly, we are able to use standard and architecture-specific magic-state distillation protocols like \cite{game-of-codes} in order to ensure universal computation.
Magic-state distillation is a critical component of error-corrected algorithms so any improvement will directly speed up algorithms including Shor's and Grover's.

We discuss several important features of any proposed error correction code, such as the threshold error rate (below which the code is able to correct more errors than its execution causes), the code distance, and the number of physical qubits to encode a logical qubit.
In many codes, the number of physical qubits can be quite large.
We develop an embedding from the standard representation to this new architecture which reduces the required number of physical transmon qubits by a factor of approximately $k$, the number of resonant modes per cavity.
We also develop a Compact variant saving an additional 2x.
This is significant because we can obtain a code distance $\sqrt{2k}$ times greater or use hardware with only $\frac{1}{2k}$ the required physical transmons for a given algorithm.
In the near-to-intermediate term, when qubits are a highly constrained resource, this will accelerate a path towards fault-tolerant computation.
In fact, the smallest instance of Compact requires only 11 transmons and 9 cavities for $k$ logical qubits.

We evaluate variants of our architecture by comparing against the surface code on a larger 2D device.
Specifically, we determine the error correction threshold rates via simulation for each and find they are all close to the baseline threshold.
This shows the additional error sources do not significantly impact the performance.
We explore the sensitivity of the threshold to many different sources of error, some of which are unique to the memory used in this architecture.
We end by evaluating magic-state distillation protocols which have a large impact on overall algorithm performance and find a 1.22x speedup normalized by the number of transmon qubits.

In summary, we make the following contributions:
\begin{itemize}
    \item We introduce a 2.5D architecture where qubit-local memory is used for random access to error-corrected, logical qubits stored across different memories.
        This allows a simple virtual and physical address scheme that exemplifies exposing native data locality to the application level.
        Error correction is performed continuously by loading each from memory.
    \item We give two efficient adaptations of the surface code in this architecture, Natural and Compact.
        Unlike a naive embedding, both support fast transversal CNOTs in addition to lattice surgery operations with improved connectivity between logical qubits.
    \item We develop an error correction implementation optimized for Compact and designed to maximise parallelism and minimize the spread of errors.
    \item Via simulation, we determine the surface code adapted to our 2.5D architecture is still an effective error correction code while greatly reducing hardware requirements.
\end{itemize}

\FloatBarrier{}

\section{Background}%
\label{sec:memory-background}

In this section we briefly introduce the basics of quantum computation. We review current superconducting qubit architectures and memory technology our proposed design takes advantage of.  We then discuss the noise present in these physical systems. Next, we introduce the basics of quantum error correction and give a detailed introduction to the surface code and lattice surgery. We conclude with a review of the basic procedure for decoding physical errors.

\subsection{Basics of Quantum Computing}%
\label{sec:memory-basics}

The fundamental unit of quantum computing is the qubit. Like the classical bit, it can exist in the $\ket{0}$ or $\ket{1}$ state, but it may also exist in a coherent superposition of the two states and $n$ qubits may exist in a superposition of all $2^n$ bit strings. For example, a single qubit state is $\ket{\psi} = \alpha\ket{0} + \beta\ket{1}$ where $\abs{\alpha}^2 + \abs{\beta}^2 = 1$ and $\alpha, \beta \in \mathbb{C}$. To manipulate these bits we apply quantum operations, often called gates. Single qubit gates like X (bit flip), Z (phase flip), H (Hadamard basis change), and T ($\frac{\pi}{4}$ phase) and two-qubit gates like CNOT (reversible XOR with output $b'=a\oplus b$) are unitary and reversible (invertible). We may measure a qubit to obtain either a 0 or a 1 outcome with probabilities $\abs{\alpha}^2$ and $\abs{\beta}^2$, respectively. Multi-qubit operations like CNOT can create entanglement between qubits. Using only CNOT and single qubit gates, universal computation is possible, meaning any reversible multi-qubit operation is possible.  The three-qubit Toffoli (reversible AND gate with output $c'=(a\wedge b)\oplus c$), a common primitive in error-corrected algorithms, can be implemented by performing a few CNOT, H, and T gates.  See~\cite{mikeike} for a more comprehensive background.

\subsection{Superconducting Qubit Architectures}%
\label{sec:memory-architectures}

\begin{figure}[h]
    \makebox[0.9\columnwidth][c]{\scalebox{\figscale}{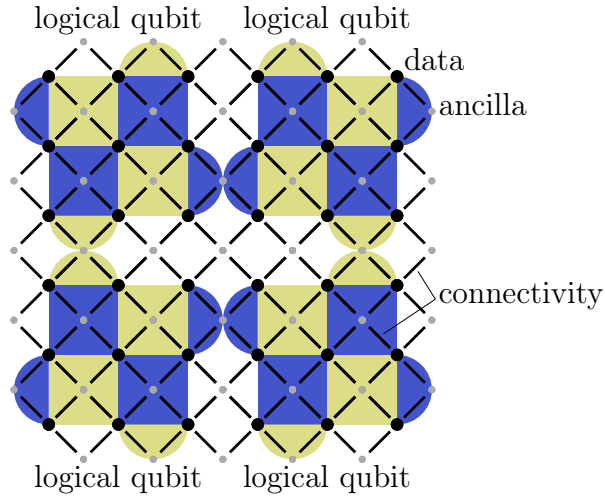}}
    \caption{A typical 2D superconducting qubit architecture.  The dots are transmon qubits where black are used as data and gray are used as ancilla for error correction.  The lines indicate physical connections between qubits that allow operations between them.  Four logical qubits, each consisting of 9 error-prone data qubits, are shown here in the rotated surface code with distance 3.  Z parity checks are shaded yellow (light) and X parity checks are shaded blue (dark) where checks on only 2 data are drawn as half circles.
    }%
    \label{fig:regular-surface-code}
\end{figure}

In contrast to other leading qubit technologies such as trapped ion devices with one or more fully-connected qubit chains, superconducting qubits are typically connected in nearest-neighbor topologies, often a 2D mesh on a regular square grid as shown in Figure~\ref{fig:regular-surface-code}. For near-term computation, this limitation makes engineering these devices easier but results in high communication costs, increasing the chance of errors on NISQ devices and communication congestion for error corrected operations.  This is a leading technology in industry, used by Rigetti, IBM, and Google.

\subsection{Qubit Memory Technology}

\begin{figure}[h]
    \centering
    \scalebox{\figfirstextrascale}{%
    \scalebox{\figscale}{%
\begingroup%
  \makeatletter%
  \providecommand\color[2][]{%
    \errmessage{(Inkscape) Color is used for the text in Inkscape, but the package 'color.sty' is not loaded}%
    \renewcommand\color[2][]{}%
  }%
  \providecommand\transparent[1]{%
    \errmessage{(Inkscape) Transparency is used (non-zero) for the text in Inkscape, but the package 'transparent.sty' is not loaded}%
    \renewcommand\transparent[1]{}%
  }%
  \providecommand\rotatebox[2]{#2}%
  \newcommand*\fsize{\dimexpr\f@size pt\relax}%
  \newcommand*\lineheight[1]{\fontsize{\fsize}{#1\fsize}\selectfont}%
  \ifx\svgwidth\undefined%
    \setlength{\unitlength}{104.02734512bp}%
    \ifx\svgscale\undefined%
      \relax%
    \else%
      \setlength{\unitlength}{\unitlength * \real{\svgscale}}%
    \fi%
  \else%
    \setlength{\unitlength}{\svgwidth}%
  \fi%
  \global\let\svgwidth\undefined%
  \global\let\svgscale\undefined%
  \makeatother%
  \begin{picture}(1,1.23130448)%
    \lineheight{1}%
    \setlength\tabcolsep{0pt}%
    \put(0,0){\includegraphics[width=\unitlength,page=1]{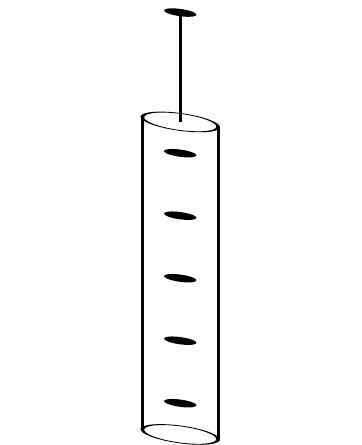}}%
    \put(0.62880196,0.7812369){\color[rgb]{0,0,0}\makebox(0,0)[lt]{\lineheight{1.25}\smash{\begin{tabular}[t]{l}mode 0\end{tabular}}}}%
    \put(0.73694661,0.41787092){\color[rgb]{0,0,0}\makebox(0,0)[lt]{\lineheight{1.25}\smash{\begin{tabular}[t]{l}\vdots\end{tabular}}}}%
    \put(0.62880196,0.08911121){\color[rgb]{0,0,0}\makebox(0,0)[lt]{\lineheight{1.25}\smash{\begin{tabular}[t]{l}mode k\end{tabular}}}}%
    \put(0.2827391,0.86775261){\color[rgb]{0,0,0}\makebox(0,0)[rt]{\lineheight{1.25}\smash{\begin{tabular}[t]{r}cavity\end{tabular}}}}%
    \put(0.41251267,1.17055761){\color[rgb]{0,0,0}\makebox(0,0)[rt]{\lineheight{1.25}\smash{\begin{tabular}[t]{r}transmon\end{tabular}}}}%
  \end{picture}%
\endgroup%
    }}
    \caption{A close-up representation of the qubit memory technology we use.  On top is a superconducting transmon qubit physically connected to a resonant superconducting cavity.  This cavity has many resonant modes each used to store a qubit.  These qubits can be loaded and stored (with random access) via the transmon.
    }%
    \label{fig:single-cavity}
\end{figure}

Recently studies, including \cite{Quantum_RAM,yale-arch}, have demonstrated random access memory for quantum information. Qubit states can be stored in the resonant modes of physical superconducting cavities attached to a transmon qubit as depicted in Figure~\ref{fig:single-cavity}. In these devices, transmon-transmon interactions are essentially the same as other superconducting transmon technology and transmon-cavity interactions are expected to perform similarly. Currently demonstrated error rates are promising, and there is nothing fundamental preventing this technology from becoming competitive with other transmon devices. We expect operation error rates to improve, cavity sizes and coherence times to increase and in general expect performance to improve as it has with other quantum technologies.

Local memory is not free. Stored qubits cannot be operated on directly. Instead, operations on this information are mediated through the transmon. Furthermore, to operate on qubits stored in memory, we first load the qubit from memory. Then we perform the desired operation on the transmons, and store the qubit back in its original location. A two-qubit operation such as a CNOT can also be performed directly between the transmon and a qubit in its connected cavity by manipulating higher states of the transmon.  We use this transmon-mode CNOT later.

In this architecture, qubits stored in the same cavity cannot be operated on in parallel. For example, consider two qubits stored in different modes of the same cavity (two virtual addresses corresponding to the same physical address). If we want to perform an H gate on each of them in parallel, this would not be possible. Instead, we serialize these operations. There are two primary benefits of this technology. First, we are able to quickly perform two-qubit interactions between any pair of qubits stored in the same cavity because we have star-graph connectivity between the transmon and its cavity modes. Second, qubits stored in the cavity are expected to have longer coherence times by about one order of magnitude i.e.\ there will be 10x fewer idle errors when qubits are stored in the cavity.

\subsection{Quantum Errors}%
\label{sec:memory-errors}

Quantum systems are inherently noisy, subject to a variety of coherent and non-coherent error. For example, when attempting to apply some gate $U$ to a qubit we may actually apply some other gate $U'$ which is close to the desired operation but may include an additional undesired operation. Fortunately, this type of coherent error is fairly easy to model. Since every single-qubit unitary can be expressed as a linear combination of the Pauli matrices\footnotemark{} $I, X, Y, Z$ we can express this coherent error as a combination of bit flip (X) and phase flip (Z) errors where $I$ is no error and $Y$ is simultaneous bit and phase errors ($Y=iXZ$). For a quantum error correcting code this will play a part in digitizing errors, meaning we will be able to simply detect and correct $X$ and $Z$ errors.
\footnotetext{The Pauli matrices along with $I$ form a complete basis over complex matrices so any single-qubit unitary $U=aI+bX+cY+dZ$ where $X=\begin{bmatrix}0&1\\1&0\end{bmatrix}$, $Y=\begin{bmatrix}0&-i\\i&0\end{bmatrix}$, $Z=\begin{bmatrix}1&0\\0&-1\end{bmatrix}$.}

Errors such as decoherence errors can be attributed to interaction with the environment. These errors are inevitable because manipulating qubits requires they not be perfectly isolated. When modeling and simulating this type of error we require the use of full density matrix simulation. In this paper, we opt not to model coherence errors in this way because simulation of this class of errors is hard (density matrices have size exponential in the number of qubits), we instead also model storage errors as Pauli errors. This is a common simplification and a conservative overestimate for the error causing our error threshold estimation to be slightly more conservative. For example, when decoherence resets a qubit to $\ket{0}$, this causes an error to a qubit in the $\ket{1}$ state but not to a qubit already in the $\ket{0}$ state whereas a Pauli X error causes a bit flip which is an error on either state.

The above errors apply to all superconducting systems and we often assume consistent error rates across the device. We treat all two-qubit interactions equally so gates like a CNOT incur some fixed error cost, a fixed chance of some error $U_1 \otimes U_2$ is applied to $\ket{\psi}$ where $U_1, U_2 \in \{I, X, Y, Z\}$. In traditional superconducting architectures (our baseline), we consider a few error sources --- storage error, one and two-qubit gate error, and measurement error. In superconducting architectures with resonant cavities such as our design, there is more nuance. We consider cavity storage and transmon storage error rates separately since each has its own coherence time and we separate transmon-transmon two-qubit gates and transmons-cavity two-qubit gates. We detail this and our other assumptions for simulation in experimental setup.

%
%
%

\begin{figure*}
    \centering
    \vspace{0.5em}
    \makebox[\textwidth][c]{%
        \scalebox{\figscale}{%
        \def\svgwidth{\textwidth}
        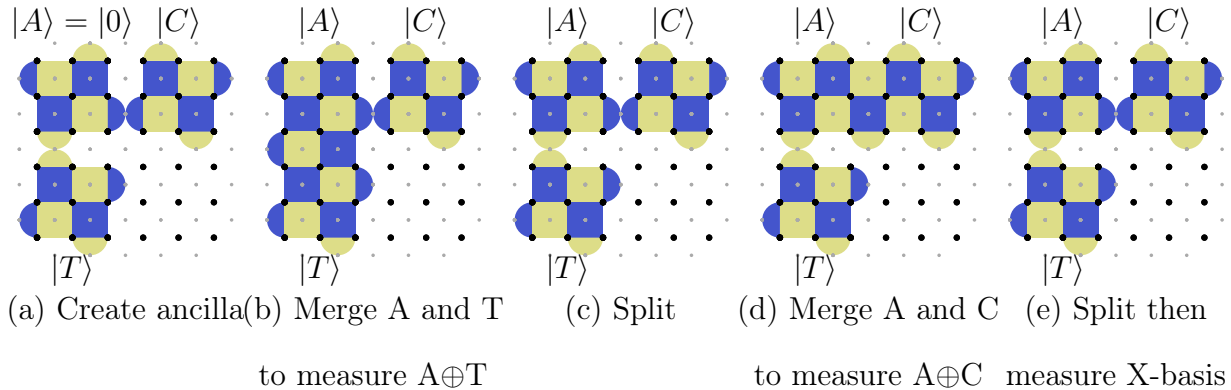\hspace{1.5em}}
    }\vspace{3em}
    \caption{The lattice surgery operations to perform a logical CNOT on the standard surface code (and directly supported in our architecture).  Given control and target qubits $\ket{C}$ and $\ket{T}$, a CNOT is performed by enabling and disabling the parity checks as shown across 6 timesteps ((e) is two steps).  We show this complex process to contrast with the fast transversal CNOT enabled by our architecture (described later in Section~\ref{sec:memory-cnot}).
    }%
    \label{fig:lattice-cnot}
\end{figure*}

\subsection{Surface Codes, Error Decoding, and Lattice Surgery}%
\label{sec:memory-qec}

The surface code by \cite{fowler-braid} is one of the most promising quantum error correction protocols because it requires only nearest neighbor connectivity between physical qubits. The surface code is implemented on a two-dimensional array of physical qubits. These qubits are either data, where the state of the logical qubit is stored, or ancilla used for syndrome extraction (parity checks). These ancilla qubits are measured to stabilize the entangled state of the data. These ancilla fall into two categories, measure-Z and measure-X for Z syndromes and X syndromes designed to detect bit and phase errors respectively. Data qubits not on the boundary are adjacent to two measure-Z and two measure-X qubits.

In Figure~\ref{fig:regular-surface-code} we show four logical qubits with code distance 3 mapped to a 2D lattice of superconducting qubits. Dark physical qubits are used as data and light qubits are used as measure qubits. In this paper, we opt to explicitly indicate qubits in order to make clear how logical qubits, formed of many square and half-circle plaquettes, are mapped directly to hardware.  In our diagrams however, we use customary notation by shading X-plaquettes blue (dark) and Z-plaquettes yellow (light). Half-plaquettes contain only 2 data qubits and are shown as half circles.

Each X (Z) plaquette corresponds to a single measure-X (Z) qubit and the four data which it interacts with. The corners of each plaquette are the data qubits. For the baseline, we use standard Z and X syndrome extraction (parity measurement) circuits where the qubits of this circuit are physical qubits. The Z-syndrome measures the bit-parity of its corner qubits and the X-syndrome measures their phase-parity.  By repeatedly performing syndrome extraction and detecting parity changes we are able to locate errors. This repeated syndrome extraction collapses any error to a correctable Pauli error and forces the data to remain in what is called the code, or quiescent, state. Once the qubits are in this state, subsequent syndrome extraction should result in the same outcomes. If errors occur, we detect them as changes in measurement outcomes.

Errors are decoded by running a classical algorithm on the measured syndromes as specified by~\cite{fowler-qec-matching}. In the surface code, when an error occurs on a data qubit, for example a single X bit-flip error, we see this as a change in the measurement outcome of \textit{both} of the Z-syndrome ancilla adjacent to it. If an error occurs on every data qubit in a chain of neighbors, only the two syndromes at the ends will detect a change.  The standard way of performing error decoding is to collect all of these changed syndromes into a complete graph with edge weights given by the log-probability of that chain of errors occurring. We perform a maximum likelihood perfect matching of this graph to find the most probable set of error locations which we correct or track in the classical control. If errors are sufficiently low these error chains will be well isolated and this decoding algorithm will be able to determine the correct set of corrections to be made. If errors are less sparse, this matching algorithm may misidentify which error chains have actually occurred and this can result in a logical error, that is a \textit{logical} bit flip or phase flip is applied. These logical errors cannot be detected because they result from misidentifying the physical errors.

There are two primary ways to manipulate the logical qubits of the surface code to perform desired logical operations --- braiding and lattice surgery. In this paper we will primarily consider lattice surgery which has been shown to have some advantages over braiding like using fewer physical qubits. For a more thorough introduction to lattice surgery we refer the reader to~\cite{fowler-lattice,game-of-codes,lao2018mapping}. In our proposed scheme, all primitive lattice surgery operations can be used such as split and merge which together perform a logical CNOT as shown in Figure~\ref{fig:lattice-cnot}. For universal quantum computation in surface codes we allow for the creation and use of magic states such as $\ket{T}$ or $\ket{CCZ}$. These states are necessary because the T and CCZ operations cannot be done transversely (using physical gates on the data in parallel to reliably perform the logical gate) in this type of code. However, high fidelity versions of these states can be generated via distillation as in~\cite{bravyi-distillation,game-of-codes} where many error-prone copies of the state are combined to generate the state with low error probability. Our scheme permits the use of these methods in the same way as other surface code schemes and also allows more efficient implementations.

\FloatBarrier{}

\section{Virtualized Logical Qubits}%
\label{sec:memory-vlq}

In this section we describe in detail our proposed architecture, an embedding of the surface code which virtualizes logical qubits, saving over 10x in required number of transmons.  This takes advantage of quantum resonant cavity memory technology described above to store \textit{logical} qubits, in the form of surface code patches, in memory local to the computational transmons. In this section we describe how we can embed surface code tiles in two variations, Natural and Compact.  We show the hardware operations needed to perform efficient syndrome extraction for both in our new fault-tolerant architecture. We then describe how typical lattice surgery operations are translated into operations in this new scheme, and finally how our system supports fault-tolerant transversal interactions between logical qubits sharing the same virtual address. We verify these operations via process tomography. We briefly describe how magic state distillation, an important primitive for algorithms, is translated to our system.

\subsection{Natural Surface Code Embedding}%
\label{sec:memory-natural}

Our goal here is to take logical qubits stored in a plane and find an embedding of that plane in 3D where the third dimension (our transmon-local memory) is a limited size, $k$.  The intuitive answer is to simply fold the surface $k$ times.  While this works, it does not have the benefits of a more clever embedding.  We propose slicing the plane into many pieces, storing them flat in memory to enable them to stitch together on-demand.  This embedding enables the fast transversal CNOT and high connectivity we will describe later.

Consider the high-level three dimensional view of the quantum memory architecture presented in~\cite{Quantum_RAM}. For every transmon in this architecture (the compute qubits in the top layer of Figure~\ref{fig:stacked-surfaces}) there is a cavity attached with a fixed number of resonant modes, $k$. Each cavity can store $k$ qubits, one per mode. Each transmon can load and store qubits from its attached cavity by performing a transmon mediated iSWAP\@.
We assume all transmons can be operated on in parallel as is the case in most superconducting hardware (i.e.\ from IBM or Google). For example, we can load qubit $q_{iz}$ to transmon $t_i$ and load $q_{jz}$ to transmon $t_j$ in parallel, simultaneously execute single qubit operations on each qubit, then store in parallel. Any other qubits stored in cavities $i$ or $j$ will be unaffected by these operations.
We expect this technology to allow cavity size $k$ on the order of 10 to 100 qubits and it will likely not be practical to scale $k$ along with the size of the 2D grid as hardware improves so we cannot implement a true 3D code such as~\cite{3d-code}. For our analysis, we conservatively assume $k=10$ and view this as a 2.5D architecture where we expect the width and height of the grid to scale while the depth, $k$, remains small.

We demonstrate how our system is sensitive to the length of these cavities in Section~\ref{sec:memory-sensitivity-results} where the amount of time between error correction cycles is directly a function of this cavity size $k$. As the size of the cavity becomes very large, the physical qubits stored are expected to be subject to more and more decoherence errors which will reduce our ability to properly decode the errors.

Consider the rotated surface code of Figure~\ref{fig:regular-surface-code} and the high level view of this architecture in Figure~\ref{fig:stacked-surfaces}. We imagine mapping each of the physical qubits of this logical qubit $q_{L, 1}$ to the same mode $z$ of each cavity in this memory architecture. Another logical qubit $q_{L, 2}$ can be mapped to mode $z_2 \neq z$ of the same set of cavities. We view this as stacking the surface code patches, the logical qubits, together under the same set of transmon qubits. The transmons themselves are only used for logical operations and error correction cycles performed on the patches.

For logical qubits with code distance $d$ we define patches on the architecture, contiguous grids of size $d \times d$ data qubits and $d \times d$ ancilla qubits. Logical qubits are mapped to multiples of $d$ coordinates on the grid and a specific mode, $z$, for storage. For example, logical qubit $q_L$ is mapped to a pair $(P_{xy}, z)$ where $P_{xy}$ refers to the square patch of data transmons $q_{d\cdot x,d\cdot y}$ to $q_{d\cdot x+d-1,d\cdot y+d-1}$ and $z$ indicates which cavity mode it is stored in. A \textit{virtual memory address} of a logical qubit refers to exactly the pair (transmon patch, index). We sometimes refer to all pairs with the same transmon patch collectively as a stack where \textit{transmon patch} is the physical memory address where a patch is loaded.

In this memory architecture, recall we are unable to operate on qubits stored in the same cavity in parallel, however we \textit{are} permitted to operated on qubits stored in different cavities in parallel. This implies for two logical qubits $q_{L, 1}$ and $q_{L, 2}$ stored in the same stack we are only able to perform syndrome extraction on at most one of these qubits at a time. In order to detect measurement errors, we typically require $d$ rounds of syndrome extraction before we perform our decoding algorithm and correct errors. If all indices are occupied by logical qubits and we want to perform $d$ rounds of correction to each one we have two primary strategies. We can load a logical qubit (meaning load all data in parallel to each transmon), perform all $d$ rounds of extraction, then store the qubit.

\begin{figure}
    \centering
    \makebox[\linewidth][c]{%
        \scalebox{\figscale}{%
        \scalebox{1.3}{%
            $\vcenter{\hbox{\scalebox{1.1418309227}{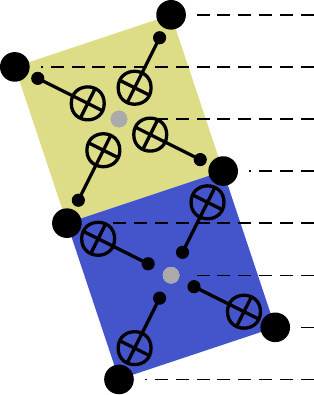}}}$%
            \hspace*{2.7pt}%
            $\vcenter{\hbox{\includegraphics[clip=false]{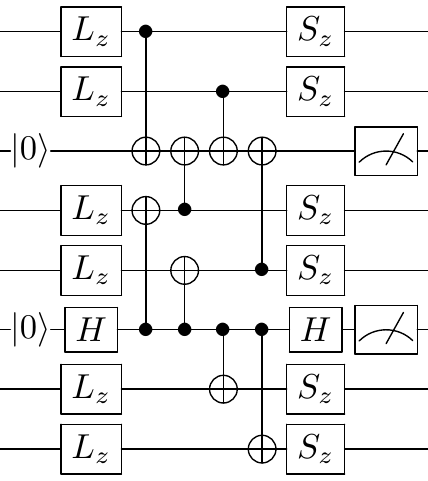}}}$%
        }}%
    }
    \caption{%
        Circuit showing how to execute our Natural embedding on hardware.  Left: The layout of six data (black) and two ancilla (gray) in hardware.  CNOT operations between qubits are drawn between.  Right: A circuit diagram of the operations applied over time where each horizontal line corresponds to a qubit and each box or symbol is an operation.  The steps are $L_z$: load \added{from memory mode $z$}, $\ket{0}$: reset ancilla, CNOTs: compute the Z or X parity, Meter: measure the result, $S_z$: store back to memory.
    }%
    \label{fig:zx-plaq-cnots}
\end{figure}

Alternatively, we can Interleave the extraction cycles by loading the logical qubit in index 0, performing one syndrome extraction step, then storing. We execute this same procedure for every logical qubit in the stack and repeat $d$ times. We expect this latter procedure to be less efficient, subjecting the data qubits to $d$ load and store errors per $d$ cycles as opposed to performing exactly one set of loads and stores when collecting all $d$ measurements at once. We study the effect of this choice of syndrome extraction on the error threshold in Section~\ref{sec:memory-threshold-results}. We detail these extraction protocols for each syndrome in Figure~\ref{fig:zx-plaq-cnots}. Here we use $L_z$ ($S_z$) to indicate loading (storing) the data from (to) index $z$ of the attached cavity.

Intuitively, this scheme is stacking many different logical tiles together in a single location. This includes mapping measure-Z/X ancilla to cavity modes. However, this is unnecessary, because measure ancilla do not actually store any data and are reset before every extraction step. Therefore, we can reduce the number of cavities required for this system by simply omitting any cavity where ancilla are stored. Instead, every patch in the same stack shares the same ancilla, the transmons at the top layer with no attached cavity.

In our system, up to $k$ logical qubits share the same set of transmons, more efficiently storing these qubits than on a single large surface. In order to interact logical qubits in different stacks we load them in parallel to the transmons then interact them via lattice surgery operations like the CNOT shown in Figure~\ref{fig:lattice-cnot}. In these cases, all of the other stacks' transmons between the interacting logical qubits act as a single (possibly large) logical ancilla. In typical planar architectures, we are unable to execute transversal two-qubit operations due to limited connectivity. We can perform physical operations between qubits in the same cavity, mediated by the transmon. Therefore, in our system, we \textit{are} able to perform transversal two-qubit interactions if the logical qubits are co-located in the same stack.  We describe this next.

\subsection{Transversal CNOT}%
\label{sec:memory-cnot}

\begin{figure}
    \centering
    \makebox[\linewidth][c]{%
        \scalebox{\figscale}{
\begingroup%
  \makeatletter%
  \providecommand\color[2][]{%
    \errmessage{(Inkscape) Color is used for the text in Inkscape, but the package 'color.sty' is not loaded}%
    \renewcommand\color[2][]{}%
  }%
  \providecommand\transparent[1]{%
    \errmessage{(Inkscape) Transparency is used (non-zero) for the text in Inkscape, but the package 'transparent.sty' is not loaded}%
    \renewcommand\transparent[1]{}%
  }%
  \providecommand\rotatebox[2]{#2}%
  \newcommand*\fsize{\dimexpr\f@size pt\relax}%
  \newcommand*\lineheight[1]{\fontsize{\fsize}{#1\fsize}\selectfont}%
  \ifx\svgwidth\undefined%
    \setlength{\unitlength}{255.26073694bp}%
    \ifx\svgscale\undefined%
      \relax%
    \else%
      \setlength{\unitlength}{\unitlength * \real{\svgscale}}%
    \fi%
  \else%
    \setlength{\unitlength}{\svgwidth}%
  \fi%
  \global\let\svgwidth\undefined%
  \global\let\svgscale\undefined%
  \makeatother%
  \begin{picture}(1,0.65227424)%
    \lineheight{1}%
    \setlength\tabcolsep{0pt}%
    \put(0,0){\includegraphics[width=\unitlength,page=1]{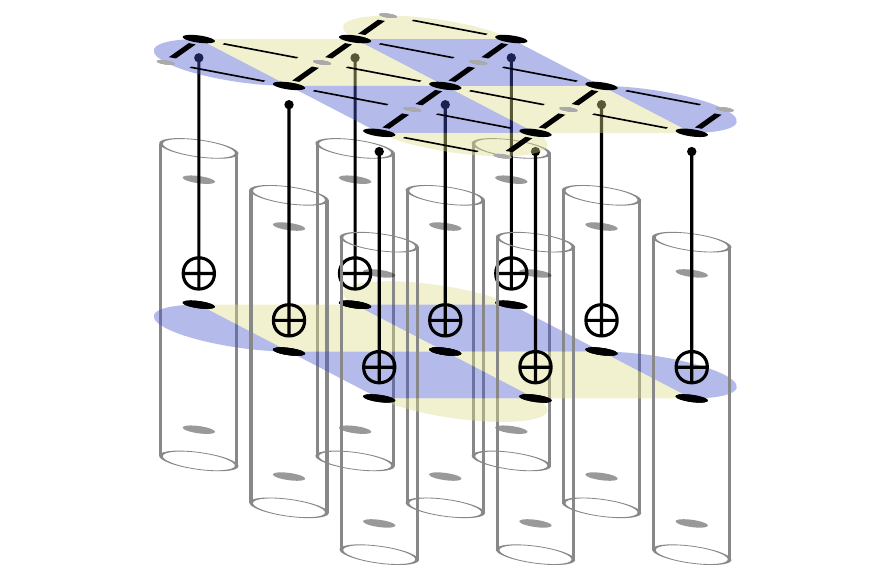}}%
    \put(0.79104318,0.41428228){\color[rgb]{0,0,0}\makebox(0,0)[lt]{\lineheight{1.25}\smash{\begin{tabular}[t]{l}CNOT gate\end{tabular}}}}%
    \put(0.17143546,0.29793067){\color[rgb]{0,0,0}\makebox(0,0)[rt]{\lineheight{1.25}\smash{\begin{tabular}[t]{r}mode $z$\end{tabular}}}}%
    \put(0,0){\includegraphics[width=\unitlength,page=2]{transversal-cnot.pdf}}%
    \put(0.87742544,0.56236617){\color[rgb]{0,0,0}\makebox(0,0)[lt]{\lineheight{1.25}\smash{\begin{tabular}[t]{l}logical\end{tabular}}}}%
    \put(0.87742544,0.5271081){\color[rgb]{0,0,0}\makebox(0,0)[lt]{\lineheight{1.25}\smash{\begin{tabular}[t]{l}control\end{tabular}}}}%
    \put(0,0){\includegraphics[width=\unitlength,page=3]{transversal-cnot.pdf}}%
    \put(0.87742544,0.2626726){\color[rgb]{0,0,0}\makebox(0,0)[lt]{\lineheight{1.25}\smash{\begin{tabular}[t]{l}logical\end{tabular}}}}%
    \put(0.87742544,0.22741452){\color[rgb]{0,0,0}\makebox(0,0)[lt]{\lineheight{1.25}\smash{\begin{tabular}[t]{l}target\end{tabular}}}}%
  \end{picture}%
\endgroup%
}
    }
    \caption{The transversal CNOT enabled by our 2.5D architecture.  The data qubits for the control logical qubit are loaded into the transmons.  Transmon-mediated CNOTs to mode $z$ for every data qubit perform the logical operation.  This takes one timestep to perform, 6x better than a lattice surgery CNOT.%
    }%
    \label{fig:transversal-cnot}
\end{figure}

A major advantage of this 2.5D architecture, enabled by our embedding of patches across memories, is the ability to do two-qubit operations transversely using the third dimension.  The logical operation is performed directly by doing the same physical gate to every data qubit and correcting any resulting errors. On typical 2D architecture error correcting codes like the surface code, the only transversal operations are single-qubit Clifford operations like X or Z. Two-qubits operations are not possible because the corresponding data qubits of two logical patches cannot all be made adjacent. However, with memory, it is possible to load one patch into the transmons and apply two-qubit gates mediated by each transmon onto the data qubits for a second qubit stored in one mode of the cavities. This works in both Natural and Compact (described later).

Figure~\ref{fig:transversal-cnot} demonstrates this for the transversal CNOT gate which we verified via process tomography (\cite{neeley2010,mikeike}) to apply the expected CNOT unitary in simulation. This can be performed in a single round of $d$ error correction cycles while the lattice surgery CNOT shown in Figures~\ref{fig:lattice-cnot} (and later~\ref{fig:lattice-cnot-compact}) takes 6 rounds. This can translate to major savings in runtime for algorithms.

The transversal CNOT is not limited to logical qubits currently stored in the same 2D address.  With an extra step it is possible to transversely interact any two logical qubits. To do this one of the qubits must be \textit{moved} to the same 2D address as the other using the move operation described in~\cite{game-of-codes}. The move operation involves growing the patch toward the move target in one step by adding new plaquettes along the entire path and performing $d$ cycles, one timestep, of error correction. Once grown, the patch can be shrunk from the other end back to its original size. The data qubits freed during the shrink are measured and used to determine any fixup operation. Once the two qubits are in the same 2D address, the transversal CNOT can be applied.  It can then be moved back, left where it is, or moved somewhere else as determined during compilation. This process takes 2 timesteps or 3 if including the second move.

\subsection{Compact Surface Code Embedding}%
\label{sec:memory-compact}

In the previous scheme, half of the transmons did not have attached cavities (or they did not make use them). An ancilla and data qubit could share a transmon because the data are stored in the cavity the majority of the time and the ancilla are reset every cycle. This leads to a more efficient, Compact embedding which halves the required number of transmons. We will see that this comes at the cost of additional loads and stores from memory due to contention during error correction, effectively trading some error and time for significant space savings.

\begin{figure}[h]
    \centering
    \scalebox{\figscale}{%
    \scalebox{\figcompactextrascale}{%
\begingroup%
  \makeatletter%
  \providecommand\color[2][]{%
    \errmessage{(Inkscape) Color is used for the text in Inkscape, but the package 'color.sty' is not loaded}%
    \renewcommand\color[2][]{}%
  }%
  \providecommand\transparent[1]{%
    \errmessage{(Inkscape) Transparency is used (non-zero) for the text in Inkscape, but the package 'transparent.sty' is not loaded}%
    \renewcommand\transparent[1]{}%
  }%
  \providecommand\rotatebox[2]{#2}%
  \newcommand*\fsize{\dimexpr\f@size pt\relax}%
  \newcommand*\lineheight[1]{\fontsize{\fsize}{#1\fsize}\selectfont}%
  \ifx\svgwidth\undefined%
    \setlength{\unitlength}{445.50000161bp}%
    \ifx\svgscale\undefined%
      \relax%
    \else%
      \setlength{\unitlength}{\unitlength * \real{\svgscale}}%
    \fi%
  \else%
    \setlength{\unitlength}{\svgwidth}%
  \fi%
  \global\let\svgwidth\undefined%
  \global\let\svgscale\undefined%
  \makeatother%
  \begin{picture}(1,0.31360233)%
    \lineheight{1}%
    \setlength\tabcolsep{0pt}%
    \put(0,0){\includegraphics[width=\unitlength,page=1]{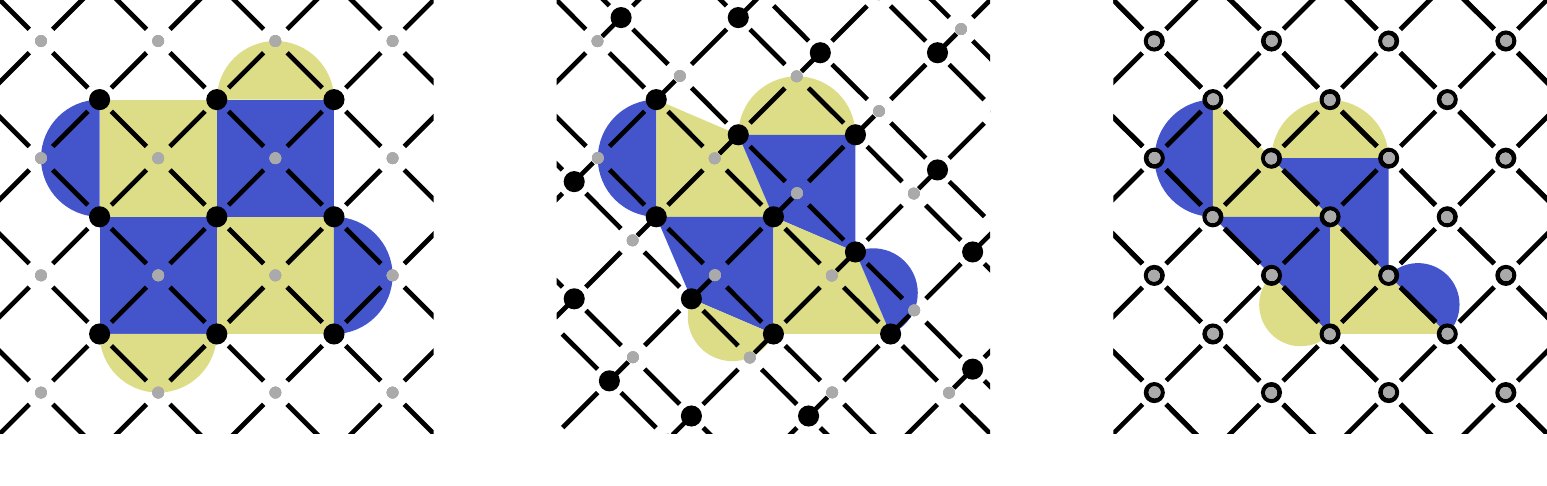}}%
    \put(0.14015152,0.00299627){\makebox(0,0)[t]{\lineheight{1.25}\smash{\begin{tabular}[t]{c}(a)\end{tabular}}}}%
    \put(0.5,0.00299627){\makebox(0,0)[t]{\lineheight{1.25}\smash{\begin{tabular}[t]{c}(b)\end{tabular}}}}%
    \put(0.85984848,0.00299627){\makebox(0,0)[t]{\lineheight{1.25}\smash{\begin{tabular}[t]{c}(c)\end{tabular}}}}%
    \put(0,0){\includegraphics[width=\unitlength,page=2]{compact-transform.pdf}}%
  \end{picture}%
\endgroup%
}}
    \caption{Transformation from Natural to Compact. (a)
    Natural embedding: Only data have attached cavities (not shown).
    (b) The transformation:
    Z ancilla (over yellow/light areas) merge with the upper-right data transmon and X ancilla (over blue/dark areas) merge with the lower-left data transmon.  The opposite parings are key to keeping 4-way grid connectivity.
    (c) Compact embedding: All ancilla transmons without attached cavities have been removed.  All remaining transmons have cavities and are used as both data and ancilla.
    }%
    \label{fig:compact-transform}
\end{figure}

\begin{figure}
    \centering
    \vspace{0.5em}
    \makebox[\textwidth][c]{%
        \scalebox{\figscale}{%
        \def\svgwidth{\textwidth}
        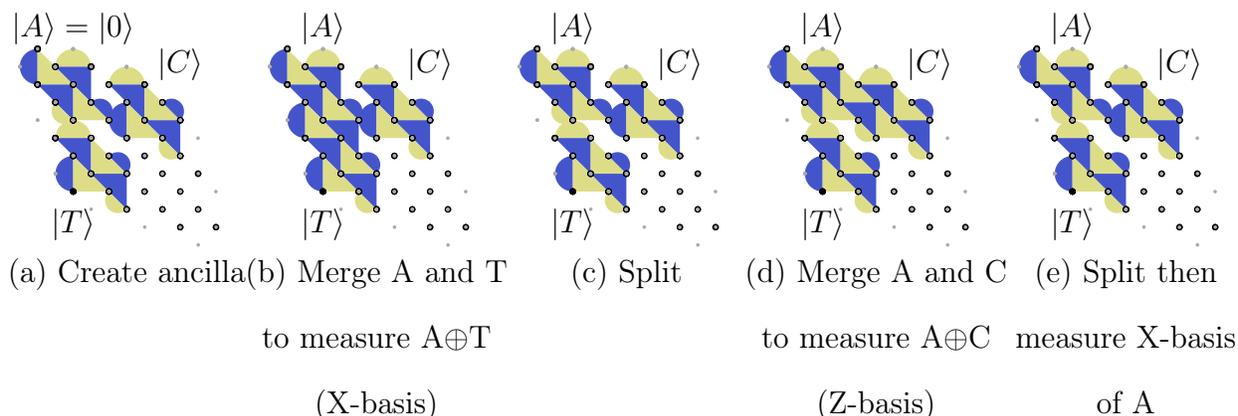\hspace{1.5em}}
    }\vspace{3em}
    \caption{The Compact lattice surgery operations to perform a CNOT\@. The logical operations performed are identical to Figure~\ref{fig:lattice-cnot} but the corresponding physical operations are arranged as shown in Figure~\ref{fig:compact-transform}.  This uses half as many transmons as Natural.  As before, it takes 6 timesteps of $d$ error correction cycles each.}%
    \label{fig:lattice-cnot-compact}
\end{figure}

In the above memory architecture, because we do not store any logical qubits in the transmon layer, these qubits can act as the measurement ancilla, rather than have separate transmons only there to act as the syndrome measurement ancilla. With this observation, we can pack the data qubits of the surface code patch of Figure~\ref{fig:compact-transform}a more efficiently with \textit{every} transmon having a cavity attached. Each plaquette of the rotated surface code has a single ancilla at its center, interacting with each data qubit. For $Z$ plaquette (yellow or light) in this mapping scheme we colocate the upper-right data and the ancilla; the upper-right data is located in the cavity attached to the transmon corresponding to the ancilla. Similarly, for each $X$ plaquette (blue or dark) we colocate the lower-left data and the ancilla; the lower-left data is located in the cavity attached to the transmon corresponding to the ancilla.

\begin{figure}[h]
    \centering
    \scalebox{\figscale}{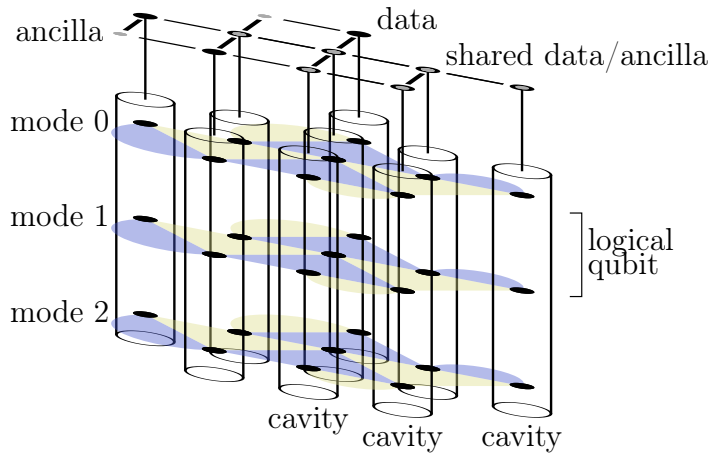}
    \caption{A 3D view of our Compact embedding.  Shown at the top is the 2D grid of transmon qubits.  Attached below every transmon is a resonant cavity.  Compact surface code patches are shown stored, one in each mode.  This deformed patch can be tiled in 2D.
    }%
    \label{fig:stacked-compact-surfaces}
\end{figure}

This mapping results in plaquettes which resemble triangles rather than squares, where the center of the hypotenuse of each triangle corresponds to both the ancilla qubit and the data qubit, stored ``beneath'' in its cavity. Every data qubit is still mapped to the \textit{same} index. Notice in this scheme every data (sans the boundary) is still adjacent to two measure-Z and two measure-X ancilla where adjacent means either in the cavity of the ancilla or in a cavity adjacent to the ancilla. We illustrate this transformation from our undistorted Natural surface code patch to Compact in Figure~\ref{fig:compact-transform} and a diagram of this architecture with a cavity for every transmon in Figure~\ref{fig:stacked-compact-surfaces}.  If a different ancilla location were chosen, for example all sharing with the upper-right data, some of the syndrome extraction gates in the resulting arrangement would require six-way connectivity, two diagonal to the grid, which would be much more difficult to engineer with low noise. This scheme where X and Z ancilla share with data in opposite directions is the best scheme we found to satisfy the hardware connectivity.

\begin{figure}
    \centering
        \scalebox{\figscale}{%
            $\vcenter{\hbox{\scalebox{1.1418309227}{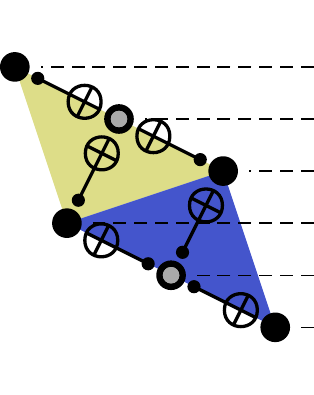}}}$%
            \hspace*{2.7pt}%
            $\vcenter{\hbox{\includegraphics[clip=false]{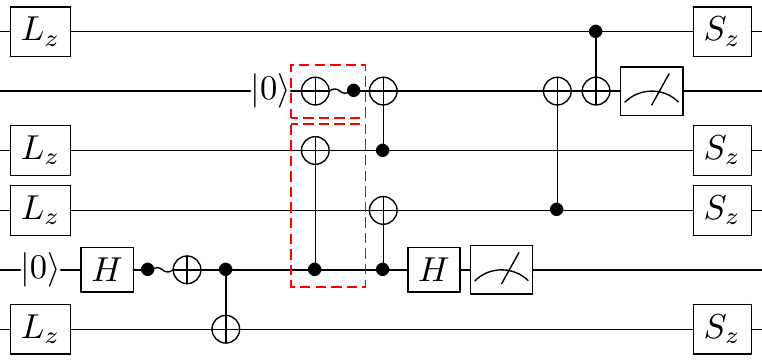}}}$%
        }\\%
        \scalebox{\figscale}{%
            $\vcenter{\hbox{\scalebox{0.77778}{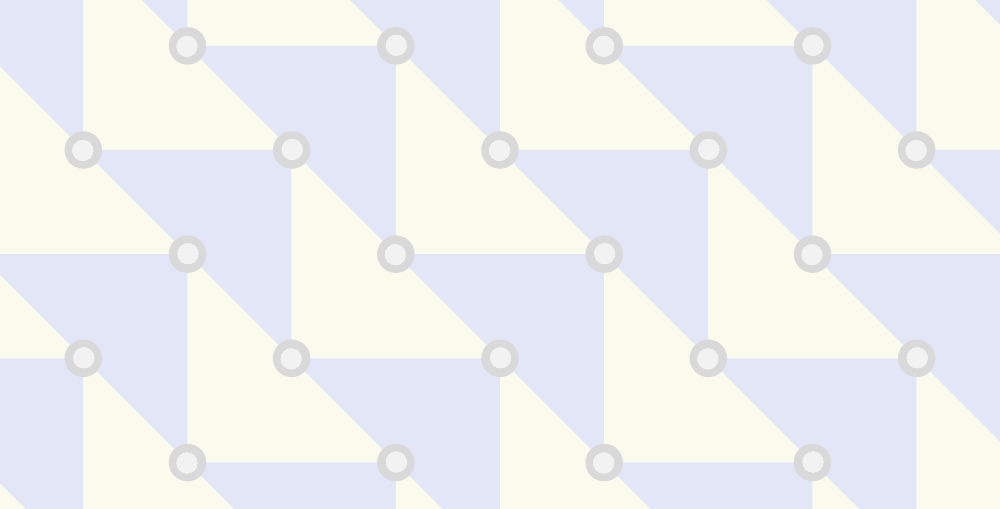}}}$%
        }%
    \caption{The CNOT sequence for parity checks in Compact.  Top: A quantum circuit showing the hardware operations over time.  Bottom: The CNOT execution order repeats $A_0D_2$, $A_1D_3$, $A_2C_0$, $A_3C_1$, $B_0C_2$, $B_1C_3$, $B_2D_0$, $B_3D_1$.  The $AB$ and $CD$ sequences run in parallel but offset to ensure ancilla and data use do not conflict.  CNOTs for $A_0D_2$ are marked in red where an isolated circle indicates a transmon-mediated CNOT.%
    }%
    \label{fig:compact-plaq-order}
\end{figure}

In Natural, we assign square patches to predetermined square patches on the hardware. In Compact, we assign square patches to predetermined rhombus or diamond patches on the hardware. Previously, operations on the virtualized patches closely resembled the original operations because the shape was unchanged, except with the addition of loads and stores to retrieve the logical qubit from memory. The same operations apply here. We can examine the original, unmapped surface code patch and perform the same sequence of operations modulo loads and stores, on the transformed coordinates of the mapped version.

This new mapping also requires a new syndrome extraction procedure because data cannot be loaded while a transmon is in use as an ancilla. A single round of syndrome extraction can be executed by dividing the plaquettes into four groups, with each group containing non-interfering plaquettes.  Two plaquettes are non-interfering if they do not share their ancilla with any data qubits of the other plaquette. This process is detailed explicitly in Figure~\ref{fig:compact-plaq-order}. It is imperative this process use both the minimum number of loads and stores and keep data qubits loaded for as short a time as possible as the error incurred during this circuit directly impacts the error threshold for the code. This has a similar cost as Natural, Interleaved where a higher numbers of load and store gates were also required.

Error correction can be performed Interleaved or All-at-once just as with Natural.  This should be chosen dependent on how likely storage errors and gate errors are. For example, if storage errors are expected to be significant, we may opt to use Interleaved syndrome extraction. This will cost more loads and stores so if gate errors are more significant than storage errors we may opt for All-at-once.

\FloatBarrier{}

\subsection{Architectural Considerations}

When compiling and executing programs in our system there are several important architectural features to keep in mind. First, it is always possible to execute a transversal two-qubit interaction, rather than requiring use of split and merge. In surface code architectures, the logical qubits are not bound to a specific hardware location and are free to move around on the grid. This qubit movement is fairly cheap requiring only a single round of $d$ error correction cycles (usually referred to as a single timestep) to move any distance. However, we require a clear area of unused patches to move through; typically, this requires about $1/3$ to $1/2$ of the total area to be kept as open channels to allow for distant qubit interactions. In our architecture this translates to keeping one of the resonant modes in every stack unused ($1/k$ of total qubits for cavity depth $k\approx 10$) and loading this mode along a path when a logical qubit needs to move, i.e.\ there is an index in the stack which has no logical qubit mapped to it.  This enables our system to transport logical qubits between stacks to execute more time and space efficient transversal CNOTs. The empty mode is necessary for Compact because data is always stored back to the cavity during syndrome extraction but not required for Natural, All-at-once where the transmons themselves can act as the unused qubits to move the logical qubit through.

Unfortunately, this qubit movement is not entirely free. During the compilation process if we request many logical qubits to move in parallel this can be expensive due to serialization of intersecting move paths. Just as in current quantum systems without error correction where it is imperative to map and schedule multi-qubit interactions in a way which minimizes total execution time, it is also important in our system that logical qubits which interact heavily be located close by for similar reasons. The mapping problem on the system presented here is interesting because there is now a tradeoff between locality and serialization between operations with qubits sharing the same 2D address.

Second, we stress even though the logical qubits are stored in memory, they are still subject to errors and it is critical that every logical qubit be error corrected regularly. In the case of Interleaved syndrome extraction, every logical qubit of a stack will be roughly guaranteed to get a round of correction every $k$ time steps, where $k$ is the cavity depth. This rate is during steady state, when qubits are idle. When logical operations are being executed, this rate may be reduced slightly. When compiling and executing on this system, we may need to delay some operations in order to ensure stored logical qubits get the required amount of error correction and are not left so long that errors accumulate and error correction becomes less likely to succeed.

Finally, many lattice surgery operations require the use of ancilla logical qubits, for example to measure specific stabilizers which are done to execute a particular set of operations in~\cite{game-of-codes}. This restriction requires our architecture and any compiler to guarantee one free mode of every stack be allocated to temporarily obtain large logical qubits. This free mode may be shared with qubit movement or separate if many ancilla logical qubits are used.

\FloatBarrier{}

\section{Evaluation}%
\label{sec:memory-evaluation}

In this section, we outline our error model and experimental setup used to determine error thresholds for our mapping and syndrome extraction schemes. We compare to the surface code on a typical 2D architecture. Our goal is to demonstrate the error thresholds for various error correction schemes, i.e.\ to determine the necessary \textit{physical} error rate required to begin obtaining exponentially better \textit{logical} error rate as the code distance increases. Currently, neither transmon devices nor transmon-memory devices used for our schemes have consistently achieved physical error rates below this threshold and instead the threshold serves as a goal or checkpoint.

\subsection{Error Model and Noise Assumptions}%
\label{sec:memory-error-model}

For our experiments we make the following further assumptions about how noise and errors behave in both a typical 2D architecture and our 2.5D cavity memory architecture since both have the same fundamental underlying transmon technology:
\begin{itemize}
    \item The error rates in the device do not fluctuate appreciably over time.
    \item Transmon qubits can be actively reset and reinitialized to $\ket{0}$ efficiently and without significant error.
    \item All errors are independent.  No leakage errors and no correlated noise.
    \item All classical processing of the syndromes is instantaneous and error-free.
    \item Every $n$-qubit gate with the same $n$ is equally error-prone. For example, every one qubit operation has the exact same chance of failure regardless of which actual physical qubit it is applied to.
    \item All errors are Pauli, i.e.\ drawn from the set ${\{I, X, Y, Z\}}^{\otimes n}$. For example, if a one-qubit error occurs with probability $p$ then we apply an $X$, $Y$, or $Z$ with probability $p/3$ and $I$ (no error) with probability $1 - p$.
    \item We detect and correct $X$ and $Z$ errors independently.  A $Y$ error is both an $X$ and $Z$ error.
\end{itemize}

For each of our experiments we rely on realistic device data for current superconducting devices, provided by \cite{ibmq}. For the memory hardware, we use experimental data from~\cite{Quantum_RAM}. These parameters are listed in Table~\ref{tab:gate_and_coherence_times}, where $T_{1, c}$ is the coherence time of the cavity, $T_{1, t}$ is the coherence time of the transmon, $\Delta_{t}$ is the single qubit gate time, $\Delta_{t-t}$ is the two-qubit transmon-transmon gate time, $\Delta_{t-m}$ is the two-qubit gate time of transmon-mode interactions, and $\Delta_{l/s}$ is the load and store times. In every experiment, the gate durations for one- and two-qubit interactions is fixed. In a first set of experiments, we vary all gate errors and coherence times together, all derived from a single probability of error $p$ given as the probability of an SC-SC (Transmon-Transmon gates) two-qubit gate error. We consider $T_1$ times of both cavities and transmons to determine the probability of storage error given as $\lambda = 1 - \exp{-\Delta t / T_1}$, where $\Delta t$ is the duration stored. We consider the same potential gate error rates for each of these devices since the underlying technology behaves very similarly. While typical coherence errors are not generally Pauli, we model them as Pauli errors here as a worst-case approximation since correcting Pauli errors is harder than correcting coherence errors in general.

\begin{table}[h]
    \caption{Starting point coherence times and constant gate times \added{for the hardware models}.}%
    \label{tab:gate_and_coherence_times}
    \centering
    \begin{tabular}{ccc}
        Hardware Parameter & Baseline Transmons & Transmons with Memory
        \\\toprule
        $T_{1,t}$ & 100 $\mu$s & 100 $\mu$s \\\midrule
        $T_{1,c}$ & --- & 1 ms \\\midrule
        $\Delta_{t-t}$ & 200 ns & 200 ns \\\midrule
        $\Delta_{t}$ & 50 ns & 50 ns \\\midrule
        $\Delta_{t-m}$ & --- & 200 ns \\\midrule
        $\Delta_{l/s}$ & --- & 150 ns \\
    \end{tabular}
\end{table}

\subsection{Experimental Setup}%
\label{sec:memory-experimental-setup}

In every experiment, we run 2,000,000 simulated trials per data point with each trial consisting of a round of error correction. We compute the logical error rate as the number of logical errors (misidentified error chains) over the total number of trials. The large number of trials is required to estimate logical error rates to $10^{-5}$. To determine the error threshold values for different surface code schemes, we vary the physical error rate over several different code sizes. The goal is to find an intersection point for each of these lines which gives a physical error rate below which we expect our logical error rate to get better as the physical error rate improves.  Below the threshold we also expect the logical error rate to get better exponentially in the code distance $d$.

We study 5 setups to determine initial error thresholds.
\begin{itemize}
    \item The surface code on a 2D superconducting architecture as our baseline.
    \item Our Natural embedding with either the All-at-once or Interleaved syndrome extraction.
    \item Our Compact embedding with either the All-at-once or Interleaved syndrome extraction.
\end{itemize}

In our designs, the possible sources of error are more nuanced and we study the thresholds' sensitivity to variation in the parameters.
In all threshold experiments, we assume cavity depth of 10 but later study sensitivity to cavity size.
\added{The simulation code used to generate our results is available on GitHub (\cite{Git-vlq}).}

\section{Error Threshold Results}%
\label{sec:memory-threshold-results}

\begin{figure}
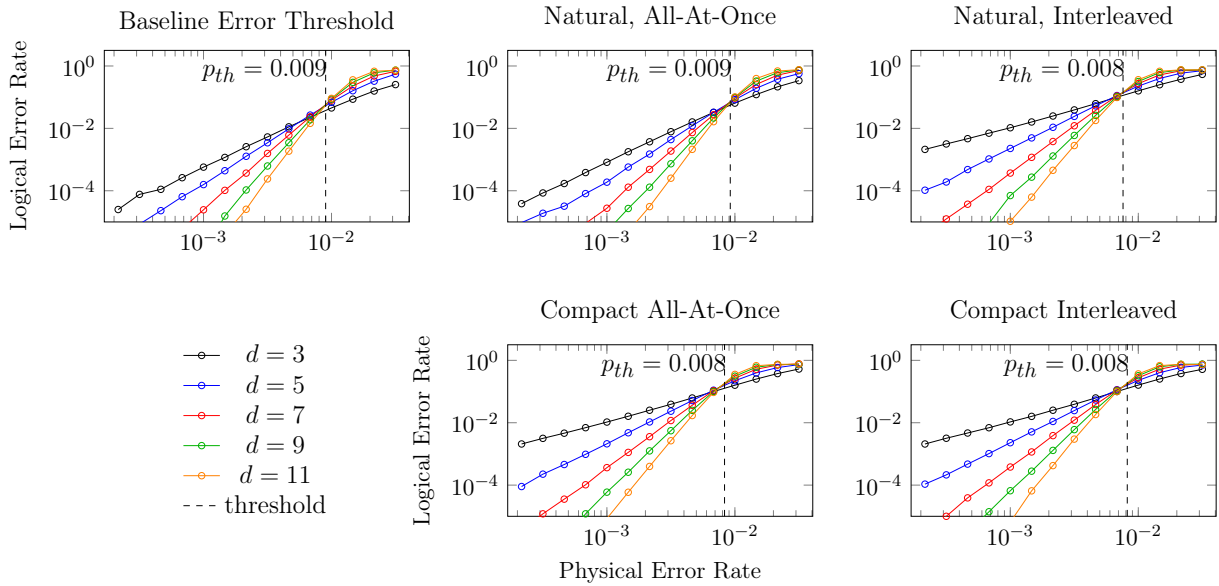

    \centering
    \scalebox{\figscaleplot}{%
    \makebox[0.99\textwidth][r]{%
        \input{chapters/memory/figs/thresholds.tikz}%
    }}\\\vspace{1em}
    \scalebox{\figscaleplot}{%
    \makebox[0.99\textwidth][r]{%
        \input{chapters/memory/figs/thresholds2.tikz}%
    }}\vspace{0.5em}
    \caption{Error thresholds for the baseline 2D architecture and Natural and Compact variants of our 2.5D architecture. The thresholds are comparable to the baseline indicating the space savings obtained in our system does not substantially reduce the error thresholds. The slopes of the lines in this figure indicate, post-threshold, how much improvement in physical error rates improve logical error rate.  Except for the baseline, all use a cavity size of 10.}%
    \label{fig:base_thresholds}
\end{figure}

We detail our threshold results in Figure~\ref{fig:base_thresholds}. We study 5 different code distances in order to obtain the physical error threshold value. The threshold value indicates at which point increasing the code distance, $d$, improves the logical error rate instead of hurting it. This threshold is a function of both the physical system model, the chosen syndrome extraction circuit, and the specific decoding procedure. For example, decoding procedures which do not accurately represent the probability of certain error chains occurring will do a poor job of correcting those errors. The decoding process should be directly informed by the error model. In systems with more complicated error models, the decoder should be aware of these further details to inform its decision about which types of errors occurred and the proper way to correct them. We use the usual maximum likelihood decoder because we use standard assumptions in our error model.

The major difference in each procedure is the additional error sources and different syndrome extraction procedures. For example, the baseline is not subject to any of the effects related to cavity storage or transmon-mode operations. These syndrome extraction procedures differ by the amount of storage time of data qubits in different locations (cavity vs.\ transmon) as well as the number of different physical gate operations applied to them. These differences however, do not cause substantial variation in the error threshold for the different setups which is extremely promising. Second, the slopes for each code distance compared across the various schemes is stable, indicating each scheme improves at a similar rate, post error threshold, and showing that the logical error rate decays exponentially with $d$ as desired. This is significant because it means we will be able to save on total number of transmons without major shifts in the error threshold. Since transmon memory technology is expected to perform as well as other competing transmon technology, we obtain higher distance codes, and hence better logical error rate, with fewer total transmons.

\section{Error Sensitivity Results}%
\label{sec:memory-sensitivity-results}

In this section, we study the effects of different sources of error on the thresholds obtained in Section~\ref{sec:memory-threshold-results}. Specifically, we show how different system-level details affect the threshold of the code. Here we focus on Compact, Interleaved as the most efficient physical qubit mapping and subject to a wide variety of errors. In these studies, the physical error rates of all but a single error source are fixed at a typical operating point below the threshold obtained previously, $2\times 10^{-3}$ and the cavity depth is fixed at 10. \added{Gate} times are fixed \added{while} we vary the physical error rate of SC-SC gates, SC-Cavity gates, Load-Store gates \added{or} the coherence times of the cavity and the transmon. \added{We additionally study the duration of load/store, the gates unique to memory technology. We note the effect of the SC-Cavity gate duration will be a similar, smaller effect since it occurs only once per qubit per error correction cycle.} Finally, we study the effect of cavity size by varying the number of modes per cavity, \added{causing a proportional} delay between error correction cycles.

\begin{figure}
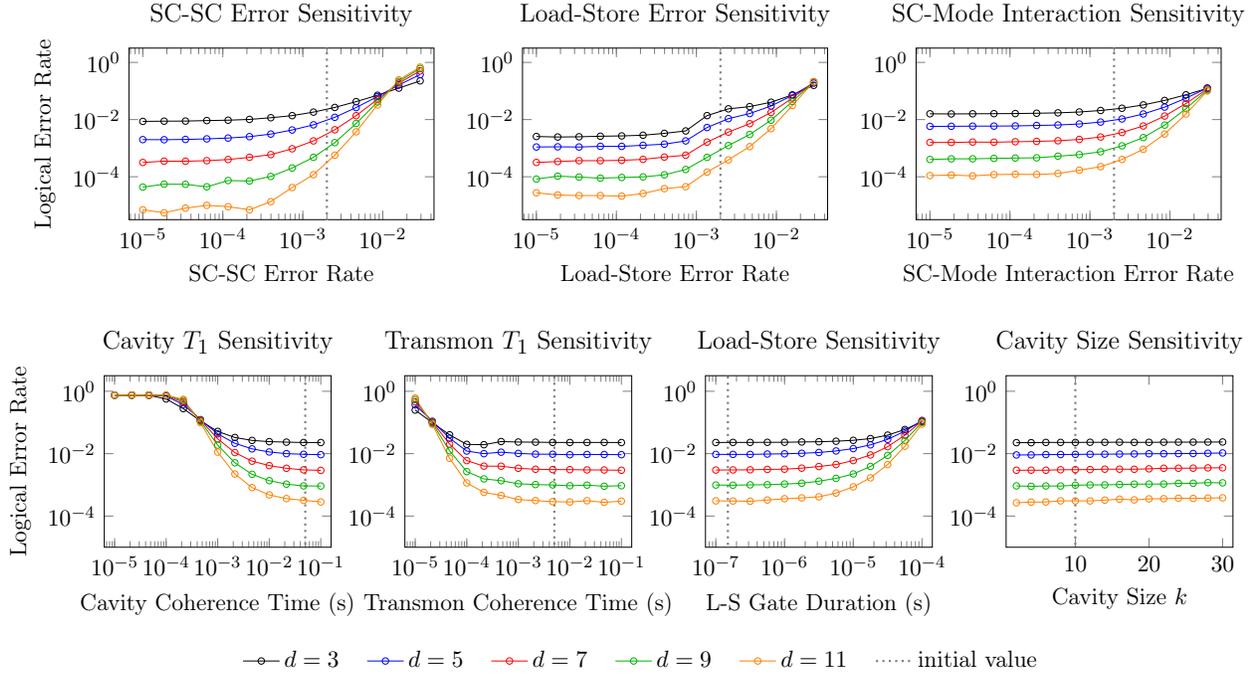

    \centering
    \scalebox{\figscaleplot}{%
    \makebox[0.99\textwidth][r]{%
        \input{chapters/memory/figs/sensitivity_plots.tikz}%
    }}\\\vspace{1em}
    \scalebox{\figscaleplot}{%
    \makebox[1.01\textwidth][r]{%
        \input{chapters/memory/figs/sensitivity_plots2.tikz}%
    }}\\\vspace{0.5em}
    \scalebox{\figscaleplot}{%
    \makebox[\textwidth][c]{%
        \begin{tikzpicture}[baseline,scale=0.85,trim axis left,trim axis right]
\pgfplotsset{every tick label/.append style={font=\small}}
\pgfplotsset{every axis label/.append style={font=\small}}
    \begin{axis}[
        name=plot2,
        title={  },
        xlabel={},
        ylabel={},
        width={0.8\textwidth},
        height={50pt},
        legend style={
            draw=none,
            at={(0.5,0)},
            anchor=south,
            font=\small},
        legend columns=-1,
        axis line style={draw=none},
        tick style={draw=none},
        xticklabels={},
        yticklabels={},
        xmin=-2, xmax=-1, ymin=-2, ymax=-1, clip=true,
    ]
        \addplot[color=black, mark=o, mark size=1.5pt] coordinates{(0, 0)};
        \addlegendentry{$d=3$~~~~}
        \addplot[color=blue, mark=o, mark size=1.5pt] coordinates{(0, 0)};
        \addlegendentry{$d=5$~~~~}
        \addplot[color=red, mark=o, mark size=1.5pt] coordinates{(0, 0)};
        \addlegendentry{$d=7$~~~~}
        \addplot[color=green!70!black, mark=o, mark size=1.5pt] coordinates{(0, 0)};
        \addlegendentry{$d=9$~~~~}
        \addplot[color=orange, mark=o, mark size=1.5pt] coordinates{(0, 0)};
        \addlegendentry{$d=11$~~~~}
        \addplot[color=gray, mark=none, dotted, line width=1pt] coordinates{(0, 0)};
        \addlegendentry{\added{initial value}}
    \end{axis}
\end{tikzpicture}%
    }}
    \caption{Sensitivity of logical error rate to various error sources in Compact, Interleaved. The logical error \added{rates are} most sensitive to physical error of Loads/Stores and SC-SC gates. The logical error rate is \added{less sensitive to the coherence times and} mostly insensitive to effects of \added{load-store duration and} cavity size.}%
    \label{fig:sens_results}
\end{figure}

The results of these sensitivity studies are found in Figure~\ref{fig:sens_results}. The logical error rate is sensitive to a particular error source's probability if the slope of the line is pronounced \added{at the marked reference value}. The logical error rate for Compact, Interleaved \added{is sensitive to all changes in system-level details to some degree. The gate error rates show the highest sensitivity, indicating improvement in these will give the greatest benefit.
Coherence times are not quite as sensitive but the slightly over 10x offset between the cavity and transmon plots shows that there is no benefit in transmon $T_1$ being longer than $1/10$ cavity $T_1$ when the cavity size is 10.}
\added{The} lines taper off, indicating other errors sources \added{eventually dominate}. Initially, we expected the cavity size to have a large impact on the logical error rate. However, when coherence times are high and gate error rates are fairly low below the threshold, the logical error rate does increase proportional to the length of the cavity but the effect is very minor. This indicates, given cavities with good coherence times, our proposed system will be able to scale smoothly into the future as cavity sizes increase.

While larger cavity sizes will make this architecture even more advantageous, there will be a point at which it has a vanishing benefit because the delay between error correction becomes too long and decoherence error dominates.  For the error rates used in the evaluation, we find that cavity decoherence error starts dominating after cavity size $k\approx 150$.  After this point, it would be more beneficial to improve cavity coherence time.

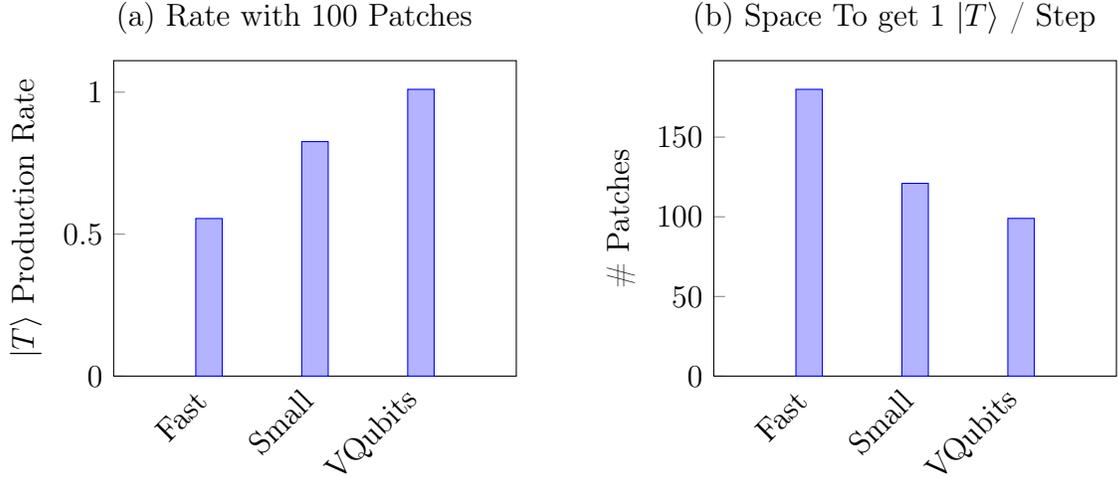
\begin{figure}[t]
    \centering
    \scalebox{1}{%
    \makebox[\linewidth][c]{%
        \begin{tikzpicture}
  \begin{axis}[
    ybar,
    ylabel={$\ket{T}$ Production Rate},
    title={(a) Rate with 100 Patches~~~~~},
    symbolic x coords={Fast, Small, VQubits},
    x tick label style={rotate=45,anchor=east},
    width=0.42*\linewidth,
    height=0.35*\linewidth,
    xtick style={draw=none},
    ytick pos=left,
    enlarge x limits=0.45,
    ymin=0,
    ]
    \addplot coordinates {(Fast, .555)(Small, .826)(VQubits, 1.010)};
  \end{axis}
\end{tikzpicture}%
\hspace{2em}
\begin{tikzpicture}
  \begin{axis}[
    ybar,
    title={(b) Space To get 1 $\ket{T}$ / Step~~~~~},
    ylabel={\# Patches},
    symbolic x coords={Fast, Small, VQubits},
    x tick label style={rotate=45,anchor=east},
    width=0.42*\linewidth,
    height=0.35*\linewidth,
    xtick style={draw=none},
    ytick pos=left,
    enlarge x limits=0.45,
    ymin=0,
    ]
    \addplot coordinates {(Fast, 180)(Small, 121)(VQubits, 99)};
  \end{axis}
\end{tikzpicture}%
    }}\\
    \vspace{-1em}
    \caption{(a) The T-state generation rates of three different circuits.  Higher generation rate is better.  (b) The space, in terms of number of patches, required to produce a single $\ket{T}$ per time step. Lower is better.  Fast~\cite{fast-distillation} and Small~\cite{game-of-codes} work in the surface code and do not use memory. VQubits is implemented with transversal CNOTs in our 2.5D architecture. All are based on~\cite{bravyi-distillation}.}%
    \label{fig:distilation-rate}
\end{figure}

\section{Magic State Distillation Resource Estimates}%
\label{sec:memory-resource-estimates}

Now that we have shown error correction is effective in our virtualized qubit architecture, we analyze how the transversal CNOT and memory connectivity can benefit the performance of an algorithm overall.  In error-corrected quantum algorithms, the dominating cost (commonly $>$ 90\%) in both space and time resources is magic state distillation (\cite{other-magic-estimates,adam-magic-estimates,gidney2019factor}).  For this analysis we consider how T-state distillation, a commonly used magic state, is improved.  Any improvements here will translate directly to improvements in important algorithms like Shor's and Grover's.

We take the 15-to-1 distillation circuit of~\cite{bravyi-distillation} to generate a T magic state using a single patch of transmons with 6 logical qubits stored in the attached cavities.  This circuit consists of 16 qubit initializations, 15 measurements, 35 CNOT gates and a few other operations.  It takes a total of 110 surface code timesteps to generate a T-state using only a single patch of transmons.  If pairs of these circuits are executed in lock-step, they only take 99 timesteps.

In Figure~\ref{fig:distilation-rate} we compare the T-state generation rate with memory against two representative extremes designed for speed or size, Fast Lattice from~\cite{fast-distillation} and Small Lattice from~\cite{game-of-codes} \added{(also based on~\cite{bravyi-distillation})}.  Fast Lattice generates a T-state every 6 timesteps but uses 30 patches of space whereas Small Lattice, generates a T-state every 11 timesteps using only 11 patches of space.  We compare these results by computing the T-state generation rate per timestep if we filled 100 patches with copies of the circuit running in parallel.
Table~\ref{tab:magic-cost} show the qubit cost of each and chip area will be proportional to the number of transmons.  Using our VQubits protocol generates 1.82x as many T-states as Fast Lattice and 1.22x as many as Small Lattice.
This improvement allows an algorithm like Shor's to run roughly 1.22x faster or work on smaller hardware.

\begin{table}[t]
    \caption{Transmon, depth-10 cavity, and total qubit costs of each T-state generation protocol for $d=5$.}%
    \label{tab:magic-cost}
    \centering
    \small
    \begin{tabular}{cccc}
        Protocol & \# transmons & \# cavities & total qubits
        \\\toprule
        Fast Lattice from~\cite{fast-distillation} & 1499 & --- & 1499 \\
        Small Lattice from~\cite{game-of-codes} & 549 & --- & 549 \\
        \textbf{VQubits (Natural)} & \textbf{49} & \textbf{25} & \textbf{299} \\
        \textbf{VQubits (Compact)} & \textbf{29} & \textbf{25} & \textbf{279} \\
    \end{tabular}
\end{table}

\FloatBarrier{}

\section{Summary}%
\label{sec:memory-conclusion}

Realizable quantum error correction protocols are a critical step in the path towards fault-tolerant quantum computing. There has been great progress in NISQ-era devices, but it is equally critical to look towards designing architectures for QEC\@. In this paper, we introduce a system which virtualizes logical, error corrected qubits and is both space and time efficient without sacrificing in terms of fault tolerance.

By taking advantage of recent advances in quantum memory technology, we present a new architecture that substantially reduces hardware requirements by storing logical qubits distributed in spatially-local memory.
This technology allows memory to be separated but local to computation in a quantum system.
By finding a memory abstraction that keeps application data spatially local, we find application-level efficiencies in communication of logical qubit data.

We provide two direct mappings of the surface code to this new system with virtual addressing and illustrate how syndrome extraction and error correction procedures can be executed efficiently on the embedded surface code.
Our embedding, combined with the random-access nature of the memory is important for several reasons.
It enables fast transversal gates like the CNOT which can reduce program execution time by allowing faster operations and indirectly through improved magic-state distillation protocols.
It significantly reduces the total number of transmon qubits required (10x for our analysis) which allows larger code distance patches while using 10x fewer transmon qubits and classical control wires.
This allows error correction to be realized much sooner on small architectures while also enabling these devices to scale.
Our results show superconducting cavity-based technology offers a promising path towards realizing spatially local memory and quickly scaling fault-tolerant quantum computation.
This design can be evaluated with 10 logical qubits using as few as 11 transmons and 9 cavities which we hope motivates further experimental efforts and prompts industry to adopt and scale-up this architecture.

\FloatBarrier{}

\chapter{Hierarchical Program Structure}%
\label{chapter:modular}

\section{Introduction}%
\label{sec:modular-introduction}

The two previous chapters cover abstractions that allow us to conceptualize storage of quantum data in the same qubit devices or nearby.
Because quantum information is delicate and prone to error from interactions with the environment, it is challenging to move.
This is in part why the previous abstractions have shown benefit due to allowing the data-locality of the application to influence the implementation and match the data layout of the technology.

In this chapter, we explicitly look at data movement.
When we compile quantum programs with arbitrary data usage patterns, the compiler will insert explicit data movement operations.
Typically, a quantum program is flattened to its basic executable gates before data movement is determined, but this leads to ineffective heuristics or impractical optimization problems for the compiler.
This chapter takes a small example of hierarchy, the three-qubit operation called the Toffoli gate, and shows how even a little hierarchy, used properly, can have large gains in compiled program performance.

\begin{figure}
    \centering
    \begin{minipage}[][][b]{0.45\columnwidth}
        \centering
        \scalebox{1.5}{\begin{tikzpicture} [node distance=1.2cm,scale=0.6, every node/.style={scale=0.6}]
    \node[fill={rgb,190:red,72;green,132;blue,189},draw,circle,line width=0.5pt,minimum size=0.75cm,] (0) {0};
    \node[fill={rgb,190:red,72;green,132;blue,189},draw,circle,line width=0.5pt,minimum size=0.75cm,right of=0] (1) {1};
    \node[fill={rgb,190:red,180;green,56;blue,101},draw,circle,line width=0.5pt,minimum size=0.75cm,right of=1] (2) {2};
    \node[fill={rgb,190:red,72;green,132;blue,189},draw,circle,line width=0.5pt,minimum size=0.75cm,right of=2] (3) {3};
    \node[fill={rgb,190:red,72;green,132;blue,189},draw,circle,line width=0.5pt,minimum size=0.75cm,right of=3] (4) {4};
    \node[fill={rgb,190:red,72;green,132;blue,189},draw,circle,line width=0.5pt,minimum size=0.75cm,below of=0] (5) {5};
    \node[fill={rgb,190:red,180;green,56;blue,101},draw,circle,line width=0.5pt,minimum size=0.75cm,below of=1] (6) {6};
    \node[fill={rgb,190:red,72;green,132;blue,189},draw,circle,line width=0.5pt,minimum size=0.75cm,below of=2] (7) {7};
    \node[fill={rgb,190:red,72;green,132;blue,189},draw,circle,line width=0.5pt,minimum size=0.75cm,below of=3] (8) {8};
    \node[fill={rgb,190:red,72;green,132;blue,189},draw,circle,line width=0.5pt,minimum size=0.75cm,below of=4] (9) {9};
    \node[fill={rgb,190:red,72;green,132;blue,189},draw,circle,line width=0.5pt,minimum size=0.75cm,below of=5] (10) {10};
    \node[fill={rgb,190:red,72;green,132;blue,189},draw,circle,line width=0.5pt,minimum size=0.75cm,below of=6] (11) {11};
    \node[fill={rgb,190:red,72;green,132;blue,189},draw,circle,line width=0.5pt,minimum size=0.75cm,below of=7] (12) {12};
    \node[fill={rgb,190:red,72;green,132;blue,189},draw,circle,line width=0.5pt,minimum size=0.75cm,below of=8] (13) {13};
    \node[fill={rgb,190:red,72;green,132;blue,189},draw,circle,line width=0.5pt,minimum size=0.75cm,below of=9] (14) {14};
    \node[fill={rgb,190:red,72;green,132;blue,189},draw,circle,line width=0.5pt,minimum size=0.75cm,below of=10] (15) {15};
    \node[fill={rgb,190:red,72;green,132;blue,189},draw,circle,line width=0.5pt,minimum size=0.75cm,below of=11] (16) {16};
    \node[fill={rgb,190:red,72;green,132;blue,189},draw,circle,line width=0.5pt,minimum size=0.75cm,below of=12] (17) {17};
    \node[fill={rgb,190:red,72;green,132;blue,189},draw,circle,line width=0.5pt,minimum size=0.75cm,below of=13] (18) {18};
    \node[fill={rgb,190:red,180;green,56;blue,101},draw,circle,line width=0.5pt,minimum size=0.75cm,below of=14] (19) {19};

    \draw
        (0) edge[line width=2pt] node{} (1)
        (1) edge[line width=2pt] node{} (2)
        (2) edge[line width=2pt] node{} (3)
        (3) edge[line width=2pt] node{} (4)
        (0) edge[line width=2pt] node{} (5)
        (4) edge[line width=2pt] node{} (9)
        (5) edge[line width=2pt] node{} (6)
        (6) edge[line width=2pt] node{} (7)
        (7) edge[line width=2pt] node{} (8)
        (8) edge[line width=2pt] node{} (9)
        (5) edge[line width=2pt] node{} (10)
        (9) edge[line width=2pt] node{} (14)
        (7) edge[line width=2pt] node{} (12)
        (10) edge[line width=2pt] node{} (11)
        (11) edge[line width=2pt] node{} (12)
        (12) edge[line width=2pt] node{} (13)
        (13) edge[line width=2pt] node{} (14)
        (10) edge[line width=2pt] node{} (15)
        (14) edge[line width=2pt] node{} (19)
        (15) edge[line width=2pt] node{} (16)
        (16) edge[line width=2pt] node{} (17)
        (17) edge[line width=2pt] node{} (18)
        (18) edge[line width=2pt] node{} (19)
        
        (19) edge[line width=1pt,->,bend left=50,below] node{1} (18)
        (6) edge[line width=1pt,->,bend right=50,above] node{1} (5)
        (2) edge[line width=1pt,->,bend right=50,above] node{1} (1)
        
        (18) edge[line width=1pt,->,bend left=50,below] node{2} (17)
        (5) edge[line width=1pt,->,bend left=30,left] node{2} (0)
        (17) edge[line width=1pt,->,bend left=50,below] node{3} (16)
        
        (0) edge[line width=1pt,->,bend right=80,left] node{4, 10} (5)
        
        (5) edge[line width=1pt,->,bend right=80,left] node{5, 9} (10)
        (1) edge[line width=1pt,->,bend right=50,above] node{5} (0)
        
        (10) edge[line width=1pt,->,bend right=80,left] node{6, 10} (15)
        (15) edge[line width=1pt,->,bend left=30,left] node{7} (10)
        
        (16) edge[line width=1pt,->,bend left=50,below] node{8} (15)
        (10) edge[line width=1pt,->,bend left=30,left] node{8} (5);

    \path[use as bounding box] (-2,-5.4) rectangle (6.8,1.8);
\end{tikzpicture}%
}%
        \vspace{-0.25em}\\
        (a) Expensive Qiskit routing
    \end{minipage}%
    \begin{minipage}[][][b]{0.45\columnwidth}
        \centering
        \scalebox{1.5}{\begin{tikzpicture} [node distance=1.2cm,scale=0.6, every node/.style={scale=0.6}]
    \node[fill={rgb,190:red,72;green,132;blue,189},draw,circle,line width=0.5pt,minimum size=0.75cm,] (0) {0};
    \node[fill={rgb,190:red,112;green,169;blue,45},draw,circle,line width=0.5pt,minimum size=0.75cm,right of=0] (1) {1};
    \node[fill={rgb,190:red,180;green,56;blue,101},draw,circle,line width=0.5pt,minimum size=0.75cm,right of=1] (2) {2};
    \node[fill={rgb,190:red,112;green,169;blue,45},draw,circle,line width=0.5pt,minimum size=0.75cm,right of=2] (3) {3};
    \node[fill={rgb,190:red,72;green,132;blue,189},draw,circle,line width=0.5pt,minimum size=0.75cm,right of=3] (4) {4};
    \node[fill={rgb,190:red,72;green,132;blue,189},draw,circle,line width=0.5pt,minimum size=0.75cm,below of=0] (5) {5};
    \node[fill={rgb,190:red,180;green,56;blue,101},draw,circle,line width=0.5pt,minimum size=0.75cm,below of=1] (6) {6};
    \node[fill={rgb,190:red,72;green,132;blue,189},draw,circle,line width=0.5pt,minimum size=0.75cm,below of=2] (7) {7};
    \node[fill={rgb,190:red,72;green,132;blue,189},draw,circle,line width=0.5pt,minimum size=0.75cm,below of=3] (8) {8};
    \node[fill={rgb,190:red,72;green,132;blue,189},draw,circle,line width=0.5pt,minimum size=0.75cm,below of=4] (9) {9};
    \node[fill={rgb,190:red,72;green,132;blue,189},draw,circle,line width=0.5pt,minimum size=0.75cm,below of=5] (10) {10};
    \node[fill={rgb,190:red,72;green,132;blue,189},draw,circle,line width=0.5pt,minimum size=0.75cm,below of=6] (11) {11};
    \node[fill={rgb,190:red,72;green,132;blue,189},draw,circle,line width=0.5pt,minimum size=0.75cm,below of=7] (12) {12};
    \node[fill={rgb,190:red,72;green,132;blue,189},draw,circle,line width=0.5pt,minimum size=0.75cm,below of=8] (13) {13};
    \node[fill={rgb,190:red,72;green,132;blue,189},draw,circle,line width=0.5pt,minimum size=0.75cm,below of=9] (14) {14};
    \node[fill={rgb,190:red,72;green,132;blue,189},draw,circle,line width=0.5pt,minimum size=0.75cm,below of=10] (15) {15};
    \node[fill={rgb,190:red,72;green,132;blue,189},draw,circle,line width=0.5pt,minimum size=0.75cm,below of=11] (16) {16};
    \node[fill={rgb,190:red,72;green,132;blue,189},draw,circle,line width=0.5pt,minimum size=0.75cm,below of=12] (17) {17};
    \node[fill={rgb,190:red,72;green,132;blue,189},draw,circle,line width=0.5pt,minimum size=0.75cm,below of=13] (18) {18};
    \node[fill={rgb,190:red,180;green,56;blue,101},draw,circle,line width=0.5pt,minimum size=0.75cm,below of=14] (19) {19};

    \draw
        (0) edge[line width=2pt] node{} (1)
        (1) edge[line width=2pt] node{} (2)
        (2) edge[line width=2pt] node{} (3)
        (3) edge[line width=2pt] node{} (4)
        (0) edge[line width=2pt] node{} (5)
        (4) edge[line width=2pt] node{} (9)
        (5) edge[line width=2pt] node{} (6)
        (6) edge[line width=2pt] node{} (7)
        (7) edge[line width=2pt] node{} (8)
        (8) edge[line width=2pt] node{} (9)
        (5) edge[line width=2pt] node{} (10)
        (9) edge[line width=2pt] node{} (14)
        (7) edge[line width=2pt] node{} (12)
        (10) edge[line width=2pt] node{} (11)
        (11) edge[line width=2pt] node{} (12)
        (12) edge[line width=2pt] node{} (13)
        (13) edge[line width=2pt] node{} (14)
        (10) edge[line width=2pt] node{} (15)
        (14) edge[line width=2pt] node{} (19)
        (15) edge[line width=2pt] node{} (16)
        (16) edge[line width=2pt] node{} (17)
        (17) edge[line width=2pt] node{} (18)
        (18) edge[line width=2pt] node{} (19)
        
        (19) edge[line width=1pt,->,bend right=50,right] node{1} (14)
        (6) edge[line width=1pt,->,bend right=50,above] node{1} (5)
        
        (5) edge[line width=1pt,->,bend left=50,left] node{2} (0)
        (14) edge[line width=1pt,->,bend right=50,right] node{2} (9)
        
        (0) edge[line width=1pt,->,bend left=50,above] node{3} (1)
        (9) edge[line width=1pt,->,bend right=50,right] node{3} (4)
        
        (4) edge[line width=1pt,->,bend right=50,above] node{4} (3);

    \path[use as bounding box] (-2,-5.4) rectangle (6.8,1.8);
\end{tikzpicture}%
}%
        \vspace{-0.25em}\\
        (b) Efficient Trios routing
    \end{minipage}
    \caption{Example routing from Qiskit (a) vs. Trios (b) for a single Toffoli operation. Circles represent qubits and lines indicate two qubits are connected.  Input qubits are highlighted in red.
    SWAP arrows are labeled by timestep.
    The routed locations for Trios routing are highlighted in green while Qiskit moves them several times.
    Qiskit adds 16 SWAPs (=48 CNOTs), some during the Toffoli, while Trios adds only 7 SWAPs (=21 CNOTs) all before the Toffoli.
    Performing multiple passes of decomposition allows direct routing and enables this huge reduction in communication, increasing the probability of program success.
    }%
    \label{fig:swap_example}
\end{figure}
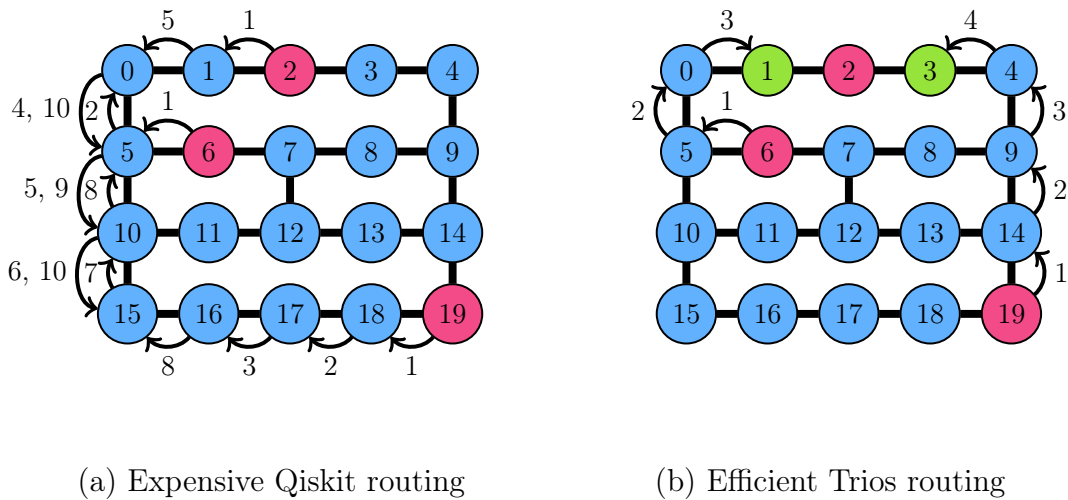


Quantum program compilation involves many passes of transformations and optimizations similar in many ways to classical compilers. Some optimizations occur at the abstract circuit level, independent of the underlying hardware, such as the gate cancellation from~\cite{cancel}. One of the first steps usually taken is to convert an input program into a gate set (ISA) supported by the target hardware. For example, on IBM devices, gates are typically rewritten using only gates in the set $\{u1, u2, u3, cx\}$ (\cite{ibmq}, single-qubit gates and the common CNOT gate described later). One critical limitation of many current available architectures is the inability to execute more complex multi-qubit operations, like the Toffoli, directly; instead, these gates must be decomposed into the supported one- and two-qubit gates. Furthermore, many current superconducting architectures only support two qubit operations on adjacent hardware qubits wired together with a coupler.  This requires the insertion of additional operations called SWAPs to move the data onto adjacent qubits (which are connected with a coupler).

\begin{figure}
    \centering
    \begin{minipage}[][][b]{0.5\columnwidth}
        \centering
        \begin{tikzpicture} [node distance=1.2cm,scale=1,inner sep=4pt]
    \node[rectangle, rounded corners, fill=black!4!white, blur shadow={shadow blur steps=5}] (0) {Input program};
    \node[arrowstyle1, below=0.1 of 0] (0a) {~~~~};
    \node[rectangle, below=0.25 of 0] (0b) {Unroll+Decompose};
    \node[rectangle, rounded corners, fill=black!5!white, blur shadow={shadow blur steps=5}, below=-0.3 of 0a] (1) {
        \begin{tabular}{c}Circuit of 1- and 2-qubit gates \\ (between any qubit pairs)\end{tabular}};
    \node[arrowstyle2, below=0.1 of 1] (1a) {~~~~};
    \node[rectangle, below=0.25 of 1] (1b) {Map and Route};
    \node[rectangle, rounded corners, fill=black!6!white, blur shadow={shadow blur steps=5}, below=-0.3 of 1a] (2) {
        \begin{tabular}{c}Circuit of 1- and 2-qubit gates \\ (between connected qubits)\end{tabular}};
    \node[arrowstyle3, below=0.1 of 2] (2a) {~~~~};
    \node[rectangle, below=0.25 of 2] (2b) {Schedule};
    \node[rectangle, rounded corners, fill=black!7!white, blur shadow={shadow blur steps=5}, below=-0.3 of 2a] (3) {Executable Circuit};
\end{tikzpicture}%
    \end{minipage}%
    \begin{minipage}[][][b]{0.5\columnwidth}
        \centering
        \begin{tikzpicture} [node distance=1.2cm,scale=1,inner sep=4pt]
    \node[rectangle, rounded corners, fill=black!4!white, blur shadow={shadow blur steps=5}] (0) {Input program};
    \node[arrowstyle1, below=0.1 of 0] (0a) {~~~~};
    \node[rectangle, below=0.25 of 0] (0b) {Unroll+Decompose to Toffoli};
    
    \node[rectangle, rounded corners, fill=black!5!white, blur shadow={shadow blur steps=5}, below=-0.3 of 0a] (1) {
        \begin{tabular}{c}Circuit of 3-qubit Toffoli \\ and other 1- and 2-qubit gates\end{tabular}};
    \node[arrowstyle2, below=0.1 of 1] (1a) {~~~~};
    \node[rectangle, below=0.25 of 1] (1b) {Map and Route};
    
    \node[rectangle, rounded corners, fill=black!6!white, blur shadow={shadow blur steps=5}, below=-0.3 of 1a] (2) {
        \begin{tabular}{c}Circuit of Toffoli gates \\ between nearby qubits\end{tabular}};
    \node[arrowstyle1, below=0.1 of 2] (2a) {~~~~};
    \node[rectangle, below=0.25 of 2] (2b) {Mapping-Aware Decompose};
    
    \node[rectangle, rounded corners, fill=black!7!white, blur shadow={shadow blur steps=5}, below=-0.3 of 2a] (3) {
        \begin{tabular}{c}Circuit of 1- and 2-qubit gates \\ (between connected qubits)\end{tabular}};
    \node[arrowstyle3, below=0.1 of 3] (3a) {~~~~};
    \node[rectangle, below=0.25 of 3] (3b) {Schedule};
    
    \node[rectangle, rounded corners, fill=black!8!white, blur shadow={shadow blur steps=5}, below=-0.3 of 3a] (4) {Executable Circuit};
\end{tikzpicture}%
    \end{minipage}

    \vspace{1em}

    \begin{minipage}[][][b]{0.5\columnwidth}
        \centering
        (a) Conventional compilation
    \end{minipage}%
    \begin{minipage}[][][b]{0.5\columnwidth}
        \centering
        (b) Trios compilation
    \end{minipage}

    \caption{(a) Typical compilation passes used by Qiskit (simplified). (b) Trios compilation passes.  The Unroll+Decompose pass is split into two parts: decompose into medium-size operations (Toffoli gates), later finish decomposition, but using information from the Map and Route pass.}%
    \label{fig:tool-flow}
\end{figure}
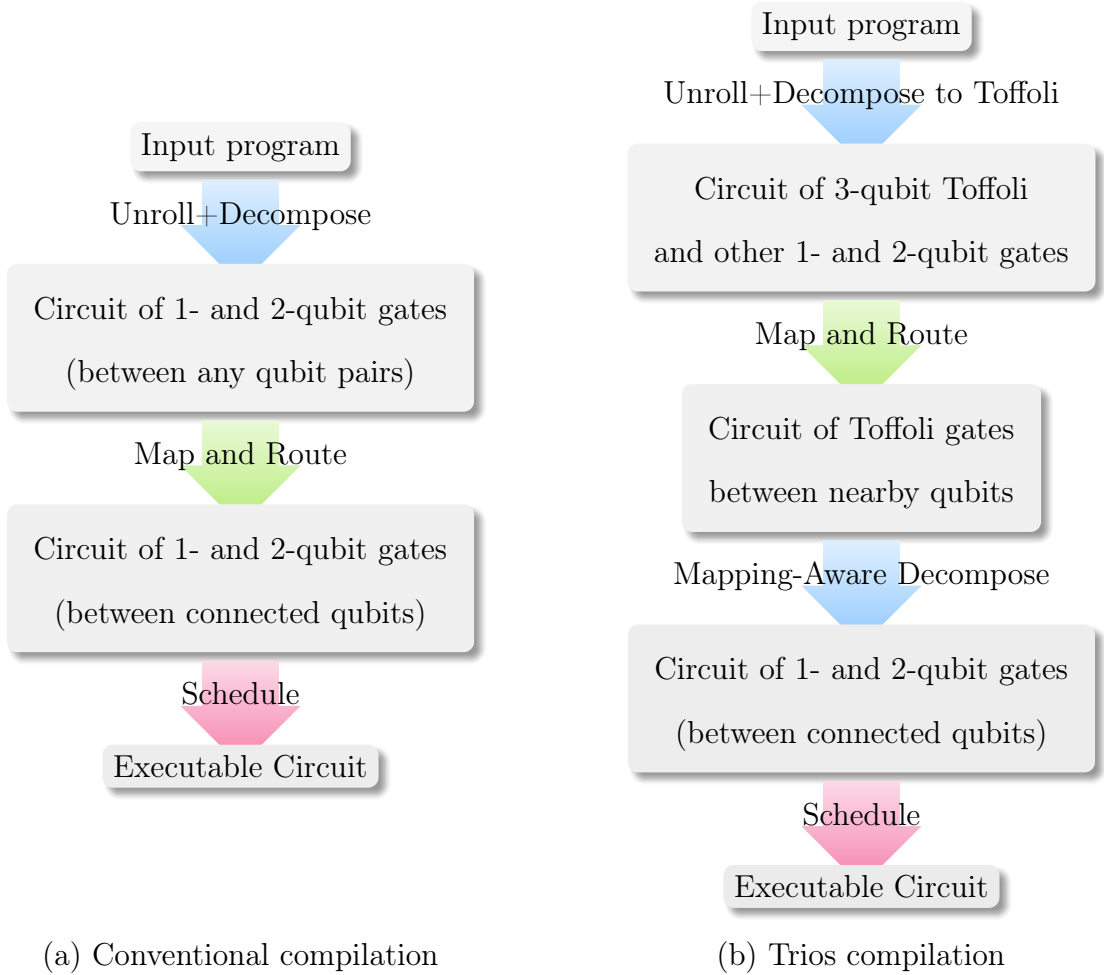

The process of transforming an optimized and decomposed program to the desired target is typically broken down into three distinct steps: decomposing the program into basic gates, mapping the abstract qubits of a program to specific hardware qubits and routing interacting quantum data so that they occupy adjacent qubits on hardware when they have an operation, and scheduling operations in order to minimize total program run time (depth) or to minimize errors due to crosstalk as in~\cite{xtalk}. Each of these steps is critical to the success of the input program. A well-mapped and well-routed program will reduce the total number of communication (SWAP) operations added and subsequently reduce the compiled program's depth, both of which will increase the chance of success. Conventionally, these three steps occur sequentially. By doing so, current strategies are unable to account for any hierarchical structure in the input program, resulting in inefficient routing of qubits.  An optimal compiler could find the best routing despite the lack of structure but at the cost of much slower, typically impractical, compilation.  Consider the SWAP sequences inserted by IBM's Qiskit compiler for a single Toffoli compiled to IBM's Johannesburg device in Figure~\ref{fig:swap_example}a. This baseline strategy adds a large number of unnecessary SWAPs as it individually routes each CNOT composing the Toffoli, dramatically reducing the probability of successful execution.

Our approach, Orchestrated Trios (originally presented in \cite{trios}\footnotemark) decomposes and routes qubits in multiple stages, as seen in Figure~\ref{fig:tool-flow}b.
Trios first decomposes, or flattens, a program into intermediate one-, two-, \textit{and} three-qubit gates (e.g.\ it does not decompose Toffoli gates).
Trios performs qubit routing as usual except for three-qubits, routing all three to a common location with minimal SWAPs. This new program can then undergo a second round of decomposition to produce a circuit containing only hardware-permitted primitive one- and two-qubit gates.
The second round may use information from previous passes (i.e.\ locations of data qubits on the device) to generate fine-tuned decompositions for the architecture.
\footnotetext{CD's contributions to the work that comprises this chapter include design of the split decompose and mapping-aware passes and the application benchmark compilation, simulation, and sensitivity.}

This layered approach has a major advantage over current routing techniques: we are better able to capture program structure by inspecting intermediate, non-primitive, operations for routing. This better informs how data should be placed and moved around the device during program execution. In Figure~\ref{fig:swap_example}, the Trios strategy reduces the total number of SWAPs added to 21: fewer than half compared to Qiskit.  This was an extreme example we selected to present the issue, not an average case.

We specifically propose a split-pass approach to circuit decomposition.  We will focus on superconducting hardware systems like IBM's cloud accessible devices, but our strategy can easily be adapted to other systems. An overview of our compilation structure is found in Figure~\ref{fig:tool-flow}b. This strategy has a substantial benefit on the overall success rate of programs. We demonstrate these improvements by executing Toffoli gates on a real IBM quantum computer and estimating success probability of a suite of benchmarks via simulation.

Our contributions are as follows:
\begin{itemize}
    \item A new compiler structure, Trios, with two passes for decomposition with a modified routing pass in between which greatly improves qubit routing.
    \item A simple method for architecture-tuned Toffoli decompositions during the second decompose pass that allows for a new kind of location-aware optimization.
    \item On Toffoli-only experiments, Trios reduces the total number of gates by 35\% geomean (geometric mean) resulting in 23\% geomean increase in success rate when run on real IBM hardware as compared to Qiskit.
    \item On near-term algorithms shown in Figure~\ref{fig:benchmark-success-norm} (4 to 20 qubit benchmarks), Trios reduces total gate count by 37\% geomean resulting in 344\% geomean increase (or 4.44x) in simulated success rate on IBM Johannesburg with noise rates of near-future hardware as compared to programs compiled without Trios.  A sensitivity analysis over four architecture types shows the benefit range from 133\% to 3020\% increase in success rate.
\end{itemize}

\FloatBarrier{}

\section{Background}%
\label{sec:modular-background}

\subsection{Quantum Computing Basics}

The most basic object in quantum computing is the quantum bit (qubit). Unlike a classical bit which is either 0 or 1, the qubit has two basis states $\ket{0}$ and $\ket{1}$ and can exist as a linear superposition over these two states, i.e.\ for a quantum state $\ket{\psi} = \alpha\ket{0} + \beta\ket{1}$ with $\alpha, \beta \in \mathbb{C}$ and $\|\alpha\|^2 + \|\beta\|^2 = 1$. In general, a quantum system consisting of $n$ qubits can exist in a linear superposition of $2^n$ basis states in contrast to a classical system of $n$ bits which can exist as exactly a single of these states. An important feature which gives quantum computing its power is the ability to \textit{entangle} qubits via two qubit operations like the CNOT\@. This, along with quantum interference between the complex amplitudes, allows quantum programs to solve certain problems faster than classical computers. Another common two-qubit gate is the SWAP gate which switches quantum data, in-place, between two qubits.

While a qubit system can exist in these superpositions during computation, at the end of the computation, the qubits are measured producing a classical binary outcome. The probability of each outcome depends on the amplitude of each basis state (the values of $\alpha, \beta, \gamma, \dots$). Consequently, since the outcome of a quantum program is a classical bitstring and because quantum systems are inherently noisy, programs are usually run thousands of times to obtain a distribution over possible answers. A comprehensive background can be found in~\cite{mikeike}.

\subsection{Quantum Circuits}

Quantum programs are typically represented as a circuit which, like a classical program, is an ordered list of instructions. Here the instructions are quantum logic gates applied to qubits. The input circuit may not be expressed in the instruction set supported by the underlying hardware or it might even be structured as hierarchical modules.

Quantum circuits have a single line for each qubit, with time flowing from left to right. Gates in a quantum circuits have the same number of inputs and outputs and gates on disjoint sets of lines can be executed in parallel. Single qubit gates are represented as boxes labeled with the indicated operations.  Controlled operations, like the CNOT and Toffoli, have one or two control qubits respectively indicated by $\bullet$ and a target qubit given by $\oplus$.

\begin{figure}
    \centering
    \begin{minipage}[][][b]{\columnwidth}
    \centering
    \scalebox{1}{%
        \input{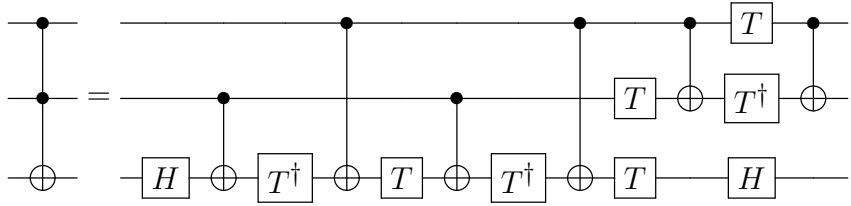}%
    }%
    \caption{The common 6-CNOT decomposition of the Toffoli gate.}%
    \label{fig:6-cnot-decomp}
    \end{minipage}\\
\end{figure}

\begin{figure}
    \begin{minipage}[][][b]{\columnwidth}
    \centering%
    \scalebox{1}{%
        \input{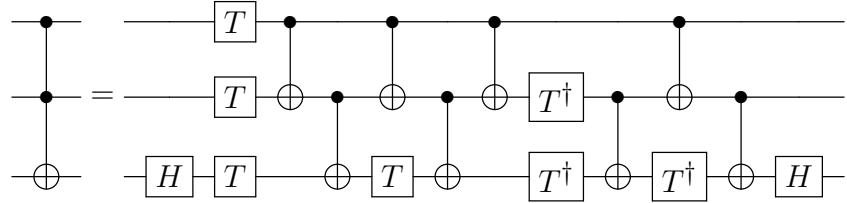}%
    }%
    \caption{An 8-CNOT decomposition of the Toffoli gate with the same behavior.}%
    \label{fig:8-cnot-decomp}
    \end{minipage}
\end{figure}

Currently available superconducting quantum hardware, like that of IBM and Rigetti, only supports one-qubit gates and two-qubit gates on specific pairs. Therefore, more complex instructions must be decomposed into multiple simpler, supported operations and SWAPs must me inserted to move quantum data around the device. For example, many quantum algorithms and subroutines make use of the Toffoli gate, a three-input gate which performs the logical AND between two control bits and writes the output onto the target bit. This gate cannot typically be executed directly on available hardware and instead is decomposed into an equivalent sequence of one- and two-qubit operations. Two such decompositions are given in Figures~\ref{fig:6-cnot-decomp} and~\ref{fig:8-cnot-decomp}. There are two key distinctions in these decompositions illustrating a more general trade off. The first, taught in~\cite{mikeike}, is the ubiquitous decomposition using the minimum 6 CNOT gates, but it requires CNOTs between all three pairs of qubits.  This would require either inserted SWAPs or a device connectivity containing a triangle. The second, originally discovered by~\cite{craig8}, uses a total of 8 CNOT gates but requires all three inputs be only linearly connected (only two of the three qubit pairs are required to be connected). While the first is apparently more efficient, this is not true if the connectivity of the underlying hardware does not directly support it.  It is more efficient to use the 8-CNOT version than to use the 6-CNOT version with SWAPs added for feasibility.

For superconducting qubits, current quantum computers support gates only between adjacent hardware qubits. In order to use qubits which are currently mapped far apart on the hardware, extra SWAP operations must be inserted, each of these SWAPs is usually decomposed as a series of 3 CNOT gates (equivalent to a classical memory in-place swap using 3 alternating XORs). In the case of the 6-CNOT Toffoli decomposition above, when mapped to a device with linear or square grid connectivity, no triangles exist so extra SWAPs will need to be inserted, resulting in a greater total number of CNOTs due to the mismatch with hardware details.


\subsection{Current Quantum Devices}%
\label{background:devices}

\begin{figure*}
    \centering
    \hfill%
    \begin{minipage}[][][b]{0.3333\textwidth}
        \centering
        \begin{tikzpicture} [node distance=1.4cm,scale=0.75, every node/.style={scale=0.75}]
        \node[fill={rgb,255:red,255;green,113;blue,0},draw,circle,line width=0.5pt,minimum size=2.2em,] (0) {0};
        \node[fill={rgb,255:red,255;green,113;blue,0},draw,circle,line width=0.5pt,minimum size=2.2em,right of=0] (1) {1};
        \node[fill={rgb,255:red,255;green,113;blue,0},draw,circle,line width=0.5pt,minimum size=2.2em,right of=1] (2) {2};
        \node[fill={rgb,255:red,255;green,113;blue,0},draw,circle,line width=0.5pt,minimum size=2.2em,right of=2] (3) {3};
        \node[fill={rgb,255:red,255;green,113;blue,0},draw,circle,line width=0.5pt,minimum size=2.2em,right of=3] (4) {4};
        \node[fill={rgb,255:red,255;green,113;blue,0},draw,circle,line width=0.5pt,minimum size=2.2em,below of=0] (5) {5};
        \node[fill={rgb,255:red,255;green,113;blue,0},draw,circle,line width=0.5pt,minimum size=2.2em,below of=1] (6) {6};
        \node[fill={rgb,255:red,255;green,113;blue,0},draw,circle,line width=0.5pt,minimum size=2.2em,below of=2] (7) {7};
        \node[fill={rgb,255:red,255;green,113;blue,0},draw,circle,line width=0.5pt,minimum size=2.2em,below of=3] (8) {8};
        \node[fill={rgb,255:red,255;green,113;blue,0},draw,circle,line width=0.5pt,minimum size=2.2em,below of=4] (9) {9};
        \node[fill={rgb,255:red,255;green,113;blue,0},draw,circle,line width=0.5pt,minimum size=2.2em,below of=5] (10) {10};
        \node[fill={rgb,255:red,255;green,113;blue,0},draw,circle,line width=0.5pt,minimum size=2.2em,below of=6] (11) {11};
        \node[fill={rgb,255:red,255;green,113;blue,0},draw,circle,line width=0.5pt,minimum size=2.2em,below of=7] (12) {12};
        \node[fill={rgb,255:red,255;green,113;blue,0},draw,circle,line width=0.5pt,minimum size=2.2em,below of=8] (13) {13};
        \node[fill={rgb,255:red,255;green,113;blue,0},draw,circle,line width=0.5pt,minimum size=2.2em,below of=9] (14) {14};
        \node[fill={rgb,255:red,255;green,113;blue,0},draw,circle,line width=0.5pt,minimum size=2.2em,below of=10] (15) {15};
        \node[fill={rgb,255:red,255;green,113;blue,0},draw,circle,line width=0.5pt,minimum size=2.2em,below of=11] (16) {16};
        \node[fill={rgb,255:red,255;green,113;blue,0},draw,circle,line width=0.5pt,minimum size=2.2em,below of=12] (17) {17};
        \node[fill={rgb,255:red,255;green,113;blue,0},draw,circle,line width=0.5pt,minimum size=2.2em,below of=13] (18) {18};
        \node[fill={rgb,255:red,255;green,113;blue,0},draw,circle,line width=0.5pt,minimum size=2.2em,below of=14] (19) {19};
    
        \draw
            (0) edge[line width=2pt] node{} (1)
            (1) edge[line width=2pt] node{} (2)
            (2) edge[line width=2pt] node{} (3)
            (3) edge[line width=2pt] node{} (4)
            (0) edge[line width=2pt] node{} (5)
            (4) edge[line width=2pt] node{} (9)
            (5) edge[line width=2pt] node{} (6)
            (6) edge[line width=2pt] node{} (7)
            (7) edge[line width=2pt] node{} (8)
            (8) edge[line width=2pt] node{} (9)
            (5) edge[line width=2pt] node{} (10)
            (9) edge[line width=2pt] node{} (14)
            (7) edge[line width=2pt] node{} (12)
            (10) edge[line width=2pt] node{} (11)
            (11) edge[line width=2pt] node{} (12)
            (12) edge[line width=2pt] node{} (13)
            (13) edge[line width=2pt] node{} (14)
            (10) edge[line width=2pt] node{} (15)
            (14) edge[line width=2pt] node{} (19)
            (15) edge[line width=2pt] node{} (16)
            (16) edge[line width=2pt] node{} (17)
            (17) edge[line width=2pt] node{} (18)
            (18) edge[line width=2pt] node{} (19);
\end{tikzpicture}%
    \end{minipage}\hfill%
    \begin{minipage}[][][b]{0.3333\textwidth}
        \centering
        \begin{tikzpicture} [node distance=1.4cm,scale=0.75, every node/.style={scale=0.75}]
        \node[fill={rgb,255:red,209;green,216;blue,0},draw,circle,line width=0.5pt,minimum size=2.2em,] (0) {0};
        \node[fill={rgb,255:red,209;green,216;blue,0},draw,circle,line width=0.5pt,minimum size=2.2em,right of=0] (1) {1};
        \node[fill={rgb,255:red,209;green,216;blue,0},draw,circle,line width=0.5pt,minimum size=2.2em,right of=1] (2) {2};
        \node[fill={rgb,255:red,209;green,216;blue,0},draw,circle,line width=0.5pt,minimum size=2.2em,right of=2] (3) {3};
        \node[fill={rgb,255:red,209;green,216;blue,0},draw,circle,line width=0.5pt,minimum size=2.2em,right of=3] (4) {4};
        \node[fill={rgb,255:red,209;green,216;blue,0},draw,circle,line width=0.5pt,minimum size=2.2em,below of=0] (5) {5};
        \node[fill={rgb,255:red,209;green,216;blue,0},draw,circle,line width=0.5pt,minimum size=2.2em,below of=1] (6) {6};
        \node[fill={rgb,255:red,209;green,216;blue,0},draw,circle,line width=0.5pt,minimum size=2.2em,below of=2] (7) {7};
        \node[fill={rgb,255:red,209;green,216;blue,0},draw,circle,line width=0.5pt,minimum size=2.2em,below of=3] (8) {8};
        \node[fill={rgb,255:red,209;green,216;blue,0},draw,circle,line width=0.5pt,minimum size=2.2em,below of=4] (9) {9};
        \node[fill={rgb,255:red,209;green,216;blue,0},draw,circle,line width=0.5pt,minimum size=2.2em,below of=5] (10) {10};
        \node[fill={rgb,255:red,209;green,216;blue,0},draw,circle,line width=0.5pt,minimum size=2.2em,below of=6] (11) {11};
        \node[fill={rgb,255:red,209;green,216;blue,0},draw,circle,line width=0.5pt,minimum size=2.2em,below of=7] (12) {12};
        \node[fill={rgb,255:red,209;green,216;blue,0},draw,circle,line width=0.5pt,minimum size=2.2em,below of=8] (13) {13};
        \node[fill={rgb,255:red,209;green,216;blue,0},draw,circle,line width=0.5pt,minimum size=2.2em,below of=9] (14) {14};
        \node[fill={rgb,255:red,209;green,216;blue,0},draw,circle,line width=0.5pt,minimum size=2.2em,below of=10] (15) {15};
        \node[fill={rgb,255:red,209;green,216;blue,0},draw,circle,line width=0.5pt,minimum size=2.2em,below of=11] (16) {16};
        \node[fill={rgb,255:red,209;green,216;blue,0},draw,circle,line width=0.5pt,minimum size=2.2em,below of=12] (17) {17};
        \node[fill={rgb,255:red,209;green,216;blue,0},draw,circle,line width=0.5pt,minimum size=2.2em,below of=13] (18) {18};
        \node[fill={rgb,255:red,209;green,216;blue,0},draw,circle,line width=0.5pt,minimum size=2.2em,below of=14] (19) {19};
    
        \draw
            (0) edge[line width=2pt] node{} (1)
            (1) edge[line width=2pt] node{} (2)
            (2) edge[line width=2pt] node{} (3)
            (3) edge[line width=2pt] node{} (4)
            (5) edge[line width=2pt] node{} (6)
            (6) edge[line width=2pt] node{} (7)
            (7) edge[line width=2pt] node{} (8)
            (8) edge[line width=2pt] node{} (9)
            (10) edge[line width=2pt] node{} (11)
            (11) edge[line width=2pt] node{} (12)
            (12) edge[line width=2pt] node{} (13)
            (13) edge[line width=2pt] node{} (14)
            (15) edge[line width=2pt] node{} (16)
            (16) edge[line width=2pt] node{} (17)
            (17) edge[line width=2pt] node{} (18)
            (18) edge[line width=2pt] node{} (19)
            (0) edge[line width=2pt] node{} (5)
            (1) edge[line width=2pt] node{} (6)
            (2) edge[line width=2pt] node{} (7)
            (3) edge[line width=2pt] node{} (8)
            (4) edge[line width=2pt] node{} (9)
            (5) edge[line width=2pt] node{} (10)
            (6) edge[line width=2pt] node{} (11)
            (7) edge[line width=2pt] node{} (12)
            (8) edge[line width=2pt] node{} (13)
            (9) edge[line width=2pt] node{} (14)
            (10) edge[line width=2pt] node{} (15)
            (11) edge[line width=2pt] node{} (16)
            (12) edge[line width=2pt] node{} (17)
            (13) edge[line width=2pt] node{} (18)
            (14) edge[line width=2pt] node{} (19);
\end{tikzpicture}%
    \end{minipage}\hfill%
    \begin{minipage}[][][b]{0.3333\textwidth}
        \centering
        \begin{tikzpicture} [node distance=1.4cm,scale=0.875, every node/.style={scale=0.75}]
        \node[fill={rgb,255:red,125;green,0;blue,255},text=white,draw,circle,line width=0.5pt,minimum size=2.2em] (0) at (0.588,-0.809) {0};
        \node[fill={rgb,255:red,125;green,0;blue,255},text=white,draw,circle,line width=0.5pt,minimum size=2.2em] (1) at (-0.588,-0.809) {1};
        \node[fill={rgb,255:red,125;green,0;blue,255},text=white,draw,circle,line width=0.5pt,minimum size=2.2em] (2) at (-0.951,0.309) {2};
        \node[fill={rgb,255:red,125;green,0;blue,255},text=white,draw,circle,line width=0.5pt,minimum size=2.2em] (3) at (0, 1) {3};
        \node[fill={rgb,255:red,125;green,0;blue,255},text=white,draw,circle,line width=0.5pt,minimum size=2.2em] (4) at (0.951, 0.309) {4};
        \node[fill={rgb,255:red,125;green,0;blue,255},text=white,draw,circle,line width=0.5pt,minimum size=2.2em] (5) at (2.588,-0.809) {5};
        \node[fill={rgb,255:red,125;green,0;blue,255},text=white,draw,circle,line width=0.5pt,minimum size=2.2em] (6) at (2.225,0.309) {6};
        \node[fill={rgb,255:red,125;green,0;blue,255},text=white,draw,circle,line width=0.5pt,minimum size=2.2em] (7) at (3.176,1) {7};
        \node[fill={rgb,255:red,125;green,0;blue,255},text=white,draw,circle,line width=0.5pt,minimum size=2.2em] (8) at (4.127,0.309) {8};
        \node[fill={rgb,255:red,125;green,0;blue,255},text=white,draw,circle,line width=0.5pt,minimum size=2.2em] (9) at (3.764,-0.809) {9};
        \node[fill={rgb,255:red,125;green,0;blue,255},text=white,draw,circle,line width=0.5pt,minimum size=2.2em] (10) at (0.588,-2) {10};
        \node[fill={rgb,255:red,125;green,0;blue,255},text=white,draw,circle,line width=0.5pt,minimum size=2.2em] (11) at (-0.588,-2) {11};
        \node[fill={rgb,255:red,125;green,0;blue,255},text=white,draw,circle,line width=0.5pt,minimum size=2.2em] (12) at (-0.951,-3.1) {12};
        \node[fill={rgb,255:red,125;green,0;blue,255},text=white,draw,circle,line width=0.5pt,minimum size=2.2em] (13) at (0,-3.809) {13};
        \node[fill={rgb,255:red,125;green,0;blue,255},text=white,draw,circle,line width=0.5pt,minimum size=2.2em] (14) at (0.951,-3.1) {14};
        \node[fill={rgb,255:red,125;green,0;blue,255},text=white,draw,circle,line width=0.5pt,minimum size=2.2em] (15) at (2.588,-2) {15};
        \node[fill={rgb,255:red,125;green,0;blue,255},text=white,draw,circle,line width=0.5pt,minimum size=2.2em] (16) at (2.225,-3.1) {16};
        \node[fill={rgb,255:red,125;green,0;blue,255},text=white,draw,circle,line width=0.5pt,minimum size=2.2em] (17) at (3.176,-3.809) {17};
        \node[fill={rgb,255:red,125;green,0;blue,255},text=white,draw,circle,line width=0.5pt,minimum size=2.2em] (18) at (4.127,-3.1) {18};
        \node[fill={rgb,255:red,125;green,0;blue,255},text=white,draw,circle,line width=0.5pt,minimum size=2.2em] (19) at (3.764,-2) {19};
        \draw
        (0) edge[line width=2pt] node{} (1)
        (0) edge[line width=2pt] node{} (2)
        (0) edge[line width=2pt] node{} (3)
        (0) edge[line width=2pt] node{} (4)
        (1) edge[line width=2pt] node{} (2)
        (1) edge[line width=2pt] node{} (3)
        (1) edge[line width=2pt] node{} (4)
        (2) edge[line width=2pt] node{} (3)
        (2) edge[line width=2pt] node{} (4)
        (3) edge[line width=2pt] node{} (4)
        (0) edge[line width=2pt] node{} (5)
        (0) edge[line width=2pt] node{} (10)
        
        (0) edge[line width=2pt] node{} (15)
        (10) edge[line width=2pt] node{} (5)
        
        (5) edge[line width=2pt] node{} (6)
        (5) edge[line width=2pt] node{} (7)
        (5) edge[line width=2pt] node{} (8)
        (5) edge[line width=2pt] node{} (9)
        (6) edge[line width=2pt] node{} (7)
        (6) edge[line width=2pt] node{} (8)
        (6) edge[line width=2pt] node{} (9)
        (7) edge[line width=2pt] node{} (8)
        (7) edge[line width=2pt] node{} (9)
        (8) edge[line width=2pt] node{} (9)
        (5) edge[line width=2pt] node{} (15)
        
        (10) edge[line width=2pt] node{} (11)
        (10) edge[line width=2pt] node{} (12)
        (10) edge[line width=2pt] node{} (13)
        (10) edge[line width=2pt] node{} (14)
        (11) edge[line width=2pt] node{} (12)
        (11) edge[line width=2pt] node{} (13)
        (11) edge[line width=2pt] node{} (14)
        (12) edge[line width=2pt] node{} (13)
        (12) edge[line width=2pt] node{} (14)
        (13) edge[line width=2pt] node{} (14)
        (10) edge[line width=2pt] node{} (15)
        
        (15) edge[line width=2pt] node{} (16)
        (15) edge[line width=2pt] node{} (17)
        (15) edge[line width=2pt] node{} (18)
        (15) edge[line width=2pt] node{} (19)
        (16) edge[line width=2pt] node{} (17)
        (16) edge[line width=2pt] node{} (18)
        (16) edge[line width=2pt] node{} (19)
        (17) edge[line width=2pt] node{} (18)
        (17) edge[line width=2pt] node{} (19)
        (18) edge[line width=2pt] node{} (19);
\end{tikzpicture}%
    \end{minipage}\hfill%
    \vspace{0.75em}\\
    \begin{minipage}[][][b]{0.3333\textwidth}
        \centering
        (a) IBM Johannesburg
    \end{minipage}\hfill%
    \begin{minipage}[][][b]{0.3333\textwidth}
        \centering
        (b) 2D Grid
    \end{minipage}\hfill%
    \begin{minipage}[][][b]{0.3333\textwidth}
        \centering
        (c) 4$\times$ 5-qubit clusters
    \end{minipage}\hfill%
    \vspace{1.5em}\\
    \begin{minipage}[][][b]{\textwidth}
        \centering
        \vspace{1em}
        \begin{tikzpicture} [node distance=1.1cm,scale=0.75, every node/.style={scale=0.75}]
        \node[fill={rgb,255:red,0;green,205;blue,110},draw,line width=0.5pt,circle,minimum size=2.2em,] (0) {0};
        \node[fill={rgb,255:red,0;green,205;blue,110},draw,line width=0.5pt,circle,minimum size=2.2em,right of=0] (1) {1};
        \node[fill={rgb,255:red,0;green,205;blue,110},draw,line width=0.5pt,circle,minimum size=2.2em,right of=1] (2) {2};
        \node[fill={rgb,255:red,0;green,205;blue,110},draw,line width=0.5pt,circle,minimum size=2.2em,right of=2] (3) {3};
        \node[fill={rgb,255:red,0;green,205;blue,110},draw,line width=0.5pt,circle,minimum size=2.2em,right of=3] (4) {4};
        \node[fill={rgb,255:red,0;green,205;blue,110},draw,line width=0.5pt,circle,minimum size=2.2em,right of=4] (5) {5};
        \node[fill={rgb,255:red,0;green,205;blue,110},draw,line width=0.5pt,circle,minimum size=2.2em,right of=5] (6) {6};
        \node[fill={rgb,255:red,0;green,205;blue,110},draw,line width=0.5pt,circle,minimum size=2.2em,right of=6] (7) {7};
        \node[fill={rgb,255:red,0;green,205;blue,110},draw,line width=0.5pt,circle,minimum size=2.2em,right of=7] (8) {8};
        \node[fill={rgb,255:red,0;green,205;blue,110},draw,line width=0.5pt,circle,minimum size=2.2em,right of=8] (9) {9};
        \node[fill={rgb,255:red,0;green,205;blue,110},draw,line width=0.5pt,circle,minimum size=2.2em,right of=9] (10) {10};
        \node[fill={rgb,255:red,0;green,205;blue,110},draw,line width=0.5pt,circle,minimum size=2.2em,right of=10] (11) {11};
        \node[fill={rgb,255:red,0;green,205;blue,110},draw,line width=0.5pt,circle,minimum size=2.2em,right of=11] (12) {12};
        \node[fill={rgb,255:red,0;green,205;blue,110},draw,line width=0.5pt,circle,minimum size=2.2em,right of=12] (13) {13};
        \node[fill={rgb,255:red,0;green,205;blue,110},draw,line width=0.5pt,circle,minimum size=2.2em,right of=13] (14) {14};
        \node[fill={rgb,255:red,0;green,205;blue,110},draw,line width=0.5pt,circle,minimum size=2.2em,right of=14] (15) {15};
        \node[fill={rgb,255:red,0;green,205;blue,110},draw,line width=0.5pt,circle,minimum size=2.2em,right of=15] (16) {16};
        \node[fill={rgb,255:red,0;green,205;blue,110},draw,line width=0.5pt,circle,minimum size=2.2em,right of=16] (17) {17};
        \node[fill={rgb,255:red,0;green,205;blue,110},draw,line width=0.5pt,circle,minimum size=2.2em,right of=17] (18) {18};
        \node[fill={rgb,255:red,0;green,205;blue,110},draw,line width=0.5pt,circle,minimum size=2.2em,right of=18] (19) {19};
    
        \draw
            (0) edge[line width=2pt] node{} (1)
            (1) edge[line width=2pt] node{} (2)
            (2) edge[line width=2pt] node{} (3)
            (3) edge[line width=2pt] node{} (4)
            (4) edge[line width=2pt] node{} (5)
            (5) edge[line width=2pt] node{} (6)
            (6) edge[line width=2pt] node{} (7)
            (7) edge[line width=2pt] node{} (8)
            (8) edge[line width=2pt] node{} (9)
            (9) edge[line width=2pt] node{} (10)
            (10) edge[line width=2pt] node{} (11)
            (11) edge[line width=2pt] node{} (12)
            (12) edge[line width=2pt] node{} (13)
            (13) edge[line width=2pt] node{} (14)
            (14) edge[line width=2pt] node{} (15)
            (15) edge[line width=2pt] node{} (16)
            (16) edge[line width=2pt] node{} (17)
            (17) edge[line width=2pt] node{} (18)
            (18) edge[line width=2pt] node{} (19);
\end{tikzpicture}%
    \end{minipage}
    \vspace{0.25em}\\
    \begin{minipage}[][][b]{\textwidth}
        \centering
        \vspace{0.5em}
        (d) Linear
        \vspace{0.5em}
    \end{minipage}
    \caption{Example topologies of near-term quantum devices. Orange (a): IBM Johannesburg. Yellow (b): 2D Grid. Purple (c): four groups of five fully connected clusters. Green (d) Linear. Our real experiments run on Johannesburg and our simulations explore all of these topologies. Colors correspond with the bars in Figures~\ref{fig:benchmark-success},~\ref{fig:benchmark-gate-ratio}, and~\ref{fig:benchmark-success-norm}.}%
    \label{fig:device-diagrams}
\end{figure*}

In this paper we focus primarily on currently available superconducting quantum devices. This type of hardware is the primary focus of many industry players like IBM, Rigetti, and Google (\cite{rigetti, ibmq, bristlecone}). We show some representative topologies for superconducting devices in Figure~\ref{fig:device-diagrams}abd.  For completeness, we include a clustered device shown in Figure~\ref{fig:device-diagrams}c representative of a QCCD ion trap device such as~\cite{honeywell}. These systems exhibit all of the properties previously discussed. They have a small universal supported gate set which all programs must be transformed into and only support local two-qubit operations. The connectivity of these devices is given as a \textit{coupling graph} specifying which pairs of qubits can execute CNOTs.

Furthermore, these systems are subject to a wide variety of noise which cause programs to fail. Some noise is due to manufacturing imperfections or calibration error. Some is inherent to quantum program execution resulting from the imperfect physical isolation of the qubits from the environment required to manipulate the quantum state (\cite{sc_errors}). In IBM machines, the experimental devices of this work, single-qubit gate errors are small, occurring on average 1 in 2000 operations. CNOT gate errors are more significant occurring on roughly 1 in 100 gates. Measurement error is also significant, with errors on the same order of magnitude as CNOT gates. Finally, qubit lifetimes (coherence times) are relatively short, allowing on the order of 50 CNOT gate durations before the qubit state is lost (but gates can often run in parallel while imposing additional crosstalk error, \cite{ibm_errors}). Therefore, quantum compilation is essential to reduce both of these sources of error: add as few extra gates as possible and minimize total execution time.

\subsection{The Compilation Problem}%
\label{background:compilation}

In the NISQ era, quantum programs are highly optimized in order to reduce the effect on errors and maximize the probability the correct answer is observed. Similar to many classical programs, compilation uses a pass structure, where a set of transformation and optimizations are applied in a fixed order resulting in the compilation of an input quantum program to an executable for the target hardware (\cite{scaffcc, cancel}). For the most part, these optimizations take place at the circuit-gate level. Some optimizations are hardware independent, for example, reducing total number of gates via commutativity-aware gate cancellation or find-and-replace with circuit identities. Other passes are focused on decomposing gates into the hardware's ISA (\cite{opt1, opt2, opt3}).

One of the most important parts of this compilation process is mapping and routing the optimized program to one executable on the target hardware, typically done post-decomposition.  Quantum mechanics imposes new constraints on these, different from classical compilation or logic synthesis.  By the no cloning theorem, quantum states cannot be copied, only entangled, which prevents fan-out or fan-in.  Instead, the data must be routed sequentially (i.e.\ moved with SWAP gates) to each place it is needed.

Compilation involves three main steps. First, mapping program qubits to hardware qubits in order to minimize the total distance between qubits that will need to be close by in the future (\cite{map1, map2, map3}). Second, routing pairs of CNOT inputs to be adjacent by inserting SWAPs (\cite{routing1, routing2}). Finally, scheduling operations to minimize total execution time (\cite{scheduling1, xtalk}). In general, the compilation problem is computationally hard and, while some attempts at optimal solutions have been pursued by~\cite{tan2020optimal, siraichi2018qubit, wille2014optimal}, the dominant approach is heuristics. In this work we focus on two pieces of this compilation problem: decomposition and routing.

IBM's Qiskit compiler, the standard for compiling programs to execute on an IBM device, has a default sequence of passes. First, all high level optimization and analysis passes are performed and all gates are unrolled and decomposed to the target gate set. Then single passes of mapping, routing, and scheduling are performed (\cite{qiskit}).

\subsection{Evaluation Metrics}%
\label{background:metrics}

When evaluating compiler methods, we use a few metrics to compare our results.  Our primary metric is program success rate, the fraction of circuit executions that result in the correct output.  Others use fidelity, which can stand-in for success rate when evaluating sub-circuits where the output is not measured.  When executing a quantum algorithm, the corresponding quantum circuit is typically executed thousands of times to gather output statistics or identify the error-free result.

Program success rate is highly dependent on the noise characteristics of the quantum computer the program runs on.  The rates of these device errors can fluctuate day-to-day so we also use the simpler metric of two-qubit gate count.  The number of two-qubit operations in the final compiled circuit is inversely correlated with the success rate because they are usually the largest source of noise.

\subsection{Simulation}%
\label{background:simulation}

Simulating general quantum systems is exponentially expensive in the size of the system and therefore it is difficult to realistically model all of the errors during the execution of a quantum program. We use a simplified model for simulation to predict, specifically to obtain a close upper bound on, the success rate of a program with specified gate error rates and qubit coherence times. In our simplified model, we compute the probability of a program succeeding as the probability that no gate errors occur: ${(p_{gate})}^{n_{gates}}$ times the probability no coherence errors occur, $p_{coherence}$, where the latter is computed as $e^{\Delta / T_1 + \Delta / T_2}$, $\Delta$ is the total program duration, and $T_1$ and $T_2$ are the relaxation and dephasing times (collectively decoherence).

Current error rates, while rapidly improving, are still insufficient to obtain high probabilities of success, making it difficult to compare our mid-size benchmarks that are large enough to need many SWAPs. For our simulations we use error rates 20x improved over current IBM Johannesburg error rates to obtain reasonable success rates and we study sensitivity to this choice later.

\FloatBarrier{}

\section{Motivation: Conventional Compilation}

In this section we motivate the need for a split decomposition pass with routing in between. We look closely at the Qiskit compiler which does not effectively account for the structure in programs.  It often produces circuits with an excessive number of swaps, suggesting room for improvement.

The default compilation framework in Qiskit, used to transform input circuits to be executed on their hardware, ensures a fully decomposed circuit before mapping, routing, and scheduling occur. As a simple example, consider three qubits placed fairly distant on IBM's Johannesburg device, but for which we need to execute a Toffoli gate on them as in Figure~\ref{fig:swap_example}a. Qiskit decomposes this Toffoli as Figure~\ref{fig:6-cnot-decomp} with 6 CNOTs. Each CNOT acts on distant qubits so the many SWAPs inserted for all 6 CNOTs gets expensive quickly. When routing, we first SWAP the first interacting pair together (usually by adding SWAPs from control to target or the reverse, but a meet-in-the-middle strategy is also possible) and the qubit mapping is updated. The next CNOT is also distant so we add SWAPs to move them together and there is an even chance that the SWAPs for the second CNOT separate the two qubits that were just brought together.

Ideally, we move the third qubit to the already adjacent pair, but Qiskit cannot recognize this situation and could just as well move the other way. This is clearly sub-optimal and could continue on for the other four CNOTs. Even in the case where it makes the correct decision to move the distant third qubit, there are problems. Because each pair of qubits needs to interact, we may need single additional SWAPs as the qubits compete to be neighbors.  This causes the 6-CNOT Toffoli decomposition to use many more than 6 CNOTs when there is not a triangle in the qubit connectivity graph. The core idea is that the routing strategy fails to take advantage of two things. First, it has effectively forgotten the desired operation is a Toffoli (which requires all three qubits be adjacent) and second, that a more efficient Toffoli decomposition could be chosen that was more suitable for the underlying device architecture. In the example, inefficient compilation adds a total of 16 SWAPS or 48 CNOTs in total.

Some approaches in the past have attempted to solve the first of these problems, for example \cite{look1, look2} use lookahead when choosing routing strategies and while this helps to treat the symptoms of pre-decomposing all operations it does not remedy the underlying problem.

\FloatBarrier{}

\section{Orchestrated Trios}%
\label{sec:modular-trios}

In this section we describe our proposed compilation structure compared to the conventional one as outlined in Figure~\ref{fig:tool-flow} earlier. Specifically, we focus on improving the routing and decomposition stages of compilation. Previously, we identified a key problem in current methods: decomposing the program to one- and two-qubit gates up front hinders the ability of heuristic-based compilers to effectively minimize the communication cost, i.e.\ the number of SWAPs added, and eliminates the possibility of location-aware decompositions.

We propose a new pass structure. Rather than performing a single round of decomposition and routing, we propose a split approach. Any program processing prior to decomposition stays the same.  The decomposition pass is then divided so the majority of decomposition occurs next but any Toffoli gates are left as-is before moving on to mapping and routing.

The mapping and routing passes come next like normal but must be modified slightly to handle three-qubit gates.  The mapper can simply treat the non-decomposed Toffoli as it would the equivalent 6 CNOTs for the purposes of determining which qubits most need to be placed nearby.
We then do the modified routing pass, moving \textit{groups} of qubits together instead of only pairs where all or all-but-one of the group are moved into a single neighborhood via SWAPs.  This greatly improves the effectiveness of the routing heuristics when applied to this modified routing pass.  There are some subtleties when coordinating the routing of multiple qubits to the same place to ensure the paths don't overlap.  For the purposes of our evaluations we do the following but many similar heuristic strategies are possible.

Taking the next operation to apply, we first find the shortest paths (using any shortest path algorithm on a graph) between all the pairs of qubits.  We choose the qubit with the shortest sum of paths to the other two qubits as the destination.  SWAPS following these two paths are then inserted into the circuit.  The two shortest paths are checked for overlap.  If the ending points overlap, the second is only routed to the penultimate hardware location along the swap path and the first becomes the middle qubit adjacent to both others.  This can save one valuable SWAP but doesn't affect the correctness. Once they are adjacent, the Toffoli gate is now on adjacent qubits and routing can continue to the next operation.

Finally, the second decomposition pass is run.  This is different from normal decomposition as there are only Toffoli gates to decompose and they are already mapped to neighboring qubits.  We could use the default 6-CNOT decomposition and still get the above benefit of improved routing but now that we have more information, this can be exploited to further reduce SWAPs due to a mismatch between the decomposition and the hardware connectivity.  If all three pairs of qubits are connected, then the 6-CNOT Toffoli of Figure~\ref{fig:6-cnot-decomp} is best, otherwise use the 8-CNOT Toffoli of Figure~\ref{fig:8-cnot-decomp}, ensuring the middle qubit is used for the middle of the decomposition (Any of the three qubits can be the target by simply moving the two H gates to that qubit).

When routing complex operations like the Toffoli, we recognize the underlying hardware does not usually support triangles in the connectivity graph but linear connectivity is sufficient for a decent decomposition.  Since we are creating operations on three qubits, the qubits must be routed into a valid linear connectivity.  That is, a configuration where each qubit is connected with at least one of the other qubits.

This method can be easily extended to be noise-aware like previous work, \cite{map1, map2}, by using a noise-aware mapper with the simple modification described earlier where the path-finding graph has weighed edges with the $\text{--}\log$ value of the CNOT success rate. The path distance represents the $\text{--}\log$ probability of success of that particular path where lower values indicate a higher success rate and the shortest path can be found just as before and the routing steps are unchanged.  Any routing strategy designed for one and two-qubit gates can be modified to work for one, two, and three-qubit gates and used as the first routing step of Trios.

In programs where there are no three qubit gates as in the typical NISQ benchmark, \cite{bv}, which is specified directly as CNOT gates, our strategy will have no effect. Many benchmarks, however, are written using Toffoli gates because they are the quantum analog of the AND gate, ubiquitous in arithmetics and other common subroutines.

Trios can naturally be extended to any multi-qubit operation of three or more qubits but this introduces the challenges of simultaneously routing many qubits and of designing decompositions that are efficient with whichever grouping the simultaneous router can achieve. It is not obvious how to route more than three qubits into a line or other desired shape.  As many NISQ benchmarks are not typically written with more complex structures and usually phrase them in terms of one-, two-, and three-qubit gates, this extension may only be desirable for larger-scale quantum computing.

\FloatBarrier{}

\section{Evaluation}%
\label{sec:modular-evaluation}

\subsection{Toffoli Only Circuits}

We first evaluate the effect of our new compilation strategy by studying simple circuits containing only a single Toffoli gate. In these experiments, we place the three input qubits at random locations on the target hardware to emulate the potential locations of the qubits at some intermediate point in the execution of a more complex circuit.

We study these circuits on a real IBM device, namely IBM Johannesburg, a 20-qubit device with limited connectivity, shown in Figure~\ref{fig:device-diagrams}a. We use the default Qiskit compiler which decomposes the Toffoli gates before doing shortest path routing compared to our proposed method where we do shortest path routing first and then decompose the Toffoli. We study the use of two different Toffoli implementations: a 6 CNOT decomposition with full qubit connectivity and an 8 CNOT decomposition with linear qubit connectivity.

In all four configurations, we compare the total compiled CNOT counts which correlates with the total success probability of a program. For execution on Johannesburg, we prepare the qubits in the states $\ket{110}$, perform the compiled Toffoli, then measure the three qubits of interest and compute the success rate as the probability of obtaining the correct answer (here the $\ket{111}$ state), where each experiment is performed with 8192 trials.

\subsection{NISQ Benchmarks and Quantum Subroutines}

We also study Trios on real quantum benchmarks of moderate size using simulation only. The error rates of current devices are still too high to run benchmarks of these sizes but are expected to run on current devices as errors improve in the near future.  We choose error rates 20x better than Johannesburg rates as this make the estimated success probabilities within a reasonable range and is a realistic near-term estimate.  We discuss sensitivity to this choice later.

We study four implementations of the many-controlled-NOT (CnX) gate. This subroutine has many use cases from Grover's algorithm to various arithmetics. The implementations take advantage of differing numbers of ancilla and are chosen based on the number of available qubits on hardware. We study three adder implementations: Cuccaro, Takahashi, and QFT\@. The first two have many uses of the Toffoli gate while the latter has no such gates, for comparison. We study a small version of Grover's algorithm as well which makes use of the \verb|cnx_logancilla| subroutine. Finally, we compile two common NISQ benchmarks: QAOA for Max-Cut and Bernstein Vazirani (BV). We expect no gain on these benchmarks since they do not contain any Toffoli gates. A summary of our benchmarks is found in Table~\ref{tab:benchmark_details} using implementations found in~\cite{ourbenchmarks}.

\begin{table}
    \centering
    \begin{tabular}{cccc}
    Benchmark & Qubits & Toffolis & CNOTs* \\
    \midrule
    \verb|cnx_dirty|,~\cite{toff_any}               & 11 &  16 & 128 \\
    \verb|cnx_halfborrowed|,~\cite{GidneyBlogPost}        & 19 &  32 & 256 \\
    \verb|cnx_logancilla|,~\cite{barenco}   & 19 &  17 & 136 \\
    \verb|cnx_inplace|,~\cite{GidneyBlogPost}             &  4 &  54 & 490 \\
    \verb|cuccaro_adder|,~\cite{cuccaro}           & 20 &  18 & 190 \\
    \verb|takahashi_adder|,~\cite{takahashi}         & 20 &  18 & 188 \\
    \verb|incrementer_borrowedbit|,~\cite{GidneyBlogPost} &  5 &  50 & 448 \\
    \verb|grovers|,~\cite{grover}                  &  9 &  84 & 672 \\
    \verb|qft_adder|,~\cite{qft}               & 16 &   0 &  92 \\
    \verb|bv|,~\cite{bv}                      & 20 &   0 &  19 \\
    \verb|qaoa_complete|,~\cite{qaoa}           & 10 &   0 &  90 \\
    \end{tabular}
    \caption{Details about our benchmarks both NISQ programs and other quantum subroutines. We consider circuits with and without Toffoli gates where we expect advantage only for circuits containing Toffoli gates. For BV we assume the all 1-bit string. The different CnX (many-controlled-NOT) gates use various numbers of ancilla. *The total number of CNOT gates is after decomposition with the 8-CNOT Toffoli but does not including any SWAPs for routing.}%
    \label{tab:benchmark_details}
\end{table}

As noted previously, the connectivity of the underlying hardware has a significant impact on the number of required SWAPs. For example, on a completely connected set of qubits, no SWAPs are ever needed. In architectures with greater connectivity, we may opt for a more efficient Toffoli decomposition using 6 CNOTs. With simulation we study the effect of connectivity on the overall expected success rates and gate counts. We study four different connectivity models, all shown in Figure~\ref{fig:device-diagrams}, each with 20 qubits, the topology of IBM's Johannesburg device containing four connected rings, a 2D mesh, a line, and a small clustered architecture representative of a QCCD ion trap.

We use error rates reported by IBM obtained via randomized benchmarking on a daily basis; for simulations we use error numbers obtained from Johannesburg obtained on 8/19/2020 with an average T1 time of $70.87\mu s$, T2 time of $72.72\mu s$, two qubit gate time of $0.559 \mu s$, a one qubit gate time of $0.07\mu s$, two qubit gate error of 0.0147, one qubit gate error of 0.0004. Source code for all experiments is available on GitHub (\cite{Git-trios}).  Experiments using IBM are tested with version 0.14.0 through their Python API\@. When compiling with Qiskit for the single Toffoli experiments, we use the default settings for the \verb|transpile| function while specifying the Johannesburg backend. This means light optimization is performed: a stochastic routing policy is chosen, and some simple optimizations such as single qubit gate consolidation is performed. We fix the initial mapping to force routing to occur.

\FloatBarrier{}

\section{Results and Discussion}%
\label{sec:modular-results}

\subsection{Trios Reduces Total Number of Gates}

In both sets of experiments, the total number of gates required to make the input programs executable is much less than when using the default Qiskit compiler. When compiling our simple programs consisting of a single Toffoli gate with qubits mapped in random locations, we reduce the average number of gates by 35\% geomean.

In Figure~\ref{fig:ibm-toffoli-gates} we show 35 different triplets of hardware qubits for each of the four strategies. For each triplet, we note the total distance between the qubits on the hardware, given by the shortest path distance in the underlying topology. Even when the distance is relatively small, Trios outperforms Qiskit, reducing overall gate count. As the distance increases, this performance margin tends to increase. In the small distance cases, this can be attributed to Trios choosing the better Toffoli decomposition for a linearly connected topology. This is significant for two reasons. First, the fewer the gates, the less likely an error occurs due to qubit manipulation. Second, fewer gates, especially long sequential chains of SWAPs, often means lower circuit depth, meaning fewer chances for decoherence errors. Together this translates into faster and more successful programs.

This advantage extends to our NISQ benchmarks which contain various numbers of Toffoli gates. In Figure~\ref{fig:benchmark-gate-ratio} we note substantial reductions in total gates across all benchmarks containing Toffoli gates across all underlying topologies. The only exception is the two smallest benchmarks (on 4 and 5 qubits) for the clustered topology because they could be compiled with zero SWAPs.

An extreme of the clustered topology is a single cluster with all-to-all connected qubits.  On this device, Orchestrated Trios would have no benefit as operations can be performed between any pair of qubits so no SWAPs are needed and routing is trivial.  However, as quantum technologies scale to more than a few qubits, fully-connected architectures hits physical limitations and must be re-engineered.  As trapped ion qubit chains get longer, for example, gate operations become slower and lower fidelity.  \cite{trapsize} showed that the optimal trap size is 15--25 ions interconnected similar to our cluster model with cluster sizes of 15--25 where Trios does benefit.

On average, for Toffoli-containing programs we reduce gate count 37\%, 36\%, 48\%, 26\% for Johannesburg, Grid, Line, and Cluster topologies respectively with the maximum gain obtained for linear devices.

\begin{figure}
    \centering
    \makebox[0.95\textwidth][r]{%
        \begin{tikzpicture}[baseline,scale=1,trim axis left,trim axis right]
\pgfplotsset{every tick label/.append style={font=\small}}
\pgfplotsset{every axis label/.append style={font=\small}}

    \begin{axis}[
        name=plot0,
        title={Toffoli Experiment on IBMQ Johannesburg},
        xlabel={},
        ylabel={success probability},
        symbolic x coords={(6-17-3) 10,(16-1-8) 10,(7-18-3) 9,(17-4-11) 9,(19-2-6) 9,(1-19-8) 8,(3-15-14) 8,(7-3-19) 8,(15-0-9) 8,(19-1-7) 8,(1-2-18) 7,(6-13-2) 7,(14-5-15) 7,(16-1-18) 7,(19-10-6) 7,(0-12-15) 6,(5-3-9) 6,(9-3-5) 6,(13-10-1) 6,(19-15-13) 6,(0-6-11) 5,(8-6-19) 5,(11-15-8) 5,(14-13-16) 5,(18-7-8) 5,(2-5-3) 4,(5-1-3) 4,(8-10-6) 4,(11-7-9) 4,(17-10-5) 4,(1-3-4) 3,(9-12-14) 3,(10-11-0) 3,(3-1-2) 2,(17-16-18) 2,gap,geo-mean},
        width={0.95*\linewidth},
        height={0.35*\columnwidth},
        ybar={0.5pt},
        bar width={2pt},
        enlargelimits=0.013888888888888888,
        ymin=0, ymax=0.9161254882812501,
        xtick=data,
        ,
        legend style={draw=none, fill=none, at={(0.5,1.03)},anchor=north,font=\small},
        legend columns=-1,
        legend image code/.code={\draw[#1, draw=none] (0em,-0.2em) rectangle (0.6em,0.4em);},
        axis line style={draw=black!20!white},
        axis on top,
        y axis line style={draw=none},
        axis x line*=bottom,
        tick style={draw=none},
        ,
        clip=false,
        enlarge y limits=0,
        ,
        x tick label style={rotate=60, anchor=east},
        grid=none,
        ymajorgrids=true,
        ,
        ,
        nodes near coords always on top/.style={
            every node near coord/.append style={
                anchor=south,
                rotate=0,
                font=\small,
                inner sep=0.2em,
            },
        },
        nodes near coords always on top,
    ]

        \addplot[
            style={
                color=transparent,
                draw=none,
                fill=black,
                ,
                mark=none,
                ,
                pattern color=black,,
            }]
        coordinates {
            ((6-17-3) 10, 0.2813720703125)
            ((16-1-8) 10, 0.1943359375)
            ((7-18-3) 9, 0.4942626953125)
            ((17-4-11) 9, 0.29345703125)
            ((19-2-6) 9, 0.2926025390625)
            ((1-19-8) 8, 0.334716796875)
            ((3-15-14) 8, 0.3291015625)
            ((7-3-19) 8, 0.5450439453125)
            ((15-0-9) 8, 0.4017333984375)
            ((19-1-7) 8, 0.468017578125)
            ((1-2-18) 7, 0.3411865234375)
            ((6-13-2) 7, 0.4473876953125)
            ((14-5-15) 7, 0.434326171875)
            ((16-1-18) 7, 0.297607421875)
            ((19-10-6) 7, 0.453369140625)
            ((0-12-15) 6, 0.4942626953125)
            ((5-3-9) 6, 0.5179443359375)
            ((9-3-5) 6, 0.435546875)
            ((13-10-1) 6, 0.578125)
            ((19-15-13) 6, 0.3743896484375)
            ((0-6-11) 5, 0.5164794921875)
            ((8-6-19) 5, 0.523681640625)
            ((11-15-8) 5, 0.4920654296875)
            ((14-13-16) 5, 0.4835205078125)
            ((18-7-8) 5, 0.3135986328125)
            ((2-5-3) 4, 0.486083984375)
            ((5-1-3) 4, 0.530029296875)
            ((8-10-6) 4, 0.5513916015625)
            ((11-7-9) 4, 0.4449462890625)
            ((17-10-5) 4, 0.4534912109375)
            ((1-3-4) 3, 0.535400390625)
            ((9-12-14) 3, 0.3896484375)
            ((10-11-0) 3, 0.4132080078125)
            ((3-1-2) 2, 0.5052490234375)
            ((17-16-18) 2, 0.375732421875)
            (gap, nan)
            (geo-mean, 0.4088172304790279)
        };
        \addlegendentry{Qiskit (baseline)~~~~};

        \addplot[
            style={
                color=transparent,
                draw=none,
                fill={rgb,190:red,72;green,132;blue,189},
                ,
                mark=none,
                ,
                pattern color={rgb,190:red,72;green,132;blue,189},,
            }]
        coordinates {
            ((6-17-3) 10, 0.1986083984375)
            ((16-1-8) 10, 0.5120849609375)
            ((7-18-3) 9, 0.26171875)
            ((17-4-11) 9, 0.26318359375)
            ((19-2-6) 9, 0.3856201171875)
            ((1-19-8) 8, 0.431884765625)
            ((3-15-14) 8, 0.249267578125)
            ((7-3-19) 8, 0.414306640625)
            ((15-0-9) 8, 0.193603515625)
            ((19-1-7) 8, 0.37939453125)
            ((1-2-18) 7, 0.2774658203125)
            ((6-13-2) 7, 0.18798828125)
            ((14-5-15) 7, 0.3502197265625)
            ((16-1-18) 7, 0.276611328125)
            ((19-10-6) 7, 0.4854736328125)
            ((0-12-15) 6, 0.5345458984375)
            ((5-3-9) 6, 0.4913330078125)
            ((9-3-5) 6, 0.4803466796875)
            ((13-10-1) 6, 0.4005126953125)
            ((19-15-13) 6, 0.322509765625)
            ((0-6-11) 5, 0.4285888671875)
            ((8-6-19) 5, 0.385009765625)
            ((11-15-8) 5, 0.5081787109375)
            ((14-13-16) 5, 0.2054443359375)
            ((18-7-8) 5, 0.511962890625)
            ((2-5-3) 4, 0.396240234375)
            ((5-1-3) 4, 0.6043701171875)
            ((8-10-6) 4, 0.3795166015625)
            ((11-7-9) 4, 0.3756103515625)
            ((17-10-5) 4, 0.29736328125)
            ((1-3-4) 3, 0.5404052734375)
            ((9-12-14) 3, 0.2032470703125)
            ((10-11-0) 3, 0.5494384765625)
            ((3-1-2) 2, 0.5340576171875)
            ((17-16-18) 2, 0.47314453125)
            (gap, nan)
            (geo-mean, 0.35173713914801175)
        };
        \addlegendentry{Qiskit (8-CNOT Toffoli)~~~~};

        \addplot[
            style={
                color=transparent,
                draw=none,
                fill={rgb,190:red,112;green,169;blue,45},
                ,
                mark=none,
                ,
                pattern color={rgb,190:red,112;green,169;blue,45},,
            }]
        coordinates {
            ((6-17-3) 10, 0.479248046875)
            ((16-1-8) 10, 0.537353515625)
            ((7-18-3) 9, 0.548583984375)
            ((17-4-11) 9, 0.315185546875)
            ((19-2-6) 9, 0.4183349609375)
            ((1-19-8) 8, 0.5638427734375)
            ((3-15-14) 8, 0.3505859375)
            ((7-3-19) 8, 0.4820556640625)
            ((15-0-9) 8, 0.5438232421875)
            ((19-1-7) 8, 0.5775146484375)
            ((1-2-18) 7, 0.3597412109375)
            ((6-13-2) 7, 0.5350341796875)
            ((14-5-15) 7, 0.440673828125)
            ((16-1-18) 7, 0.54150390625)
            ((19-10-6) 7, 0.3846435546875)
            ((0-12-15) 6, 0.3994140625)
            ((5-3-9) 6, 0.467529296875)
            ((9-3-5) 6, 0.5311279296875)
            ((13-10-1) 6, 0.4718017578125)
            ((19-15-13) 6, 0.4345703125)
            ((0-6-11) 5, 0.5518798828125)
            ((8-6-19) 5, 0.5556640625)
            ((11-15-8) 5, 0.5135498046875)
            ((14-13-16) 5, 0.409912109375)
            ((18-7-8) 5, 0.611572265625)
            ((2-5-3) 4, 0.4586181640625)
            ((5-1-3) 4, 0.4739990234375)
            ((8-10-6) 4, 0.5352783203125)
            ((11-7-9) 4, 0.64794921875)
            ((17-10-5) 4, 0.4459228515625)
            ((1-3-4) 3, 0.524658203125)
            ((9-12-14) 3, 0.392333984375)
            ((10-11-0) 3, 0.5059814453125)
            ((3-1-2) 2, 0.511474609375)
            ((17-16-18) 2, 0.4376220703125)
            (gap, nan)
            (geo-mean, 0.47450881060592054)
        };
        \addlegendentry{Trios (6-CNOT Toffoli)~~~~};

        \addplot[
            style={
                color=transparent,
                draw=none,
                fill={rgb,190:red,180;green,56;blue,101},
                ,
                mark=none,
                ,
                pattern color={rgb,190:red,180;green,56;blue,101},,
            }]
        coordinates {
            ((6-17-3) 10, 0.5272216796875)
            ((16-1-8) 10, 0.5565185546875)
            ((7-18-3) 9, 0.5689697265625)
            ((17-4-11) 9, 0.3179931640625)
            ((19-2-6) 9, 0.4375)
            ((1-19-8) 8, 0.5755615234375)
            ((3-15-14) 8, 0.377197265625)
            ((7-3-19) 8, 0.5753173828125)
            ((15-0-9) 8, 0.569091796875)
            ((19-1-7) 8, 0.552978515625)
            ((1-2-18) 7, 0.514404296875)
            ((6-13-2) 7, 0.5552978515625)
            ((14-5-15) 7, 0.4522705078125)
            ((16-1-18) 7, 0.58837890625)
            ((19-10-6) 7, 0.4390869140625)
            ((0-12-15) 6, 0.5189208984375)
            ((5-3-9) 6, 0.5526123046875)
            ((9-3-5) 6, 0.5179443359375)
            ((13-10-1) 6, 0.439453125)
            ((19-15-13) 6, 0.3365478515625)
            ((0-6-11) 5, 0.491943359375)
            ((8-6-19) 5, 0.490966796875)
            ((11-15-8) 5, 0.492431640625)
            ((14-13-16) 5, 0.3438720703125)
            ((18-7-8) 5, 0.6097412109375)
            ((2-5-3) 4, 0.5042724609375)
            ((5-1-3) 4, 0.605712890625)
            ((8-10-6) 4, 0.6082763671875)
            ((11-7-9) 4, 0.7047119140625)
            ((17-10-5) 4, 0.5357666015625)
            ((1-3-4) 3, 0.5863037109375)
            ((9-12-14) 3, 0.4461669921875)
            ((10-11-0) 3, 0.539794921875)
            ((3-1-2) 2, 0.5263671875)
            ((17-16-18) 2, 0.4676513671875)
            (gap, nan)
            (geo-mean, 0.5017832670104678)
        };
        \addlegendentry{Trios (8-CNOT)~~~~~~~~~~};
    (

    \end{axis}

\end{tikzpicture}%
    }%
    \caption{Success probabilities of Toffoli gates between random triplets of qubits.  Higher is better.  The x-labels specify the three qubits and total swap distance.  The geometric mean success rates for each compiler are 41\%, 35\%, 47\%, and 50\% respectively.  Trios (8-CNOT) improves average success rate by 23\% vs.\ the Qiskit baseline.}%
    \label{fig:ibm-toffoli-success}
\end{figure}
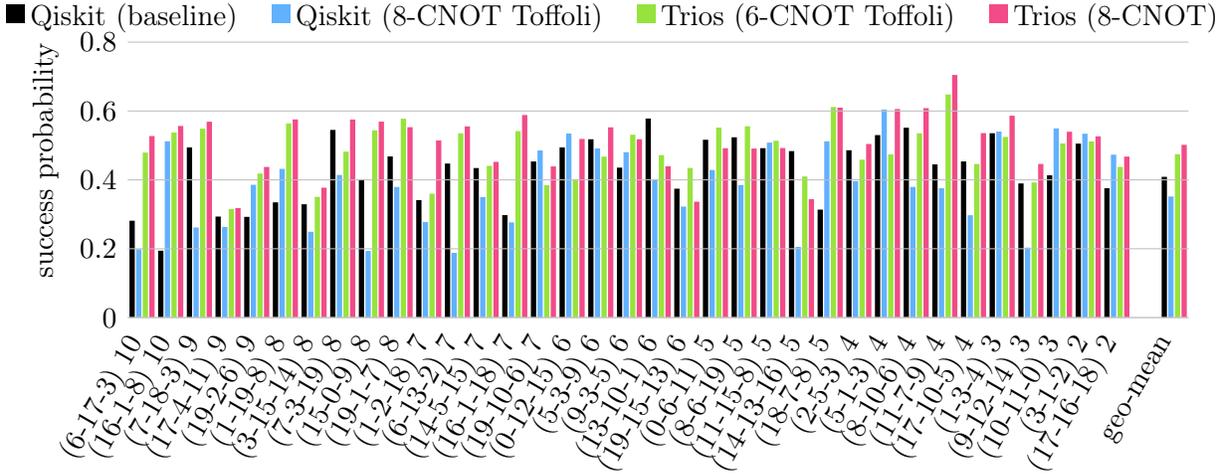
\begin{figure}
    \centering
    \makebox[0.95\textwidth][r]{%
        \begin{tikzpicture}[baseline,scale=1,trim axis left,trim axis right]
\pgfplotsset{every tick label/.append style={font=\small}}
\pgfplotsset{every axis label/.append style={font=\small}}

    \begin{axis}[
        name=plot0,
        title={Toffoli Experiment on IBMQ Johannesburg},
        xlabel={},
        ylabel={CNOT gate count},
        symbolic x coords={(6-17-3) 10,(16-1-8) 10,(7-18-3) 9,(17-4-11) 9,(19-2-6) 9,(1-19-8) 8,(3-15-14) 8,(7-3-19) 8,(15-0-9) 8,(19-1-7) 8,(1-2-18) 7,(6-13-2) 7,(14-5-15) 7,(16-1-18) 7,(19-10-6) 7,(0-12-15) 6,(5-3-9) 6,(9-3-5) 6,(13-10-1) 6,(19-15-13) 6,(0-6-11) 5,(8-6-19) 5,(11-15-8) 5,(14-13-16) 5,(18-7-8) 5,(2-5-3) 4,(5-1-3) 4,(8-10-6) 4,(11-7-9) 4,(17-10-5) 4,(1-3-4) 3,(9-12-14) 3,(10-11-0) 3,(3-1-2) 2,(17-16-18) 2,gap,geo-mean},
        width={0.95*\linewidth},
        height={0.35*\columnwidth},
        ybar={0.5pt},
        bar width={2pt},
        enlargelimits=0.013888888888888888,
        ymin=0, ymax=70.2,
        xtick=data,
        ,
        legend style={draw=none, fill=none, at={(0.5,1.03)},anchor=north,font=\small},
        legend columns=-1,
        legend image code/.code={\draw[#1, draw=none] (0em,-0.2em) rectangle (0.6em,0.4em);},
        axis line style={draw=black!20!white},
        axis on top,
        y axis line style={draw=none},
        axis x line*=bottom,
        tick style={draw=none},
        ,
        clip=false,
        enlarge y limits=0,
        ,
        x tick label style={rotate=60, anchor=east},
        grid=none,
        ymajorgrids=true,
        ,
        ,
        nodes near coords always on top/.style={
            every node near coord/.append style={
                anchor=south,
                rotate=0,
                font=\small,
                inner sep=0.2em,
            },
        },
        nodes near coords always on top,
    ]

        \addplot[
            style={
                color=transparent,
                draw=none,
                fill=black,
                ,
                mark=none,
                ,
                pattern color=black,,
            }]
        coordinates {
            ((6-17-3) 10, 37)
            ((16-1-8) 10, 54)
            ((7-18-3) 9, 24)
            ((17-4-11) 9, 45)
            ((19-2-6) 9, 54)
            ((1-19-8) 8, 30)
            ((3-15-14) 8, 40)
            ((7-3-19) 8, 21)
            ((15-0-9) 8, 39)
            ((19-1-7) 8, 39)
            ((1-2-18) 7, 43)
            ((6-13-2) 7, 30)
            ((14-5-15) 7, 27)
            ((16-1-18) 7, 48)
            ((19-10-6) 7, 33)
            ((0-12-15) 6, 21)
            ((5-3-9) 6, 22)
            ((9-3-5) 6, 33)
            ((13-10-1) 6, 30)
            ((19-15-13) 6, 37)
            ((0-6-11) 5, 18)
            ((8-6-19) 5, 28)
            ((11-15-8) 5, 22)
            ((14-13-16) 5, 33)
            ((18-7-8) 5, 27)
            ((2-5-3) 4, 18)
            ((5-1-3) 4, 16)
            ((8-10-6) 4, 21)
            ((11-7-9) 4, 27)
            ((17-10-5) 4, 21)
            ((1-3-4) 3, 13)
            ((9-12-14) 3, 10)
            ((10-11-0) 3, 21)
            ((3-1-2) 2, 9)
            ((17-16-18) 2, 18)
            (gap, nan)
            (geo-mean, 28.965511018659207)
        };
        \addlegendentry{Qiskit (baseline)~~~~};

        \addplot[
            style={
                color=transparent,
                draw=none,
                fill={rgb,190:red,72;green,132;blue,189},
                ,
                mark=none,
                ,
                pattern color={rgb,190:red,72;green,132;blue,189},,
            }]
        coordinates {
            ((6-17-3) 10, 53)
            ((16-1-8) 10, 29)
            ((7-18-3) 9, 32)
            ((17-4-11) 9, 32)
            ((19-2-6) 9, 38)
            ((1-19-8) 8, 41)
            ((3-15-14) 8, 41)
            ((7-3-19) 8, 23)
            ((15-0-9) 8, 35)
            ((19-1-7) 8, 35)
            ((1-2-18) 7, 47)
            ((6-13-2) 7, 41)
            ((14-5-15) 7, 38)
            ((16-1-18) 7, 32)
            ((19-10-6) 7, 29)
            ((0-12-15) 6, 17)
            ((5-3-9) 6, 29)
            ((9-3-5) 6, 26)
            ((13-10-1) 6, 35)
            ((19-15-13) 6, 44)
            ((0-6-11) 5, 14)
            ((8-6-19) 5, 35)
            ((11-15-8) 5, 26)
            ((14-13-16) 5, 29)
            ((18-7-8) 5, 26)
            ((2-5-3) 4, 20)
            ((5-1-3) 4, 14)
            ((8-10-6) 4, 23)
            ((11-7-9) 4, 23)
            ((17-10-5) 4, 32)
            ((1-3-4) 3, 14)
            ((9-12-14) 3, 20)
            ((10-11-0) 3, 11)
            ((3-1-2) 2, 8)
            ((17-16-18) 2, 8)
            (gap, nan)
            (geo-mean, 27.571546147540772)
        };
        \addlegendentry{Qiskit (8-CNOT Toffoli)~~~~};

        \addplot[
            style={
                color=transparent,
                draw=none,
                fill={rgb,190:red,112;green,169;blue,45},
                ,
                mark=none,
                ,
                pattern color={rgb,190:red,112;green,169;blue,45},,
            }]
        coordinates {
            ((6-17-3) 10, 34)
            ((16-1-8) 10, 33)
            ((7-18-3) 9, 30)
            ((17-4-11) 9, 36)
            ((19-2-6) 9, 34)
            ((1-19-8) 8, 25)
            ((3-15-14) 8, 25)
            ((7-3-19) 8, 30)
            ((15-0-9) 8, 30)
            ((19-1-7) 8, 27)
            ((1-2-18) 7, 33)
            ((6-13-2) 7, 33)
            ((14-5-15) 7, 27)
            ((16-1-18) 7, 33)
            ((19-10-6) 7, 30)
            ((0-12-15) 6, 27)
            ((5-3-9) 6, 27)
            ((9-3-5) 6, 24)
            ((13-10-1) 6, 27)
            ((19-15-13) 6, 24)
            ((0-6-11) 5, 16)
            ((8-6-19) 5, 25)
            ((11-15-8) 5, 21)
            ((14-13-16) 5, 21)
            ((18-7-8) 5, 16)
            ((2-5-3) 4, 19)
            ((5-1-3) 4, 24)
            ((8-10-6) 4, 15)
            ((11-7-9) 4, 18)
            ((17-10-5) 4, 21)
            ((1-3-4) 3, 18)
            ((9-12-14) 3, 12)
            ((10-11-0) 3, 15)
            ((3-1-2) 2, 9)
            ((17-16-18) 2, 18)
            (gap, nan)
            (geo-mean, 23.39548267542428)
        };
        \addlegendentry{Trios (6-CNOT Toffoli)~~~~};

        \addplot[
            style={
                color=transparent,
                draw=none,
                fill={rgb,190:red,180;green,56;blue,101},
                ,
                mark=none,
                ,
                pattern color={rgb,190:red,180;green,56;blue,101},,
            }]
        coordinates {
            ((6-17-3) 10, 29)
            ((16-1-8) 10, 29)
            ((7-18-3) 9, 26)
            ((17-4-11) 9, 29)
            ((19-2-6) 9, 29)
            ((1-19-8) 8, 26)
            ((3-15-14) 8, 26)
            ((7-3-19) 8, 23)
            ((15-0-9) 8, 26)
            ((19-1-7) 8, 26)
            ((1-2-18) 7, 23)
            ((6-13-2) 7, 23)
            ((14-5-15) 7, 23)
            ((16-1-18) 7, 23)
            ((19-10-6) 7, 23)
            ((0-12-15) 6, 17)
            ((5-3-9) 6, 20)
            ((9-3-5) 6, 20)
            ((13-10-1) 6, 20)
            ((19-15-13) 6, 20)
            ((0-6-11) 5, 14)
            ((8-6-19) 5, 17)
            ((11-15-8) 5, 17)
            ((14-13-16) 5, 17)
            ((18-7-8) 5, 17)
            ((2-5-3) 4, 14)
            ((5-1-3) 4, 14)
            ((8-10-6) 4, 14)
            ((11-7-9) 4, 14)
            ((17-10-5) 4, 14)
            ((1-3-4) 3, 11)
            ((9-12-14) 3, 11)
            ((10-11-0) 3, 11)
            ((3-1-2) 2, 8)
            ((17-16-18) 2, 8)
            (gap, nan)
            (geo-mean, 18.824704952412723)
        };
        \addlegendentry{Trios (8-CNOT)~~~~~~~~~~};
    (

    \end{axis}

\end{tikzpicture}%
    }%
    \caption{Total number of two-qubit (CNOT) gates required to execute a Toffoli gate between various distant qubits. Lower is better.  The x-labels specify the three qubits and total swap distance.  The geometric mean gate counts for each compiler are 29, 28, 23, and 19 respectively.  Trios (8-CNOT) reduces average gate count by 35\%.}%
    \label{fig:ibm-toffoli-gates}
\end{figure}
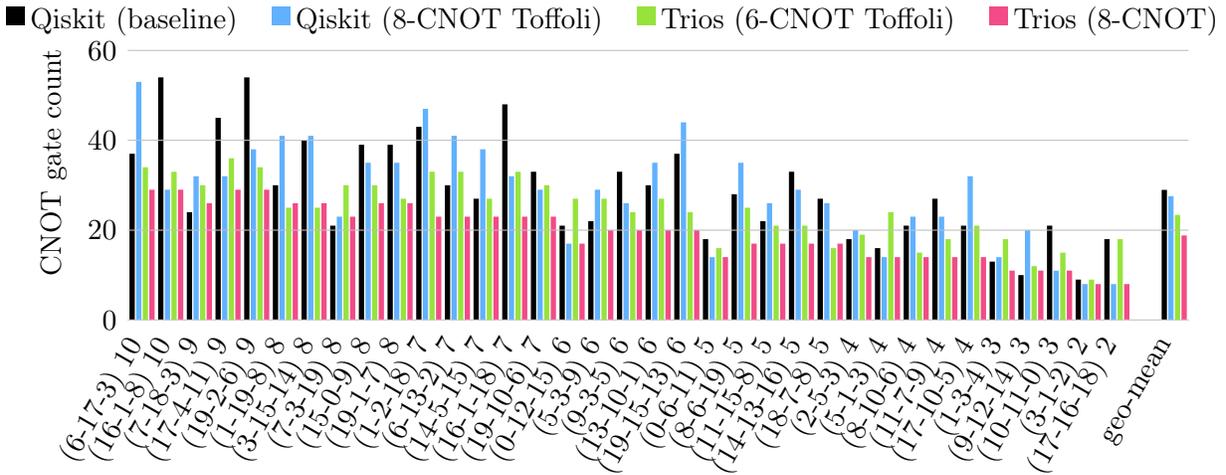
\begin{figure}
    \centering
    \makebox[0.95\textwidth][r]{%
        \input{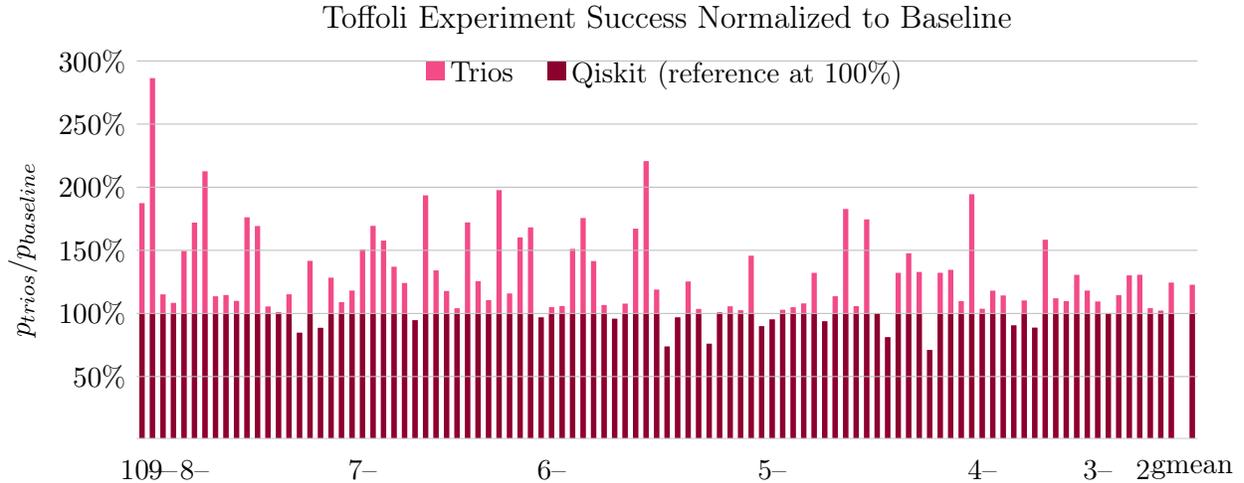}%
    }%
    \caption{Normalized success probabilities of Toffoli gates between triplets of qubits. Higher is better. Bars below 100\% indicate lower success rate for Trios.  The geometric mean increase in success rate is 23\%. The x-labels indicate the qubit distance for a range of bars.}%
    \label{fig:ibm-toffoli-success-norm}
\end{figure}

\subsection{Trios Improves Overall Success Rate}

In general, we expect programs with fewer total two-qubit gates to succeed with higher probability. In devices with limited connectivity, the addition of routing operations like SWAPs, usually decomposed to 3 CNOTs, can severely reduce the chance an input program can succeed. While success rate is inversely correlated with number of gates, gate error is not the only reason a program can fail and reducing gate counts does not \textit{guarantee} improved success rates.

In Figure~\ref{fig:ibm-toffoli-success} we show the success rates of our Toffoli-only experiments when the two controls are initialized to $\ket{1}$ and the target is initialized to $\ket{0}$ so we measure the probability of obtaining $\ket{111}$. These results are obtained from Johannesburg on 8/19/2020. The x-axes of both Figures~\ref{fig:ibm-toffoli-success} and~\ref{fig:ibm-toffoli-gates} line up to compare gate counts and resulting success rate. In general, experimentally, fewer gates results in substantial improvements to success rates. For example, a Toffoli on qubits 6, 17, and 3 compiled with Trios improves success rate from around 30\% to over 50\%. On average, we improve success rates by 23 \% geomean with max of 286\%. In Figure~\ref{fig:ibm-toffoli-success-norm}, we show improvements compiled with Trios normalized to baseline for 99 different triplets of varying total distance on Johannesburg.

Trios on average improves the probability of success for these circuits. However, there are a small number of cases where Trios performs worse despite having a smaller number of total gates. This can be attributed to several different factors. For example, the chosen edges for SWAP paths may be more noisy, or on pairs of edges with greater crosstalk, or the final qubits which are measured have worse readout error. Regardless, reducing the overall gate count of a program is an important contributing factor to improving expected success rate.

\begin{figure}
    \centering
    \makebox[0.97\textwidth][r]{%
        \input{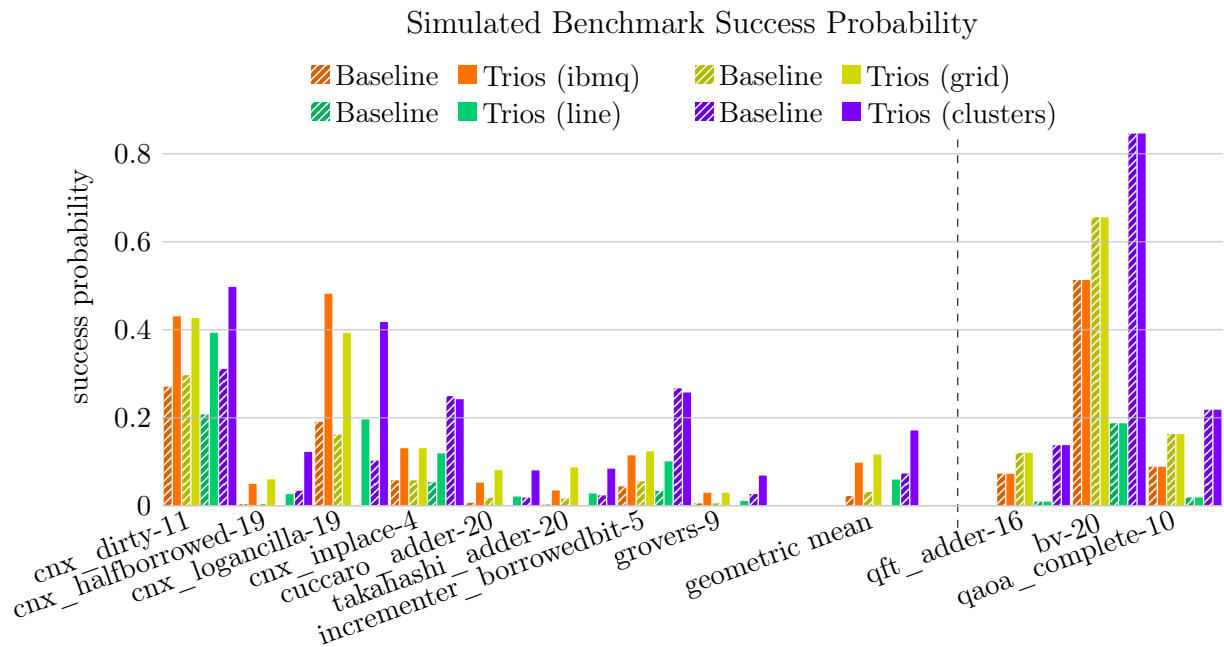}%
    }%
    \caption{Simulated upper-bounds on the program execution success probability on various hardware (using 20x lower idle and gate errors than Johannesburg).  Neighboring pairs of bars compare the baseline with Trios compiled for Johannesburg.  Higher is better when comparing pairs of bars with the same color.  The geometric mean success rates over the benchmarks that use Toffoli gate for each device type respectively are 2.2\%$\rightarrow$9.8\%, 3.2\%$\rightarrow$12\%, 0.19\%$\rightarrow$6.0\%, 7.3\%$\rightarrow$17\%.  The rightmost three benchmarks contain zero Toffoli gates so have no change vs.\ the baseline.}%
    \label{fig:benchmark-success}
\end{figure}

For our simulated NISQ benchmarks, we see even larger gains. The reduced gate counts in Figure~\ref{fig:benchmark-gate-ratio} translate to major improvements in simulated success rate in Figure~\ref{fig:benchmark-success} (normalized success rates in Figure~\ref{fig:benchmark-success-norm}). For example, in \verb|cnx_logancilla-19|, Trios more than doubles the expected success rates when compiled to each of the architectures. In many cases, the expected success rate of programs compiled with Qiskit is effectively zero while Trios has a realistic chance of obtaining the correct answer. As expected, on programs containing no Toffoli gates, Trios has no effect on success showing that it introduces no measurable overhead. This suggests Trios can easily be added to other quantum compilation tool flows.

\begin{figure}
    \centering
    \makebox[0.97\textwidth][r]{%
        \begin{tikzpicture}[baseline,scale=1,trim axis left,trim axis right]
\pgfplotsset{every tick label/.append style={font=\small}}
\pgfplotsset{every axis label/.append style={font=\small}}

    \begin{axis}[
        name=plot0,
        title={Simulated Benchmark Gate-Count Reduction over Baseline},
        xlabel={},
        ylabel={percent fewer CNOT gates},
        symbolic x coords={cnx\_dirty-11,cnx\_halfborrowed-19,cnx\_logancilla-19,cnx\_inplace-4,cuccaro\_adder-20,takahashi\_adder-20,incrementer\_borrowedbit-5,grovers-9, ,geometric mean,  ,qft\_adder-16,bv-20,qaoa\_complete-10},
        width={0.95*\linewidth},
        height={0.33*\columnwidth},
        ybar={2pt},
        bar width={4pt},
        enlargelimits=0.038461538461538464,
        ymin=0, ymax=79,
        xtick=data,
        ,
        legend style={draw=none, fill=none, at={(0.5,1.03)},anchor=north,font=\small},
        legend columns=-1,
        legend image code/.code={\draw[#1, draw=none] (0em,-0.2em) rectangle (0.6em,0.4em);},
        axis line style={draw=black!20!white},
        axis on top,
        y axis line style={draw=none},
        axis x line*=bottom,
        tick style={draw=none},
        yticklabel={\pgfmathparse{\tick*1}\pgfmathprintnumber{\pgfmathresult}\%},
        clip=false,
        enlarge y limits=0,
        ,
        x tick label style={rotate=20, anchor=east},
        grid=none,
        ymajorgrids=true,
        ,
        ,
        nodes near coords always on top/.style={
            every node near coord/.append style={
                anchor=south,
                rotate=0,
                font=\small,
                inner sep=0.2em,
            },
        },
        nodes near coords always on top,
    ]
        \addplot[
            style={
                color=transparent,
                draw=none,
                fill={rgb,255:red,255;green,113;blue,0},
                ,
                mark=none,
                ,
                pattern color={rgb,255:red,255;green,113;blue,0},,
            }]
        coordinates {
            (cnx\_dirty-11, 33.83333333333333)
            (cnx\_halfborrowed-19, 45.196969696969695)
            (cnx\_logancilla-19, 43.303030303030305)
            (cnx\_inplace-4, 29.147214854111407)
            (cuccaro\_adder-20, 37.234693877551024)
            (takahashi\_adder-20, 42.285714285714285)
            (incrementer\_borrowedbit-5, 30.525445292620866)
            (grovers-9, 33.83333333333333)
            ( , nan)
            (geometric mean, 37.17711233811081)
            (  , nan)
            (qft\_adder-16, 0.5)
            (bv-20, 0.5)
            (qaoa\_complete-10, 0.5)
        };
        \addlegendentry{ibmq-johannesburg~~~~};

        \addplot[
            style={
                color=transparent,
                draw=none,
                fill={rgb,255:red,209;green,216;blue,0},
                ,
                mark=none,
                ,
                pattern color={rgb,255:red,209;green,216;blue,0},,
            }]
        coordinates {
            (cnx\_dirty-11, 33.83333333333333)
            (cnx\_halfborrowed-19, 43.75396825396825)
            (cnx\_logancilla-19, 45.166666666666664)
            (cnx\_inplace-4, 29.147214854111407)
            (cuccaro\_adder-20, 32.97422680412371)
            (takahashi\_adder-20, 39.00267379679144)
            (incrementer\_borrowedbit-5, 27.600271002710024)
            (grovers-9, 33.83333333333333)
            ( , nan)
            (geometric mean, 35.949396509888224)
            (  , nan)
            (qft\_adder-16, 0.5)
            (bv-20, 0.5)
            (qaoa\_complete-10, 0.5)
        };
        \addlegendentry{full-grid-5x4~~~~};

        \addplot[
            style={
                color=transparent,
                draw=none,
                fill={rgb,255:red,0;green,205;blue,110},
                ,
                mark=none,
                ,
                pattern color={rgb,255:red,0;green,205;blue,110},,
            }]
        coordinates {
            (cnx\_dirty-11, 42.16666666666667)
            (cnx\_halfborrowed-19, 62.67948717948718)
            (cnx\_logancilla-19, 66.41880341880342)
            (cnx\_inplace-4, 28.05102040816326)
            (cuccaro\_adder-20, 44.40243902439025)
            (takahashi\_adder-20, 46.37378640776699)
            (incrementer\_borrowedbit-5, 32.2016317016317)
            (grovers-9, 48.984848484848484)
            ( , nan)
            (geometric mean, 47.945364137106395)
            (  , nan)
            (qft\_adder-16, 0.5)
            (bv-20, 0.5)
            (qaoa\_complete-10, 0.5)
        };
        \addlegendentry{line-20~~~~};

        \addplot[
            style={
                color=transparent,
                draw=none,
                fill={rgb,255:red,125;green,0;blue,255},
                ,
                mark=none,
                ,
                pattern color={rgb,255:red,125;green,0;blue,255},,
            }]
        coordinates {
            (cnx\_dirty-11, 34.94444444444444)
            (cnx\_halfborrowed-19, 34.94444444444444)
            (cnx\_logancilla-19, 46.7962962962963)
            (cnx\_inplace-4, 0.5)
            (cuccaro\_adder-20, 32.107629427792915)
            (takahashi\_adder-20, 30.561349693251532)
            (incrementer\_borrowedbit-5, 0.5)
            (grovers-9, 17.166666666666664)
            ( , nan)
            (geometric mean, 26.288559342717726)
            (  , nan)
            (qft\_adder-16, 0.5)
            (bv-20, 0.5)
            (qaoa\_complete-10, 0.5)
        };
        \addlegendentry{clusters-5x4};
    \addplot[draw=black,dashed,smooth]
    coordinates {(  ,-5) (  ,65)};%
    \end{axis}

\end{tikzpicture}%
    }%
    \caption{A comparison between the baseline and Trios for various hardware.  Above 0\% indicates benefit.  All two-qubit gates (for communication and computation) are counted.  The geometric mean reductions in gate counts are 37\%, 36\%, 48\%, and 26\% respectively.  The rightmost three benchmarks contain zero Toffoli gates so have no change vs.\ the baseline.}%
    \label{fig:benchmark-gate-ratio}
\end{figure}
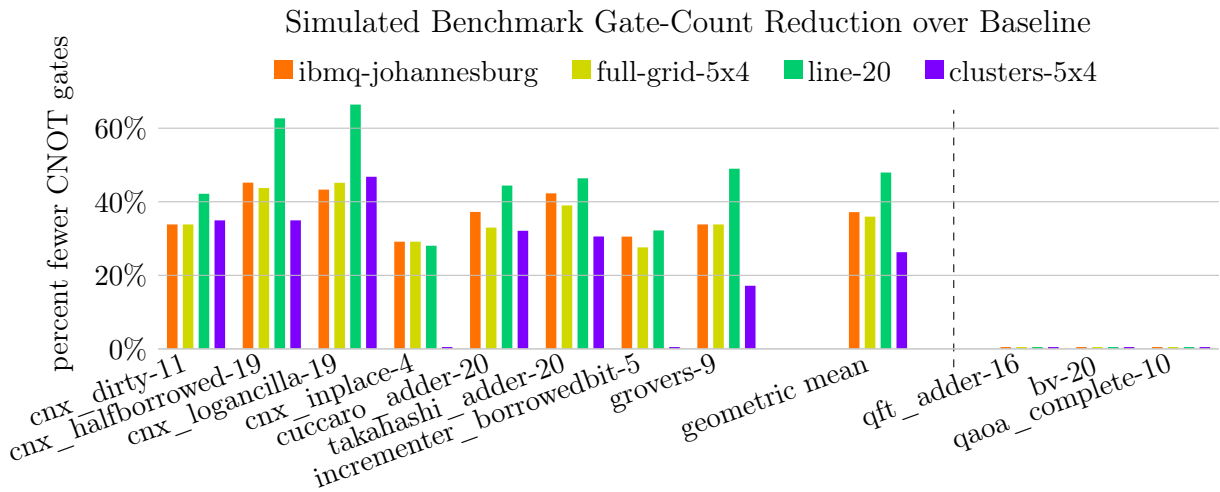
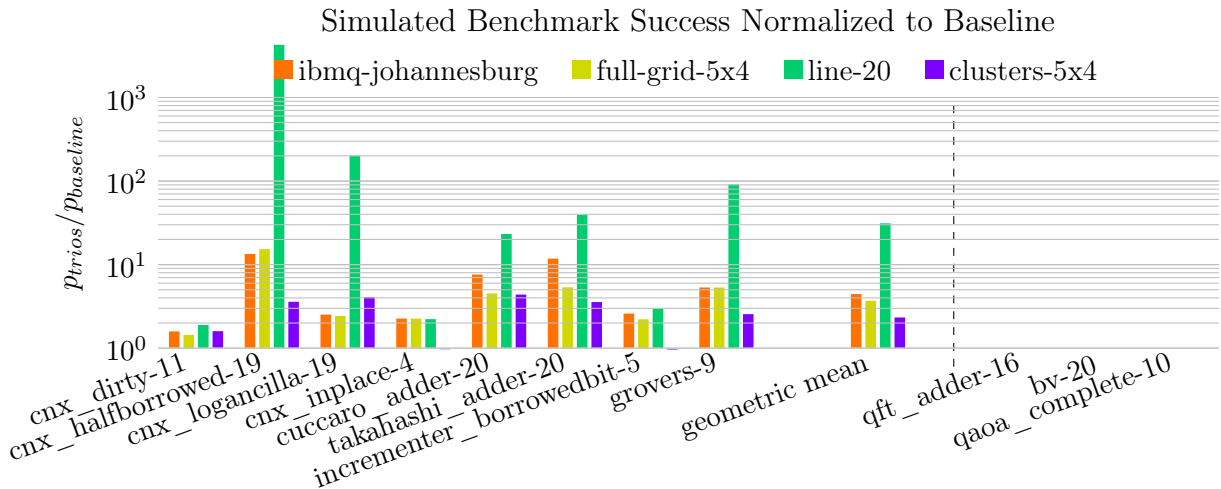
\begin{figure}
    \centering
    \makebox[0.97\textwidth][r]{%
        \begin{tikzpicture}[baseline,scale=1,trim axis left,trim axis right]
\pgfplotsset{every tick label/.append style={font=\small}}
\pgfplotsset{every axis label/.append style={font=\small}}

    \begin{semilogyaxis}[
        name=plot0,
        title={Simulated Benchmark Success Normalized to Baseline},
        xlabel={},
        ylabel={},
        symbolic x coords={cnx\_dirty-11,cnx\_halfborrowed-19,cnx\_logancilla-19,cnx\_inplace-4,cuccaro\_adder-20,takahashi\_adder-20,incrementer\_borrowedbit-5,grovers-9, ,geometric mean,  ,qft\_adder-16,bv-20,qaoa\_complete-10},
        width={0.95*\linewidth},
        height={0.33*\columnwidth},
        ybar={1.5pt},
        bar width={4pt},
        enlargelimits=0.038461538461538464,
        ymin=1, ymax=2999,
        xtick=data,
        ,
        legend style={draw=none, fill=none, at={(0.5,1.03)},anchor=north,font=\small},
        legend columns=-1,
        legend image code/.code={\draw[#1, draw=none] (0em,-0.2em) rectangle (0.6em,0.4em);},
        axis line style={draw=black!20!white},
        axis on top,
        y axis line style={draw=none},
        axis x line*=bottom,
        tick style={draw=none},
        ,
        clip=false,
        enlarge y limits=0,
        ,
        x tick label style={rotate=20, anchor=east},
        grid=none,
        ymajorgrids=true,
        yminorgrids=true, ylabel=$p_{trios}/p_{baseline}$,
        ,
        nodes near coords always on top/.style={
            every node near coord/.append style={
                anchor=south,
                rotate=0,
                font=\small,
                inner sep=0.2em,
            },
        },
        nodes near coords always on top,
    ]
        \addplot[
            style={
                color=transparent,
                draw=none,
                fill={rgb,255:red,255;green,113;blue,0},
                ,
                mark=none,
                ,
                pattern color={rgb,255:red,255;green,113;blue,0},,
            }]
        coordinates {
            (cnx\_dirty-11, 1.5896648545735392)
            (cnx\_halfborrowed-19, 13.407651193733729)
            (cnx\_logancilla-19, 2.524547907696733)
            (cnx\_inplace-4, 2.263988613645191)
            (cuccaro\_adder-20, 7.6082931085593835)
            (takahashi\_adder-20, 11.80367295443326)
            (incrementer\_borrowedbit-5, 2.597857631656044)
            (grovers-9, 5.326082341226377)
            ( , nan)
            (geometric mean, 4.441252548195526)
            (  , nan)
            (qft\_adder-16, 1.0)
            (bv-20, 1.0)
            (qaoa\_complete-10, 1.0)
        };
        \addlegendentry{ibmq-johannesburg~~~~};

        \addplot[
            style={
                color=transparent,
                draw=none,
                fill={rgb,255:red,209;green,216;blue,0},
                ,
                mark=none,
                ,
                pattern color={rgb,255:red,209;green,216;blue,0},,
            }]
        coordinates {
            (cnx\_dirty-11, 1.4377463977511804)
            (cnx\_halfborrowed-19, 15.311806031497524)
            (cnx\_logancilla-19, 2.4307232746263026)
            (cnx\_inplace-4, 2.263988613645191)
            (cuccaro\_adder-20, 4.521922465526685)
            (takahashi\_adder-20, 5.356175676310816)
            (incrementer\_borrowedbit-5, 2.2185736795499724)
            (grovers-9, 5.326082341226377)
            ( , nan)
            (geometric mean, 3.6940031823077475)
            (  , nan)
            (qft\_adder-16, 1.0)
            (bv-20, 1.0)
            (qaoa\_complete-10, 1.0)
        };
        \addlegendentry{full-grid-5x4~~~~};

        \addplot[
            style={
                color=transparent,
                draw=none,
                fill={rgb,255:red,0;green,205;blue,110},
                ,
                mark=none,
                ,
                pattern color={rgb,255:red,0;green,205;blue,110},,
            }]
        coordinates {
            (cnx\_dirty-11, 1.8933437282657255)
            (cnx\_halfborrowed-19, 4258.64023330463)
            (cnx\_logancilla-19, 201.07417925378348)
            (cnx\_inplace-4, 2.227549686978761)
            (cuccaro\_adder-20, 23.224060039305026)
            (takahashi\_adder-20, 39.4009760141775)
            (incrementer\_borrowedbit-5, 2.9791791316994827)
            (grovers-9, 90.98391900034488)
            ( , nan)
            (geometric mean, 31.19060603069423)
            (  , nan)
            (qft\_adder-16, 1.0)
            (bv-20, 1.0)
            (qaoa\_complete-10, 1.0)
        };
        \addlegendentry{line-20~~~~};

        \addplot[
            style={
                color=transparent,
                draw=none,
                fill={rgb,255:red,125;green,0;blue,255},
                ,
                mark=none,
                ,
                pattern color={rgb,255:red,125;green,0;blue,255},,
            }]
        coordinates {
            (cnx\_dirty-11, 1.599641130404093)
            (cnx\_halfborrowed-19, 3.5894014754239083)
            (cnx\_logancilla-19, 4.074161573419985)
            (cnx\_inplace-4, 0.9725673948512388)
            (cuccaro\_adder-20, 4.379170294963949)
            (takahashi\_adder-20, 3.565238909329951)
            (incrementer\_borrowedbit-5, 0.9643362774598649)
            (grovers-9, 2.554455092114703)
            ( , nan)
            (geometric mean, 2.3321217818921745)
            (  , nan)
            (qft\_adder-16, 1.0)
            (bv-20, 1.0)
            (qaoa\_complete-10, 1.0)
        };
        \addlegendentry{clusters-5x4};
    \addplot[draw=black,dashed,smooth]
    coordinates {(  ,0.6) (  ,1000)};%
    \end{semilogyaxis}

\end{tikzpicture}%
    }%
    \caption{Normalized Figure~\ref{fig:benchmark-success} to show our consistent increase in program success with Trios.  Above $10^0$ indicates benefit.  Some improvement factors are huge due to near-zero baseline success rates.  The geometric mean increases in success rate are 4.4x, 3.7x, 31x, and 2.3x respectively.  The rightmost three benchmarks contain zero Toffoli gates so have no change vs.\ the baseline.}%
    \label{fig:benchmark-success-norm}
\end{figure}

\subsection{Trios Routes Complex Interactions Better}

Trios improves gate counts, and consequently improves success rates, by routing more efficiently and choosing more appropriate Toffoli decompositions based on the underlying architecture's connectivity. Current compilers, like Qiskit, perform routing on fully decomposed and unrolled programs, and while this must eventually be done, it leads to less efficient routing policies and relies on assumptions that a theoretically good decomposition (fewest CNOTs) is the best decomposition for the hardware. Trios eliminates this by choosing a context-dependent Toffoli decomposition and routing multi-qubit gates as single units.

Trios greatly improves effectiveness compared to a \textit{heuristic-based} compiler by applying similar heuristics to the higher abstraction level Toffoli gates.  An optimal routing of the decomposed circuit would be better except it cannot select the best architecture location-specific decomposition.  This makes a huge difference specifically with Toffolis on any square-grid-based device.  One might choose to improve the solution found by an optimal compiler by always decomposing Toffolis to the 8-CNOT version before optimally routing, but this will still limit the solution.  There are multiple possible qubit orders for the decomposition and the best can only be selected after the routing pass.

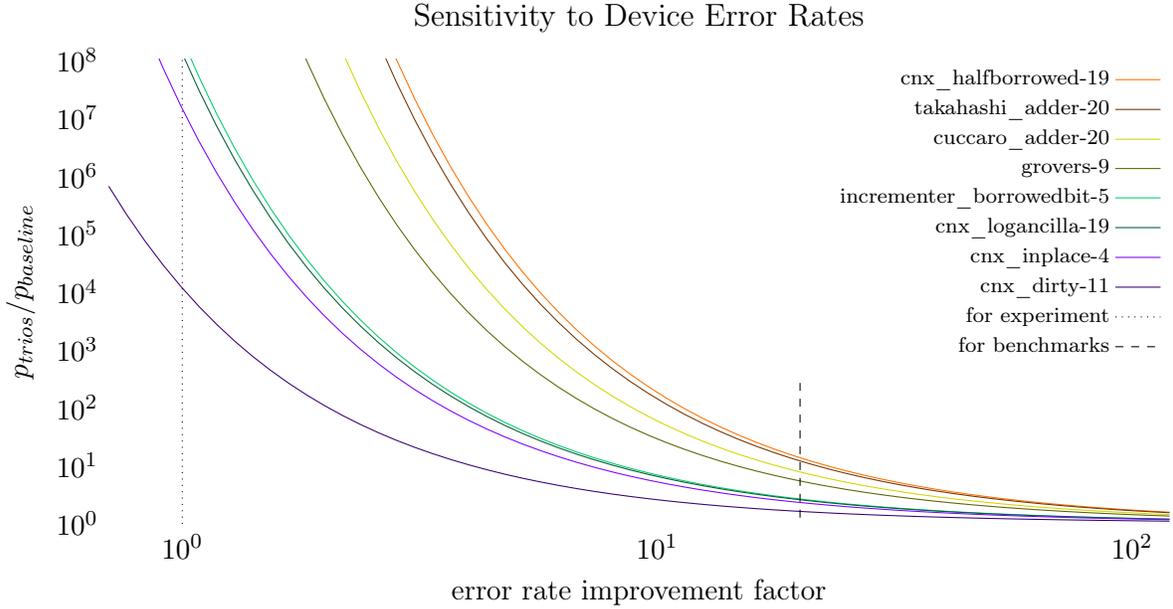
\begin{figure}
    \centering
    \makebox[0.95\columnwidth][r]{%
        \begin{tikzpicture}[baseline,scale=1,trim axis left,trim axis right]
\pgfplotsset{every tick label/.append style={font=\small}}
\pgfplotsset{every axis label/.append style={font=\small}}

    \begin{loglogaxis}[
        name=plot9,
        title={Sensitivity to Device Error Rates},
        xlabel={error rate improvement factor},
        ylabel={},
        width={0.95*\columnwidth},
        height={0.47*\columnwidth},
        xmin=0.7, xmax=120, ymin=0.99, ymax=102329299.22807537,
        ,
        legend style={
            draw=none,
            at={(1,1)},
            anchor=north east,
            font=\small},
        ,
        clip=false,
        axis line style={draw=none},
        tick style={draw=none},,
        ylabel=$p_{trios}/p_{baseline}$,
    legend style={
        cells={anchor=east},
        legend plot pos=right,
        draw=none,
        at={(1,1)},
        anchor=north east,
        fill=none,
        font=\small},
    clip=true,
    ]
        \addplot[color={rgb,255:red,255;green,113;blue,0}] table[x=error-rate-improvement-factor 0, y=scriptsizecnxhalfborrowed-19  0, col sep=comma]
            {chapters/modular/data/sensitivity-to-device-error-rates.csv}
        ;%
        \addlegendentry{\scriptsize{cnx\_halfborrowed-19}};

        \addplot[color={rgb,555:red,255;green,113;blue,0}] table[x=error-rate-improvement-factor 1, y=scriptsizetakahashiadder-20  1, col sep=comma]
            {chapters/modular/data/sensitivity-to-device-error-rates.csv}
        ;%
        \addlegendentry{\scriptsize{takahashi\_adder-20}};

        \addplot[color={rgb,255:red,209;green,216;blue,0}] table[x=error-rate-improvement-factor 2, y=scriptsizecuccaroadder-20  2, col sep=comma]
            {chapters/modular/data/sensitivity-to-device-error-rates.csv}
        ;%
        \addlegendentry{\scriptsize{cuccaro\_adder-20}};

        \addplot[color={rgb,555:red,209;green,216;blue,0}] table[x=error-rate-improvement-factor 3, y=scriptsizegrovers-9  3, col sep=comma]
            {chapters/modular/data/sensitivity-to-device-error-rates.csv}
        ;%
        \addlegendentry{\scriptsize{grovers-9}};

        \addplot[color={rgb,255:red,0;green,205;blue,110}] table[x=error-rate-improvement-factor 4, y=scriptsizeincrementerborrowedbit-5  4, col sep=comma]
            {chapters/modular/data/sensitivity-to-device-error-rates.csv}
        ;%
        \addlegendentry{\scriptsize{incrementer\_borrowedbit-5}};

        \addplot[color={rgb,555:red,0;green,205;blue,110}] table[x=error-rate-improvement-factor 5, y=scriptsizecnxlogancilla-19  5, col sep=comma]
            {chapters/modular/data/sensitivity-to-device-error-rates.csv}
        ;%
        \addlegendentry{\scriptsize{cnx\_logancilla-19}};

        \addplot[color={rgb,255:red,125;green,0;blue,255}] table[x=error-rate-improvement-factor 6, y=scriptsizecnxinplace-4  6, col sep=comma]
            {chapters/modular/data/sensitivity-to-device-error-rates.csv}
        ;%
        \addlegendentry{\scriptsize{cnx\_inplace-4}};

        \addplot[color={rgb,555:red,125;green,0;blue,255}] table[x=error-rate-improvement-factor 7, y=scriptsizecnxdirty-11  7, col sep=comma]
            {chapters/modular/data/sensitivity-to-device-error-rates.csv}
        ;%
        \addlegendentry{\scriptsize{cnx\_dirty-11}};

        \addplot[color=black, dotted] table[x=error-rate-improvement-factor 8, y=scriptsizefor-experiment  8, col sep=comma]
            {chapters/modular/data/sensitivity-to-device-error-rates.csv}
        ;%
        \addlegendentry{\scriptsize{for experiment}};

        \addplot[color=black, dashed] table[x=error-rate-improvement-factor 9, y=scriptsizefor-benchmarks  9, col sep=comma]
            {chapters/modular/data/sensitivity-to-device-error-rates.csv}
        ;%
        \addlegendentry{\scriptsize{for benchmarks}};

    \end{loglogaxis}

\end{tikzpicture}%
    }%
    \caption{Factor of improvement in success rate in Trios over baseline for scaling gate error rates. The dotted line indicates current error rates on IBM Johannesburg and the dashed line (20x improvement) indicates values of the near future used in simulation. In our approximation of success rate factors of improvement in gate error rates lead to an exponential fall off in success ratios, as expected. In the very near term, we expect Trios to drastically improve the execution of quantum programs.}%
    \label{fig:benchmark-error-sensitivity}
\end{figure}

\subsection{Simulation Sensitivity to Error Rates}

For our simulations we use an error model (20x better than current errors on Johannesburg) which is forward looking. As errors improve, we expect Trios to have a reduced impact on program success rates since gate errors will contribute less and less to program failure though Trios will never perform worse than the baseline. In Figure~\ref{fig:benchmark-error-sensitivity} we study the sensitivity of simulation results to two qubit error rates beginning with current IBM error rates. For poor error rates, the benefit of Trios is extremely large, owed to the fact that programs compiled with the baseline have probabilities of success very close to 0. In our simplified simulation framework, as error rates improve we expect an exponential drop off in improvement with the most advantage obtained with current error rates.

\FloatBarrier{}

\section{Summary}%
\label{sec:modular-conclusion}

We present a new quantum compilation structure, Trios, with a split decomposition pass to greatly reduce compiled communication cost and enable architecture-tuned decompositions. We specifically target the three-qubit Toffoli operation to capture program structure enabling more optimal compiled circuits.  Because current quantum computers are especially error prone, they require high levels of optimization to reduce gate counts and maximize the probability the compiled program will succeed.
Prior optimization strategies discarded hierarchy to better maximize flat program optimization but Trios shows that the extra structural information from program hierarchy can be more helpful than the increased flexibility of a flattened program structure.

Orchestrated Trios both greatly improves the effectiveness of qubit routing given newly exposed program structure and, additionally, improves decompositions with a connectivity-aware second pass.
These both greatly benefit the program success rate, a critical metric for today's error-prone and resource-constrained quantum computers.
We hope this inspires more hierarchically designed NISQ algorithms now that we have shown that keeping the abstraction of hierarchy and reorganizing compilation passes can help bridge the gap between these noisy quantum hardware and practical applications.

\FloatBarrier{}

\chapter{Conclusion}%
\label{chapter:conclusion}

From the three cases we present and evaluate, we find that the right abstraction enables system design, compiler design, and program design to align.
Each of the abstraction's representation of quantum computation preserved the data locality of the application in ways that reduced communication costs during program execution.
This aspect of keeping related data close-by emerged from fundamental constraints of quantum computing technology and we found useful abstractions that worked within this.
First, we treat additional logical levels as scratch space to enable program subroutines with simplified communication patterns.
Second, qubit-local memory enables a third spacial dimension, improving fault-tolerant operations and fault-tolerant data movement.
And finally, by re-introducing a small amount of hierarchical program abstraction, we can greatly improve compiler heuristics and enable smarter compiler passes.

\section{Future Abstractions}%
\label{section:future-abstractions}

The abstractions in this dissertation are just the tip of the iceberg of what will eventually become a cohesive set of high-level concepts for understanding and describing a quantum algorithm or quantum program.
Quantum programmers, quantum compiler designers, and quantum architects use the models they have to understand and manipulate quantum algorithms at an abstract level.
Below, we consider some important categories of future abstractions and current research towards new abstractions.

\subsection{Programmer Abstractions}

The job of a programmer is to describe an algorithm as a concrete program, or list of instructions.
This program describes precisely, step-by-step, how a computer should execute the algorithm.
However, computers can only follow extremely simple instructions that would be tedious for the computer programmer to write down one-by-one.
This is why modern classical programming languages represent high-level concepts such as variables, functions, loops, and threads that can be translated into simple instructions.
Abstractions like variables, functions, loops, and threads work well to express classical algorithms, making the translation process from algorithm to program as intuitive as possible.

The majority of quantum programmers right now describe their algorithms in a \emph{quantum assembly language}, a list of basic instructions that a quantum computer can execute.
Common abstractions used are named qubits and subroutines (similar to variables and non-recursive functions in classical programming) but this is not enough.
High-level concepts common to multiple quantum algorithms are a useful starting point to build new quantum program abstractions.

Many fault-tolerant algorithms like Shor's factoring and Grover's database search use classical subroutines or oracles.  Languages like Q\# (\cite{svore2018q,singhal2022q}) make this easier for quantum programmers by automatically writing the uncomputation and inverse subroutines.

The no-cloning theorem of quantum physics (\cite{mikeike}) means that quantum data cannot be copied.
Classically, data copying happens implicitly, all the time in function calls and variable assignments for example.
When a quantum programming language is based on a classical programming language, this makes it very easy to mistakenly write a quantum program the violates the no-cloning theorem.
Ideally, when the limitations of the programming language align with the limitations of the computer, programs will be easier to write and easier to understand.
Research in programming language type theory has found type systems to check if a quantum program uses its qubits correctly to not violate the no-cloning theorem of quantum physics, \cite{fu2020linear}.
Our goal should not be to have a language where no-cloning is enforced but one where the program structure implicitly assumes no-cloning.
For example, a data flow-like language where every subroutine output only connects to a single subroutine input would make is impossible to the physics of quantum information.

Some types of quantum algorithms, especially those performing Hamiltonian simulation, have many subroutines that \emph{commute}.
When two subroutines commute, either one can be executed first with the same result.
The compiler presented in \cite{lao20222qan} is told which subroutines it is allowed to reorder.
The flexibility to reorder subroutines allows the compiler to find much more efficient ways to execute the quantum program.
\cite{lao20222qan} demonstrates the advantage of reordering flexibility at the compiler level, but a valuable addition to a quantum programming language would be a syntax to concisely represent commutation relations between Hamiltonian terms or other subroutines.

\subsection{Intermediate Representations}

Intermediate representations of quantum programs are used by quantum compilers during the process of translating a high-level quantum programming language into basic instructions ready to execute on a quantum computer.
The most common intermediate representation is the quantum circuit as used throughout this dissertation.
Sometimes blocks of the circuit are annotated by the compiler to indicate high-level properties of the circuit that are invisible when considering the circuit.
An example of this is the Hamiltonian term compiler described earlier where annotations of circuit blocks that can be reordered allows the compiler to make optimizations to the circuit that would otherwise be computationally intractable to find.
Orchestrated Trios (Section \ref{chapter:modular}) relies directly on different forms of the quantum circuit as intermediate representations between each of its passes.

Quantum circuits are the de facto intermediate representation but they may actually be limiting.
A model of quantum computation called Measurement Based Quantum Computation (\cite{briegel2009measurement}) presents an alternative to executing gates on qubits.
In MBQC, a large entangled \emph{graph state} is prepared and computation is performed by executing single-qubit gates and measurements one at a time, conditioned on earlier measurements.
For this model of computation, a measurement graph represents the quantum program instead of a quantum circuit.
A measurement graph is an open graph (with ordered input edges and output edges) containing nodes and edges where each node is parameterized by an angle and a measurement basis.
Efficient algorithms exist to convert between quantum circuits and (some) measurement graphs (\cite{mhalla2008finding, backens2021there}), enabling us to use whichever representation best enables a particular compiler pass.

ZX diagrams are closely related to measurement graphs but have simpler structure and well-studied rewrite rules (a ZX Calculus) that preserve their meaning while modifying their structure (\cite{van2020zx}).
Circuit optimizations written for ZX diagrams are often simpler and produce better results than equivalent optimizations written for quantum circuits (\cite{kissinger2019reducing}).

ZX diagrams are currently only used for individual compiler passes and converted back to circuits but they may eventually replace quantum circuits as the primary representation of quantum programs.
Particularly when compiling algorithms for fault tolerant quantum computers, ZX diagrams are promising because of the close correspondence between ZX diagram nodes and Lattice Surgery merge and split operations (\cite{de2020zx}).
The ZX calculus does not represent measurement outcomes or ancilla well, which can limit its uses to ancilla-free unitary blocks of a quantum algorithm.
However, the study of the ZX calculus has motivated study of other graphical calculi that may overcome these limitations (\cite{chardonnet2022many}).
These calculi are more flexible than a quantum circuit, enabling easier optimization, but a key property to note is their lack of an order of operations, data flow, or causality.
Quantum physics itself often disregards the direction of time as exemplified in quantum teleportation where, one interpretation says, the quantum data flows backward in time through the Bell pair to reach the target.
There is much work to be done, but these diagrams bring us closer to a quantum program representation that expresses the underlying physics in an intuitive way.

%


\section{Outlook}%
\label{section:outlook}

These future ideas for abstractions, along with those evaluated in the body of this dissertation, will eventually be an entire ecosystem that fits together and integrates with the quantum computer scientist's programming languages, hardware stack, and toolchain.
As qubit technology improves, as new error correction protocols are developed, and as new algorithms are invented, our abstractions may need to be modified or replaced and co-designed with the technology stack.
As the stack evolves, the paradigm around quantum computing will continue to grow into its own, fundamentally new niche and become more than a classical computer under superposition.


\makebibliography{}


\end{document}